\pdfoutput=1

\documentclass[11pt,twoside,a4paper,cmspaper,final,collab]{cms-tdr}
\def\svnVersion{364817}\def\cmsCernNoTag{CERN-EP/2016-040}\def\cmsCernDate{\today}\def\cmsMessage{Published in the European Physical Journal C as \href{http://dx.doi.org/10.1140/epjc/s10052-016-4292-5}{\doi{10.1140/epjc/s10052-016-4292-5.}}}

\begin{document}\cmsNoteHeader{SUS-13-023}

\hyphenation{had-ron-i-za-tion}
\hyphenation{cal-or-i-me-ter}
\hyphenation{de-vices}
\RCS$HeadURL: svn+ssh://svn.cern.ch/reps/tdr2/papers/SUS-13-023/trunk/SUS-13-023.tex $
\RCS$Id: SUS-13-023.tex 364468 2016-08-12 14:28:00Z nmccoll $
\newlength\cmsFigWidth
\ifthenelse{\boolean{cms@external}}{\setlength\cmsFigWidth{0.85\columnwidth}}{\setlength\cmsFigWidth{0.4\textwidth}}
\ifthenelse{\boolean{cms@external}}{\providecommand{\cmsLeft}{top}}{\providecommand{\cmsLeft}{left}}
\ifthenelse{\boolean{cms@external}}{\providecommand{\cmsRight}{bottom}}{\providecommand{\cmsRight}{right}}
\providecommand{\NA}{\text{---}\xspace}
\newlength{\cmsTabSkip}\setlength\cmsTabSkip{1.5ex}
\ifthenelse{\boolean{cms@external}}{\providecommand{\cmsTableResize[1]}{\relax{
#1}}}{\providecommand{\cmsTableResize[1]}{\resizebox{\textwidth}{!}{#1}}}
\newcommand{\POWHEGONE} {{\POWHEG (version 1.0 r1380)}\xspace}
\newcommand{\HERWIGSIX} {{\HERWIG (version 6.520)}\xspace}
\newcommand{\MADGRAPHFIVE} {{\MADGRAPH (version 5.1.3.30)}\xspace}
\newcommand{\MCATNLOTHREE} {{\MCATNLO (version 3.41)}\xspace}
\newcommand{\PYTHIASIX} {{\PYTHIA (version 6.4.26)}\xspace}
\newcommand{\TAUOLATWEN}{{\TAUOLA (version 27.121.5)}\xspace}
\newcommand{\CORRAL} {\textsc{corral}\xspace}
\newcommand{\ptJet}{\ensuremath{\ptvec^{\,\text{jet}}}}
\newcommand{\mCHG}{\ensuremath{m_{\PSGcpm}}\xspace}
\newcommand{\mStop}{\ensuremath{m_{\PSQt}}\xspace}
\newcommand{\mLSP}{\ensuremath{m_{\PSGczDo}}\xspace}
\newcommand{\pti}[1]{\ensuremath{p_{\mathrm{T},#1}}\xspace}
\newcommand{\zll}{\ensuremath{\Z \to \ell^{+}\ell^{-}}\xspace}
\newcommand{\mt}{\ensuremath{m_{\mathrm{T}}}\xspace}
\newcommand{\nj}{\ensuremath{N_{\textrm{j}}}\xspace}
\newcommand{\nb}{\ensuremath{N_{\PQb}}\xspace}
\newcommand{\nt}{\ensuremath{N_{\PQt}}\xspace}

\cmsNoteHeader{SUS-13-XXX} 
\title{Search for direct pair production of supersymmetric top quarks decaying to all-hadronic final states in pp collisions at $\sqrt{s} = 8\TeV$}
\titlerunning{Search for direct pair production of supersymmetric top quarks\ldots}

\date{\today}

\abstract{Results are reported from a search for the pair production of top squarks, the supersymmetric partners of top quarks,
in final states with jets and missing transverse momentum. The data sample used in this search was collected by the CMS detector and corresponds to an integrated luminosity of 18.9\fbinv of proton-proton collisions at a centre-of-mass energy of 8\TeV produced by the LHC. The search features novel background suppression and prediction methods, including a dedicated top quark pair reconstruction algorithm. The data are found to be in agreement with the predicted backgrounds. Exclusion limits are set in simplified supersymmetry models with the top squark decaying to jets and an undetected neutralino, either through a top quark or through a bottom quark and chargino. Models with the top squark decaying via a top quark are excluded for top squark masses up to 755\GeV in the case of  neutralino masses below 200\GeV. For decays via a chargino, top squark masses up to 620\GeV are excluded, depending on the masses of the chargino and neutralino.}

\hypersetup{%
pdfauthor={CMS Collaboration},%
pdftitle={Search for direct pair production of supersymmetric top quarks decaying to all-hadronic final states in pp collisions at sqrt(s) = 8 TeV},%
pdfsubject={CMS},%
pdfkeywords={CMS, physics, Supersymmetry, SUSY, scalar top, stop}}

\maketitle 

\section{Introduction}
The standard model (SM) of particle physics is an extremely powerful framework for the description of the known elementary particles and their interactions.  Nevertheless, the existence of dark matter \cite{Zwicky:1933gu,Rubin:1970zza,Drees:2012ji} inferred from astrophysical observations, together with a wide array of theoretical considerations, all point to the likelihood of physics beyond the SM. New physics could be in the vicinity of the electroweak (EW) scale and accessible to experiments at the CERN LHC \cite{LHC}.  In addition, the recent discovery of a Higgs boson  \cite{Aad:2012tfa,Chatrchyan:2012ufa,Chatrchyan:2013lba} at a mass of 125\GeV \cite{Aad:2014aba,Khachatryan:2014jba,Aad:2015zhl}  has meant that the hierarchy problem, also known as the `fine-tuning' or `naturalness' problem \cite{'tHooft:1979bh,Witten:1981nf,Dine:1981za,Dimopoulos:1981au,Dimopoulos:1981zb,Kaul:1981hi}, is no longer hypothetical.

A broader theory that can address many of the problems associated with the SM is supersymmetry (SUSY) \cite{Barbieri:1982eh,Dawson:1983fw,Nilles:1983ge,Haber:1984rc,Chung:2003fi}, which postulates a symmetry  between fermions and bosons.  In particular, a SUSY particle (generically referred to as a `sparticle' or  `superpartner') is proposed for each SM particle. A sparticle is expected to have the same couplings and quantum numbers as its SM counterpart with the exception of spin, which differs by a half-integer. Spin-$1/2$ SM fermions (quarks and leptons) are thus paired with spin-0 sfermions (the squarks and sleptons). There is a similar, but slightly more complicated pairing for bosons; SUSY models have extended Higgs sectors that contain neutral and charged higgsinos that mix with the SUSY partners of the neutral and charged EW gauge bosons, respectively. The resulting mixed states are referred to as neutralinos $\PSGcz$  and charginos $\PSGcpm$.

Supersymmetry protects the mass of the Higgs boson against divergent quantum corrections associated with virtual SM particles by providing cancellations via the corresponding  corrections for virtual superpartners \cite{Barbieri:1987fn,de_Carlos:1993yy,Dimopoulos:1995mi,Barbieri:1995uv}.  Since no sparticles have been observed to date, they are generally expected to be more massive than their SM counterparts. On the other hand, sparticle masses cannot be arbitrarily large if they are to stabilise the Higgs boson mass without an unnatural level of fine-tuning.  This is particularly important for the partners of the third generation SM particles that have large Yukawa couplings to the Higgs boson \cite{Sakai:1981gr,Papucci:2011wy,Brust:2011tb,Delgado:2012eu}. The top and bottom squarks ($\PSQt$ and $\PSQb$), are expected to be among the lightest sparticles and potentially the most accessible at the LHC, especially when all other constraints are taken into consideration \cite{Feng:2013pwa,Papucci:2011wy}. With conservation of  R-parity \cite{Wess:1974tw,Farrar:1978xj}, SUSY particles are produced in pairs and the lightest SUSY particle (LSP) is stable. If the lightest weakly interacting neutralino ($\PSGczDo$) is the stable LSP, it is a leading candidate for dark matter~\cite{Feng:2010gw}. Based upon these considerations, it is of particular interest at the LHC to look for evidence of the production of $\PSQt\PASQt$ with decay chains of the $\PSQt$ and $\PASQt$ ending in SM particles and LSPs.  The latter do not interact with material in the detector and so must have their presence inferred from missing transverse momentum \ptvecmiss, which in each event is defined as the projection of the negative vector sum of the momenta of all reconstructed particles onto the plane perpendicular to the beam line. Its magnitude is referred to as \ETm.

\begin{figure}[htb]
\centering
    \includegraphics[width=0.49\textwidth]{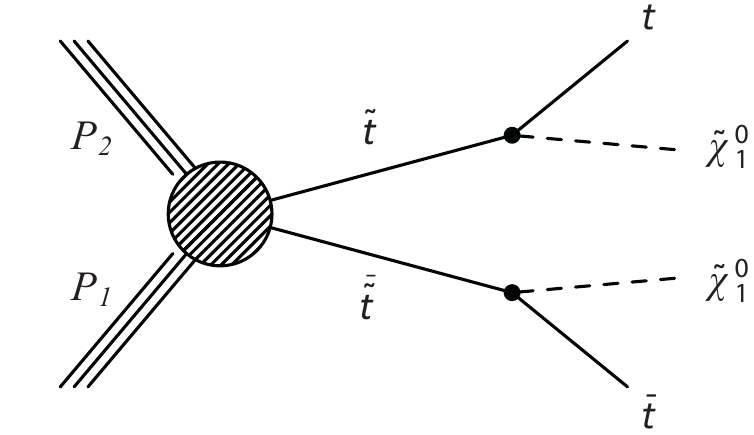}
    \includegraphics[width=0.49\textwidth]{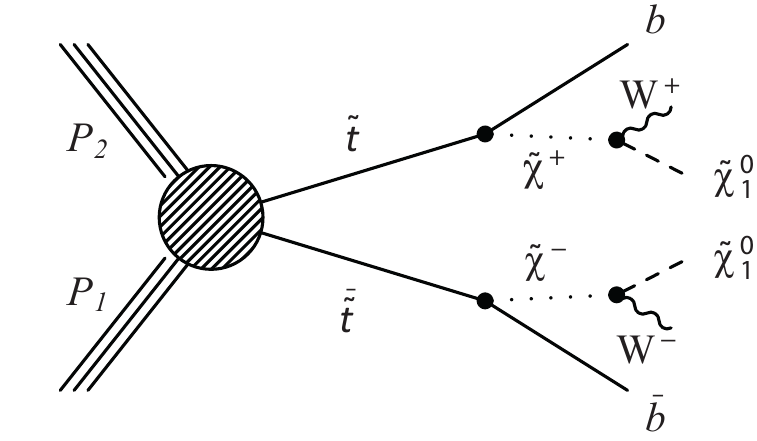}
  \caption{Diagrams representing the two simplified models of direct top squark pair production considered in this study: T2tt with top squark decay via a top quark (left) and T2bW with top squark decay via a chargino (right).}
  \label{fig:SMS}
\end{figure}

Within the Simplified Model Spectra (SMS) framework~\cite{Alwall:2008ag,Alwall:2008va,Alves:2011wf} the study presented here considers two broad classes of signals that lead to a $\bbbar\PQq\PQq\PAQq\PAQq+\ETm$ final state via decay modes denoted T2tt and T2bW. These are defined, respectively, as (i) $\PSQt$ decay to a top quark: $\PSQt\to\PQt\PSGczDo\to\PQb\PWp\PSGczDo$, and (ii) $\PSQt$ decay via a chargino: $\PSQt\to\PQb\PSGcp\to\PQb\PWp\PSGczDo$. Figure~\ref{fig:SMS} shows the diagrams representing these two simplified models. The two decay modes are not mutually exclusive, and it is possible for one of the top squarks to decay as in T2tt and the other as in T2bW. However, such a scenario is not considered in the analysis presented here.

Only the lightest $\PSQt$ mass eigenstate is assumed to be involved, although the results are equivalent for the heavier eigenstate. The polarization of the $\PSQt$ decay products depends on the properties of the SUSY model, such as the left and right $\PSQt$ mixing~\cite{Perelstein:2008zt,Low:2013aza}. Instead of choosing a specific model, each SMS scenario is assumed to have unpolarized decay products and has a 100\% branching ratio to the final state under consideration.  As such, the results can be interpreted, with appropriately rescaled branching fractions, in the context of any SUSY model in which these decays are predicted to occur.

With event characteristics of these signals in mind, we have developed a search for pair production of top squarks with decays that result in a pair of LSPs in the final state in addition to SM particles. Two selection criteria address the desire to extract a potentially very small signal from a sample dominated by top quark pair events. The first criterion comes from the \ETm signature associated with the  LSPs, which motivates the focus  on all-hadronic final states,  as this eliminates large sources of SM background events with genuine \ETm from neutrinos in leptonic W decays. The all-hadronic final state with \ETm constitutes 45\% of the signal because W bosons decay to quarks with a 67\% branching ratio. For the same reason this final state makes up an even higher proportion of the subset of events with high jet multiplicity including many jets with high transverse momentum, \pt, that is often required in SUSY searches to eliminate SM backgrounds. The second criterion relies upon the identification of top quark decay products  to eliminate such backgrounds as SM production of W bosons in association with jets. Together, these criteria define a preselection region consisting of events that pass stringent vetoes on the presence of charged leptons, and are required to have large \ETm, two tagged b quark jets, and four additional jets from the hadronisation and decay of light quarks.

In spite of these stringent requirements, the low production cross sections of new physics signals mean that they are easily overwhelmed by SM backgrounds. In the case of SUSY, for example, the cross section for the production of top squark pairs with $\mStop=800\GeV$ is predicted to be nearly five orders of magnitude smaller than that of top quark pairs~\cite{Beenakker:1997ut}. For this reason, this analysis focuses heavily on background suppression, employing several new methods that improve sensitivity to signal. The relevant SM processes contributing to this analysis fall into four main categories: (i) top quark and W boson events where the W decays leptonically, thereby contributing genuine \ETm, but the lepton is not successfully reconstructed or identified, or it is outside the acceptance of the detector; (ii) invisible decays of the Z boson when produced in association with jets, $\Z$+jets with $Z\to \nu\PAGn$;  (iii) QCD multijet production, which, due to its very high rate, can produce events with substantial \ETm in the very rare cases of either extreme mismeasurements of jet momenta or the leptonic decay of heavy-flavour hadrons with large neutrino $\pt$; and (iv) $\ttbar\Z$ production (with $\Z\to \nu\PAGn$), which is an irreducible background to signals with top squark decays via on-shell top quarks. The $\ttbar\Z$ process has a small cross section that has been measured by  ATLAS and CMS to be $176^{+58}_{-52}\fbinv$~\cite{Aad:2015eua} and $242^{+65}_{-55}\fbinv$~\cite{Khachatryan:2015sha}, respectively.

The first step in developing the search is the construction of a set of optimised vetoes for all three lepton flavours that reduce SM backgrounds for both signal types. Next, specific features of each signal type are exploited by combining several variables in a multivariate analysis (MVA) based upon Boosted Decision Trees (BDT). For T2tt, a high performance hadronic top quark decay reconstruction algorithm is developed and used to facilitate discrimination of signal from background by using details of top quark kinematics.

Previous searches in the leptonic as well as the hadronic channels place limits on models with $\mStop < 750\GeV$ for $\mLSP < 100\GeV$ and have sensitivity to some models with $\mLSP < 280\GeV$~\cite{Khachatryan:2015pwa,Aad:2015pfx}. Previous searches for top and bottom squark pair production at the LHC are presented in Refs. \cite{Aad:2012ywa,Aad:2012xqa,Aad:2012uu,Aad:2014qaa,Aad:2013ija,Aad:2014mfk,Aad:2014nra,Chatrchyan:2013xna,Chatrchyan:2013lya,Chatrchyan:2012paa,Chatrchyan:2013mya,Khachatryan:2014doa,CMS:2014dpa,Khachatryan:2015wza,Aad:2014kra,Aad:2014mha,Aad:2014bva,Khachatryan:2015pwa,Aad:2015pfx}. Previous searches at the Tevatron are presented in Refs. \cite{Aaltonen:2010uf,Aaltonen:2010dy,Aaltonen:2012tq,Aaltonen:2009sf,Abazov:2010xm,Abazov:2010wq,Abazov:2008rc,Abazov:2012cz}. The analysis reported here significantly extends the sensitivity of a previous CMS analysis~\cite{Khachatryan:2015wza} using this dataset by means of more refined background controls and enhanced signal retention techniques.

This paper is organised as follows: Section \ref{sec:detector} describes the CMS detector, while Section \ref{sec:eventselection} discusses event reconstruction, event selection, and Monte Carlo (MC) simulations of signal and background.  The top quark pair reconstruction algorithm and lepton vetoes are described in Sections \ref{sec:toptagger} and \ref{sec:leptonvetoes}, respectively. The search regions are discussed in Section \ref{sec:searchregions}, and the evaluation of backgrounds is presented in Section \ref{sec:background} along with a discussion of the method of MC reweighting. Final results and their interpretations are presented in Section \ref{sec:results}, followed by a summary in Section \ref{sec:conclusion}.

\section{CMS detector}
\label{sec:detector}
The central feature of the CMS apparatus is a superconducting solenoid of 6\unit{m} internal diameter, providing a magnetic field of 3.8\unit{T}. Within the  solenoid volume are a silicon pixel and strip tracker, a lead tungstate crystal electromagnetic calorimeter (ECAL), and a brass and scintillator hadron calorimeter (HCAL), each composed of a barrel and two endcap sections. Extensive forward calorimetry complements the coverage provided by the barrel and endcap detectors. Muons are measured in gas-ionization detectors embedded in the steel flux-return yoke outside the solenoid.

The silicon tracker measures charged particles within the range $\abs{\eta} < 2.5$. Isolated particles of $\pt=100\GeV$ emitted with $\abs{\eta} < 1.4$ have track resolutions of 2.8\% in $\pt$ and 10\,(30)\mum in the transverse (longitudinal) impact parameter \cite{Chatrchyan:2014fea}. The ECAL and HCAL measure energy deposits in the range $\abs{\eta} < 3$. Quartz-steel forward calorimeters extend the coverage to $\abs{\eta} = 5$. The HCAL, when combined with the ECAL, measures jets with a resolution $\Delta E/E \approx 100\%/ \sqrt{E} [\GeVns{}] \oplus 10\%$ \cite{1748-0221-6-11-P11002}. Muons are measured in the range $\abs{\eta} < 2.4$. Matching muons to tracks measured in the silicon tracker results in a relative \pt resolution for muons with $20 < \pt < 100\GeV$ of 1.3--2.0\% in the barrel and better than 6\% in the endcaps. The $\pt$ resolution in the barrel is better than 10\% for muons with $\pt$ up to 1\TeV \cite{Chatrchyan:2012xi}.

The events used in the search presented here were collected using the CMS two-tiered trigger system:  a hardware-based level-1 trigger and a software-based high-level trigger. A more complete description of the CMS detector, together with a definition of the coordinate system used and the relevant kinematic variables, can be found in Ref.~\cite{Chatrchyan:2008zzk}.

\section{Data sample and event selection}
\label{sec:eventselection}
This search uses data corresponding to an integrated luminosity of $18.9\pm0.5\fbinv$ collected at a centre-of-mass energy of 8\TeV~\cite{CMS:2013gfa}.  Events are reconstructed with the CMS particle-flow (PF) algorithm~\cite{CMS:2009nxa,CMS:2010eua}. Each particle is identified as a charged hadron, neutral hadron, photon, muon, or electron by means of  an optimised combination of information from the tracker, the calorimeters, and the muon systems. The energy of a photon is obtained from the ECAL measurement, corrected for zero suppression effects. For an electron the energy is determined from a combination of its estimated momentum at the primary interaction vertex as determined by the tracker, the energy of the corresponding ECAL cluster, and the energy sum of all bremsstrahlung photons spatially compatible with originating from the electron track~\cite{Khachatryan:2015hwa}. Muon momentum is obtained from the curvature of the corresponding track. The energy of charged hadrons is determined from a combination of the momentum measured in the tracker and the matching ECAL and HCAL energy deposits, corrected for zero-suppression effects and for the response function of the calorimeters to hadronic showers. Charged hadrons associated with vertices other than the primary vertex, defined as the pp interaction vertex with the largest sum of charged-track $\pt^2$ values, are not considered. Finally, the energies of neutral hadrons are obtained from the corresponding corrected ECAL and HCAL energies.

Particles reconstructed with the CMS PF algorithm are clustered into jets by the anti-\kt algorithm~\cite{Cacciari:2008gp, Cacciari:2011ma} with a distance parameter of 0.5 in the $\eta$-$\phi$ plane. For a jet, the momentum is determined as the vectorial sum of all associated particle momenta and is found from MC simulated data to be within 5--10\% of the true momentum of the generated particle from which the jet originates over the whole \pt spectrum and detector acceptance. An offset correction determined for each jet via the average \pt density per unit area and the jet area is applied to jet energies to take into account the contribution from pileup, defined as the additional proton-proton interactions within the same or adjacent bunch crossings~\cite{1748-0221-6-11-P11002}. Jet energy corrections are derived from simulated events and are confirmed with in situ measurements of the energy balance in dijet and photon+jet events. Additional selection criteria are applied to each event to remove spurious jet-like features originating from isolated noise patterns in certain HCAL regions~\cite{Chatrchyan:2009hy}.

Jets referred to as `picky jets' are the input to the Comprehensively Optimised Resonance Reconstruction ALgorithm (\CORRAL)  for top quark reconstruction. The picky jet reconstruction algorithm is not constrained to any fixed characteristic width or cutoff and therefore is optimized for clustering the particles associated with the b quark and quarks from the $\PW$ boson. This leads to an improvement in the reconstruction of top quark decays with a wide range of Lorentz boosts, as expected in signal events. The \CORRAL and picky jet algorithms are described in Section~\ref{sec:toptagger}.

Jets are identified as originating from the hadronisation of a bottom quark (b-tagged) by means of the CMS combined secondary vertex (CSV) tagger~\cite{Chatrchyan:2012jua,CMS:2013vea}.  The standard CMS ``tight" operating point for the CSV tagger is used~\cite{Chatrchyan:2012jua}, which has approximately 50\% b tagging efficiency, 0.1\% light flavour jet misidentification rate, and an efficiency of 5\% for c quark jets.

Several simulated data samples based on MC event generators are used throughout this analysis. Signal samples are produced using the \MADGRAPHFIVE \cite{madgraph} event generator with CTEQ6L~\cite{cteq} parton distribution functions (PDFs). For both the T2tt  and T2bW signals, the top squark mass ($\mStop$) is varied from 200 to 1000\GeV, while the LSP mass ($\mLSP$) is varied from 0 to 700\GeV for T2tt and 0 to 550\GeV for T2bW. The masses are varied in steps of 25\GeV in all cases. For the T2bW sample the chargino mass is defined via the fraction $x$ applied to the top squark and neutralino masses as follows: $\mCHG = x \, \mStop + (1 - x) \, \mLSP$. We consider three fractions for $x: 0.25$, 0.50, and 0.75.

Standard model backgrounds are generated with  \MADGRAPH, \POWHEGONE~\cite{Nason:2004rx,Frixione:2007vw,Alioli:2010xd,Alioli:2011as,Alioli:2009je}, \PYTHIASIX~\cite{pythia}, or \MCATNLOTHREE~\cite{Frixione:2002ik,Frixione:2003ei}. The \MADGRAPH generator is used for  the generation of $\Z$ and $\PW$ bosons accompanied by up to three additional partons as well as for diboson and $\ttbar\PW$ processes, while the single top quark and $\ttbar$ processes are generated with \POWHEG. Multijet QCD  events are produced in two samples, one generated with \PYTHIA and the other with \MADGRAPH. Two $\ttbar\Z$ event samples are used. One is generated with \MCATNLO and the other with \MADGRAPH. The decays of $\tau$ leptons are simulated with \TAUOLATWEN~\cite{Davidson:2010rw}.

The \PYTHIA generator is subsequently used to perform parton showering for all signal and background samples, except for the \MCATNLO $\ttbar\Z$ sample, which uses \HERWIGSIX~\cite{Corcella:2000bw}. The detector response for all background samples is simulated with \GEANTfour~\cite{Geant2}, while the CMS fast simulation package \cite{1742-6596-331-3-032049} is used for producing signal samples in the grid of mass points described earlier. Detailed cross checks are performed to ensure that the results obtained with the fast simulation are in agreement with those obtained with the \GEANT-based full simulation.

Events are selected online by a trigger that requires $\ETm>80\GeV$ and the presence of two central  ($\abs{\eta}<2.4$) jets with $\pt >50\GeV$.  Offline, a preselection of events common to all search samples used in the analysis has the following requirements:
\begin{itemize}
\item There must not be any isolated electrons, muons, or tau leptons in the event. This requirement is intended mainly to suppress backgrounds with genuine \ETm that arise from $\PW$ boson decays. The high efficiency lepton selection criteria used in the definitions of the lepton vetoes are described in detail in Section~\ref{sec:leptonvetoes}.
\item There must be $\ETm> 175\GeV$ and at least two jets with $\pt > 70\GeV$ and $\abs{\eta} < 2.4$, such that the online selection is fully efficient.
\item The azimuthal angular separation between each of the two highest $\pt$ jets and \ptvecmiss must satisfy $\abs{\Delta\phi} > 0.5$, while for the third leading jet, the requirement is $\abs{\Delta\phi} > 0.3$. These criteria suppress rare QCD multijet events with severely mismeasured high-$\pt$ jets.
\end{itemize}

Baseline selections for the two targeted signal types are then defined by the following additional requirements. The T2tt baseline selection requires one or more b-tagged picky jets with $\pt > 30\GeV$ and $\abs{\eta} < 2.4$, and at least one pair of top quarks reconstructed by the \CORRAL algorithm. The T2bW baseline selection requires at least five jets ($\pt > 30\GeV$ and $\abs{\eta} < 2.4$) of which at least one must be b-tagged. SM background yields, estimated as described in Section~\ref{sec:background}, and signal yields after the baseline selections are shown in Table~\ref{tab:baselinecounts}. The trigger efficiency is measured to be greater than 95\% for events passing these baseline selections.

\begin{table*}[h!tb]
\topcaption{\label{tab:baselinecounts}Estimated SM background yields as obtained with the methods described in Section~\ref{sec:background}, and the observed data yields for the T2tt and T2bW baseline selections. The T2bW yield corresponds to the simplified model point with $(\mStop, \mLSP; x) = (600\GeV, 0\GeV; 0.75)$, and the T2tt yield is for the simplified model point with $(\mStop, \mLSP) = (700\GeV, 0\GeV)$.  The uncertainties listed are statistical only.}
\centering
\newcolumntype{x}{D{,}{\,\pm\,}{5.4}}
\begin{tabular}{lxx}
\hline
& \multicolumn{1}{c}{T2tt baseline selection yield} & \multicolumn{1}{c}{T2bW baseline selection yield} \\
\hline
$\ttbar$, $\PW$+jets, and single top        & 1735,16      &1850,12        \\
$\Z$+jets                                &  263.3,3.7   &207.5,3.4     \\
$\ttbar\Z$                               &  28.14,0.57  &28.92,0.57   \\
QCD multijet                             & 176,34       &175,33        \\[\cmsTabSkip]
All SM backgrounds              & 2202,38      &2261,36        \\
Observed data                  & \multicolumn{1}{c}{2161}            &\multicolumn{1}{c}{2159}               \\                 \hline
T2tt (700, 0) & 29.47,0.17 & \multicolumn{1}{c}{\NA} \\
T2bW (600, 0; 0.75)  & \multicolumn{1}{c}{\NA} & 69.26,0.47 \\
\hline
\end{tabular}
\end{table*}

A number of data control samples are used to derive corrections to reconstructed quantities and to estimate SM backgrounds. There are four control samples involving at least one well-identified lepton and two that are high purity QCD multijet samples.  The leptonic control samples are used to understand $\ttbar$ and vector boson plus jets backgrounds and are named accordingly, as indicated below. The data are drawn from samples collected online with triggers that require the presence of at least one charged lepton. The standard CMS lepton identification algorithms operating at their tightest working points~\cite{Khachatryan:2015hwa,Chatrchyan:2012xi} are then applied offline. Each event must have at least one selected muon with $\pt > 28\GeV$ and $\abs{\eta} < 2.1$ or a selected electron with $\pt > 30\GeV$ and $\abs{\eta} < 2.4$. Additional leptons must have $\pt > 15\GeV$ and $\abs{\eta} < 2.4$. Selected leptons are not included in the jet collection. Sample names and distinguishing characteristics are as follows:

\begin{itemize}
\item The inclusive $\ttbar$ control sample: At least one identified lepton and three or more jets, of which at least one must be b-tagged.
\item The high purity $\ttbar$ control sample: This is the subset of the inclusive $\ttbar$ control sample for which the selected lepton is a muon and there are at least two b-tagged jets.
\item The inclusive W+jets control sample: There must be one identified muon. In addition, the transverse mass \mt formed from $\ptvecmiss$ and the muon momentum is required to be $\geq40\GeV$ in order to reduce QCD multijet contamination.
\item The inclusive Z+jets control sample: There must be two identified leptons of the same flavour with an invariant mass in the range $80 < m_{\ell\ell} < 100$, consistent with the mass of the Z boson.
\end{itemize}

The two additional data control samples selected to be pure in QCD mulitjet events are defined as follows:

\begin{itemize}
\item  The inclusive QCD multijet control sample: Events are required to have \HT, the scalar sum of jet $\pt$, $>$340\GeV and are collected with a set of $\HT$ triggers.
\item The high $\ETm$ QCD multijet control sample: Events are selected with the same trigger used for the baseline selection. All events must satisfy $\ETm> 175\GeV$ and have at least two jets with $\pt > 70\GeV$ in order to be fully efficient with respect to the online selection. The QCD multijet purity is increased by vetoing any events with isolated electrons, muons, or tau leptons and by inverting the baseline selection requirement on the angular separation between the three leading jets and $\ptvecmiss$.
\end{itemize}

\section{Top quark pair reconstruction for the T2tt simplified model}
\label{sec:toptagger}

The T2tt and T2bW signal modes involve the same final-state particles but differ in that only T2tt involves the decays of on-shell top quarks. The only  SM background with potentially large \ETm and a visible component that is  identical to that of T2tt is $\ttbar\Z$, with the $\ttbar$ pair decaying hadronically and the $\Z$ boson decaying invisibly to neutrinos. Efficient identification of a pair of hadronically decaying top quarks in events with large \ETm provides an important means of suppressing most other backgrounds. As mentioned in the previous section, we developed the \CORRAL dedicated top quark reconstruction algorithm for this purpose. Kinematic properties of the top quark candidates reconstructed with \CORRAL are exploited to further improve the discrimination of signal from background.

Top quark taggers are typically characterized by high efficiencies for the reconstruction of all-hadronic decays of top quarks that have been Lorentz boosted to sufficiently high momentum for their final state partons and associated showers to form a single collimated jet. Such taggers are not ideal for the regions of parameter space targeted by this search because the top quarks from top squark decays can experience a wide range of boosts in these regions and it is not uncommon for one of the top quarks to have a boost that is too low to produce such a coalescence of final-state objects. An additional problem arises with traditional jet algorithms that do not always distinguish two separate clusters of particles whose separation is smaller than their fixed distance parameter or cone radius. In addition, for low-\pt jets and those originating from hadronisation of b quarks, it is not unusual for algorithms with fixed distance metrics to miss some of the particles that should be included in the jet. These issues are addressed by making use of a variable jet-size clustering algorithm that is capable of successfully resolving six jets in the decays of top quark pairs with efficiency ranging between 25\% in the case of signal with compressed mass splitting ($\mStop=400\GeV\approx m_{\PQt}+\mLSP+75\GeV$) to 40\% in the case of large mass splitting ($\mStop=750\GeV \approx m_{\PQt}+\mLSP+550\GeV$).

{\sloppy
The algorithm starts by clustering jets with the Cambridge--Aachen algorithm~\cite{Dokshitzer:1997in,Wobisch:1998wt} with a distance parameter of 1.0 in the $\eta$-$\phi$ plane to produce what will be referred to as proto-jets. Studies based on MC simulation show that this parameter value is large enough to capture partons with \pt as low as 20\GeV. Each proto-jet is then considered for division into a pair of subjets. The N-subjettiness metric~\cite{nsubjettiness-2011}, $\tau_{\mathrm{N}}$, is used to determine the relative compatibility of particles in a proto-jet with a set of ``N'' jet axes. It is defined as the $\pt$-weighted sum of the distances of proto-jet constituents to the nearest jet axis, resulting in lower values when the particles are clustered near jet axes and higher values when they are more widely dispersed. As discussed in Ref.~\cite{nsubjettiness-2011}, the exclusive two-jet \kt algorithm~\cite{Catani1993187,Ellis:1993tq} can be used to find an initial pair of subjet axes in the proto-jet that approximately minimizes the $\tau_{2}$ metric. The exclusive two-jet algorithm differs from the inclusive \kt algorithm in that it does not have a distance parameter. It simply clusters a specified set of particles into exactly two jets. In our case, the axes are varied in the vicinity of the initial set until a local minimum in the value of $\tau_{2}$ is found. This defines the final set of axes and each particle in the proto-jet is then associated with the closest of the two axes, resulting in two candidate subjets.
\par}

An MVA `picky' metric is then used to determine if it is more appropriate to associate the particles with two subjets than with the original proto-jet. The input variables include the $\tau_{1}$ and $\tau_{2}$ subjettiness metrics, the mass of the proto-jet, the ($\eta$,$\phi$) separation of the two subjets, and a profile of the proto-jet's energy deposition. An MVA discriminator working point is defined as the threshold value at which the efficiency to correctly split proto-jets into distinct constituent subjets of top quark decays is 95\%, while incorrectly splitting fewer than 10\% of jets that are already distinct constituents.   If the discriminator value doesn't meet or exceed the threshold, the proto-jet is treated as a single jet and added to the final jet list, otherwise the two subjets enter the proto-jet list to be considered for possible further division. The algorithm runs recursively until there are no remaining proto-jets, yielding a collection of variable-size jet clusters known as `picky' jets.

\begin{figure}[htbp]
\centering
   \includegraphics[width=0.49\textwidth]{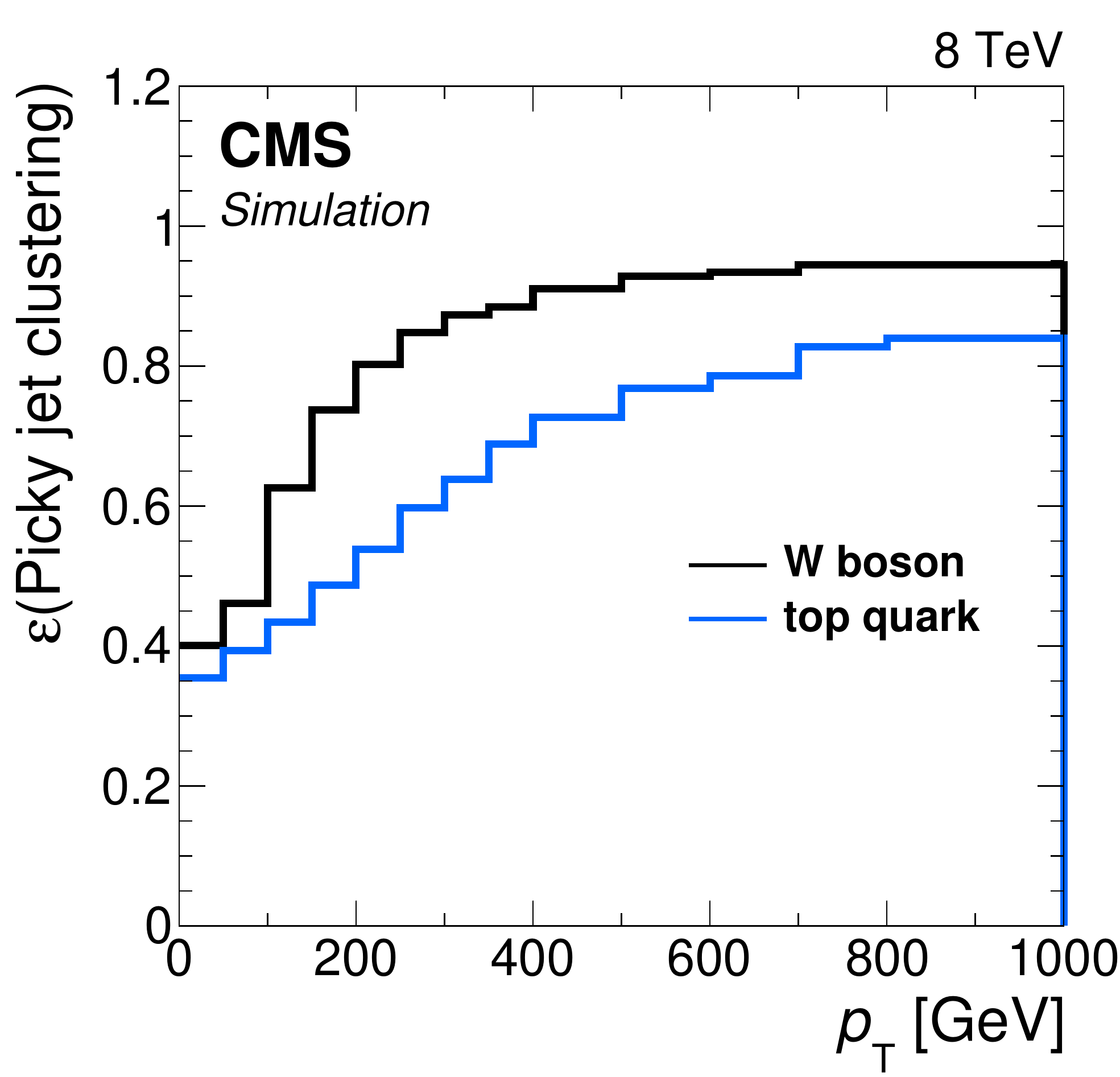}
    \includegraphics[width=0.49\textwidth]{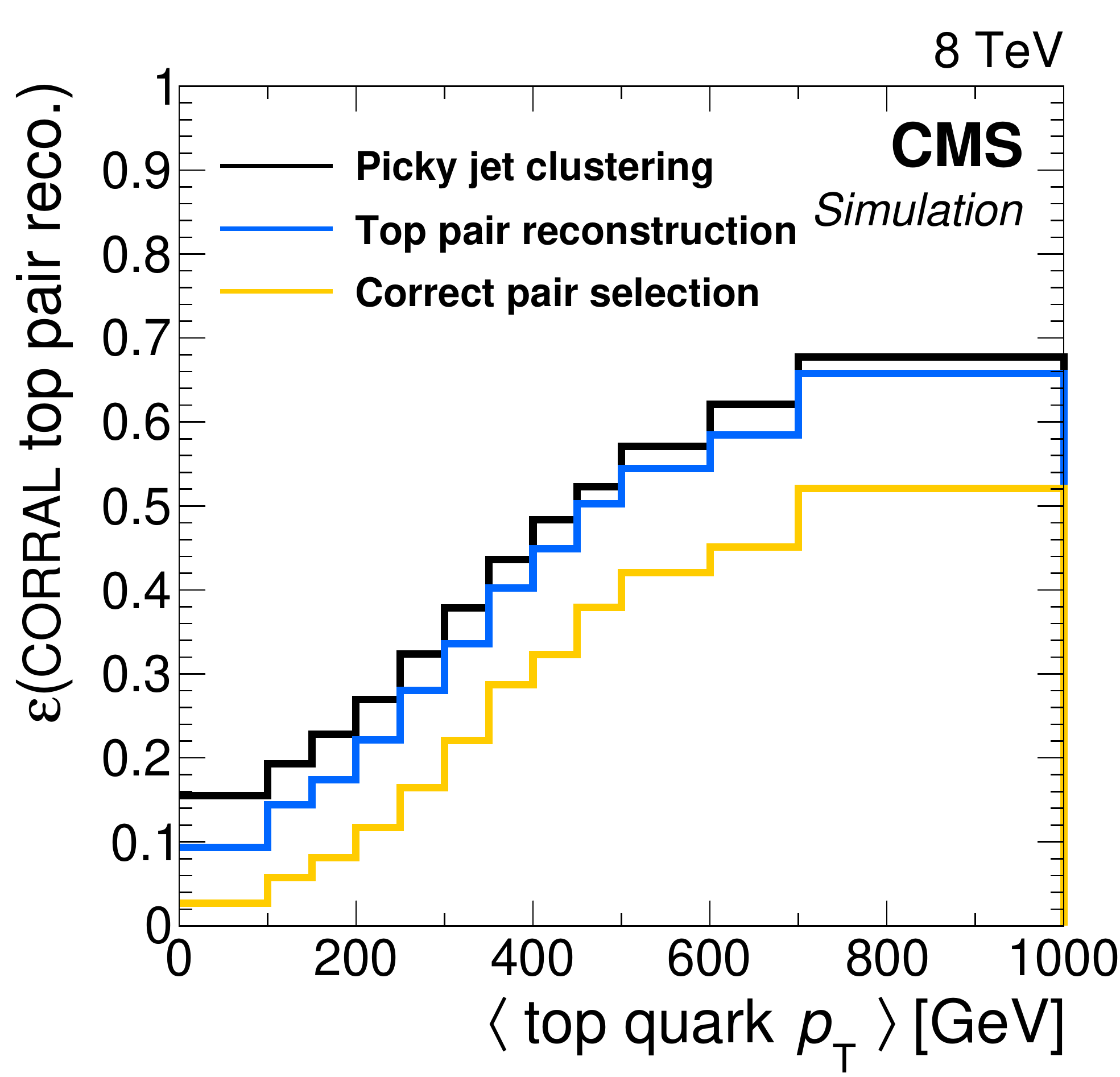}
  \caption{\label{fig:CORRAL_eff} Efficiency as a function of generator level $\pt$ for picky jet clustering and \CORRAL top quark pair reconstruction in all-hadronic T2tt events with $\mStop = 600\GeV$ and $\mLSP = 50\GeV$. \cmsLeft: The efficiency to correctly cluster  final state particles from each $\PW$ boson and top quark decay into two and three picky jets, respectively, as a function of particle (top quark or $\PW$ boson) $\pt$. \cmsRight: The efficiency at each stage of the \CORRAL algorithm to reconstruct a hadronically decaying top quark pair as a function of the average $\pt$ of the two top quarks. They are the efficiency to correctly cluster the final state particles from top quark decays into six picky jets, labelled ``Picky jet clustering''; the efficiency to both carry out picky jet clustering and reconstruct the top quark pair with these six picky jets, labelled ``Top pair reconstruction''; and finally the efficiency to carry out picky jet clustering,  top pair reconstruction, and then correctly select the reconstructed top quark pair for use in the analysis, labelled ``Correct pair selection''.}
\end{figure}

The efficiency to correctly cluster $\PW$ bosons (top quarks) into two (three) picky jets satisfying the basic acceptance requirements of $\pt > 20\GeV$ and $\abs{\eta} < 2.4$ is shown in Fig.~\ref{fig:CORRAL_eff} as a function of generated particle (top quark or $\PW$ boson) $\pt$ in all-hadronic T2tt events with $\mStop = 600\GeV$ and $\mLSP = 50\GeV$. In each event the six quarks arising from the hadronic decays of the two top quarks are matched to reconstructed picky jets by means of ghost association ~\cite{Cacciari:2008gn}. This technique associates particles produced in the fragmentation and hadronization of the quark prior to detector response simulation. The `generator-level' particles are clustered together with the full reconstructed particles used to form the picky jets as described above, but the momentum of each of the generator-level particles is scaled by a very small number so that the picky jet collection is not altered by their inclusion. A quark is then determined to be matched to the picky jet that contains the largest fraction of the quark's energy if it is greater than 15\% of the quark's total energy. In the case that two or more quarks are associated with the same picky jet, the picky jet is matched to the quark with the largest clustered energy in that jet.

\begin{figure}[htb]
\centering
   \includegraphics[width=0.49\textwidth]{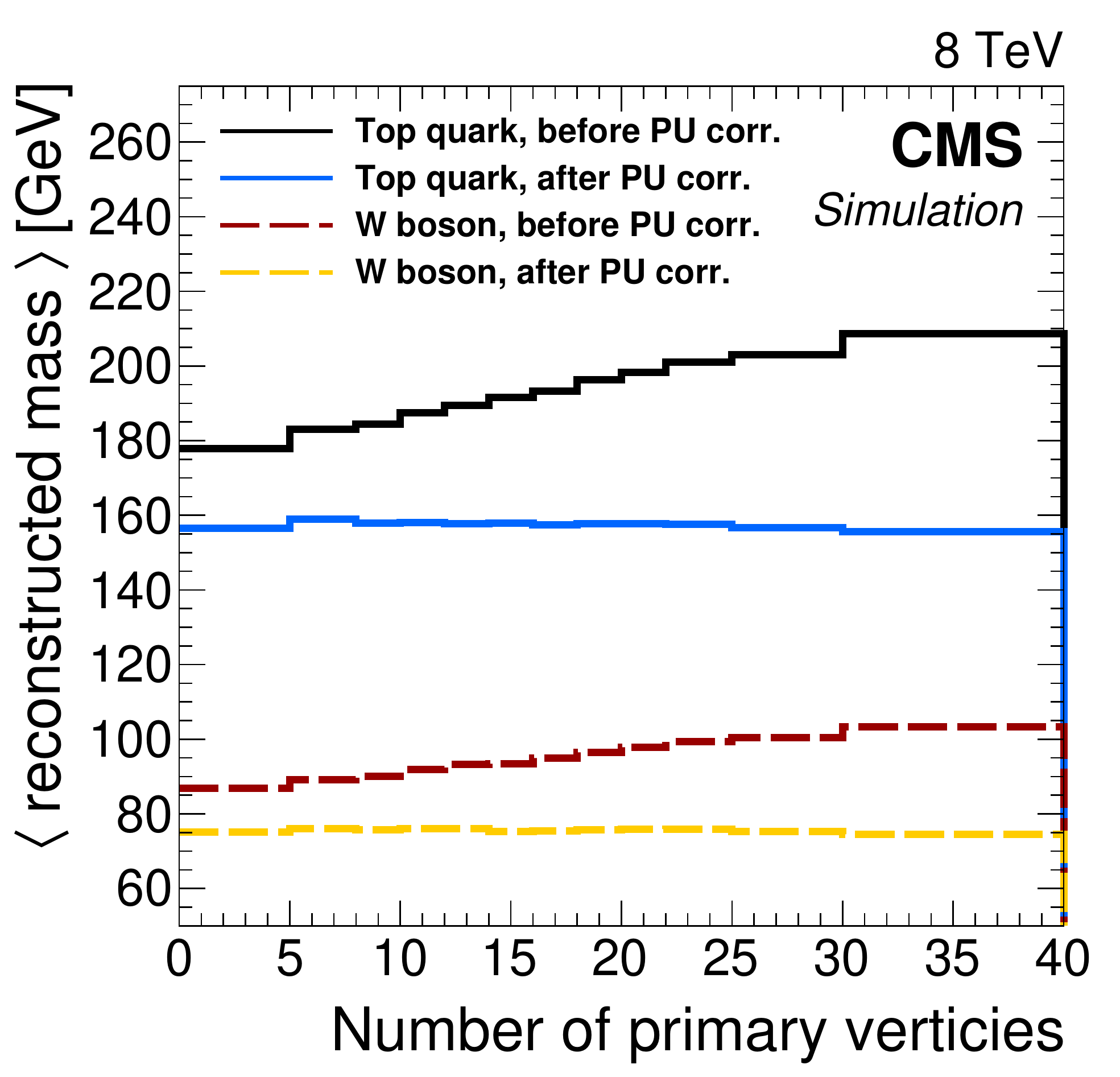}
  \caption{\label{fig:CORRAL_trimming}  Masses of the top quarks and $\PW$ bosons reconstructed with picky jets that are matched at particle level in simulation, as discussed in the text, in all-hadronic T2tt events with $\mStop = 600\GeV$ and $\mLSP = 50\GeV$. The labels ``before PU corr.'' and ``after PU corr.'' refer to  results obtained before and after application of the trimming procedure used to correct for pileup effects.}
\end{figure}

The energy of each resulting picky jet is corrected for pileup by subtracting the measured energy associated with pileup on a jet-by-jet basis by means of a trimming procedure similar to the one discussed in Ref.~\cite{jettrimming-2010}. The procedure involves reclustering of the particles associated with the jet into subjets of radius 0.1 in $\eta$-$\phi$ and then  ordering them by decreasing $\pt$. The lowest $\pt$ subjets are removed one-by-one until the summed momentum and mass of the remaining subjets have minimal differences with the same quantities after subtracting an estimate of the pileup contribution~\cite{PhysRevLett.110.162001}. The reconstructed $\PW$ boson and top quark masses as a function of the number of reconstructed primary vertices are shown in Fig.~\ref{fig:CORRAL_trimming} in all-hadronic T2tt events with $\mStop = 600\GeV$ and $\mLSP = 50\GeV$. The reconstructed mass values are seen to have no pileup dependence after the trimming procedure is applied. No additional jet energy scale corrections, other than those mentioned below, have been derived to remove the remaining 5-10\% bias in the reconstructed mass values. The \CORRAL algorithm is optimized for the uncorrected top quark and $\PW$ boson mass values.

The $\pt$ spectra of picky jets in MC data are corrected to match those observed in data in the inclusive $\ttbar$ and $\Z$+jets control samples by rescaling of individual picky jet $\pt$ values. The rescaling factors are derived separately for each of the two processes and for the flavour of parton that initiated the jet. They are found to be within 2--3\% of unity. Picky jets can also be b-tagged with the CSV algorithm by considering the tracks that have been used in their formation.

A candidate for a hadronically decaying top quark pair is a composite object constructed from six picky jets that passes every step of the \CORRAL algorithm, which will now be described. To reduce the number of jet combinations that must be considered, the algorithm involves several  stages, with progressively tighter selection criteria at each stage. First, BDTs are trained to discriminate the highest $\pt$ jet coming from a top quark decay from all other jets in the event using input variables related to jet kinematics, b tagging discrimination and jet composition information. Jets are labelled as seed jets if they have an associated discriminator value that exceeds a high efficiency cutoff value. Three-jet top quark candidates are then constructed from all combinations of three jets in the event that include at least one seed jet.  High quality top quark candidates are those that pass one of two MVA working points chosen to identify 97--99\% of those cases in which the jets are correctly matched to top quark decays and to reject 60--80\% of the candidates that are not correctly matched. The most important input variables are the $\PW$ boson and top quark invariant masses and the picky jet b tagging discriminator value. Other variables such as the angular separations of the jets are included for additional discrimination. A final list of top quark pairs contains all combinations of two high quality top quark candidates with distinct sets of three jets. The final reconstructed top quark pair used in the analysis is the one with the highest discriminator value from a BDT that is trained with variables similar to those used in the candidate selection but also including information on the correlations between the top quark candidates.

The \CORRAL algorithm reconstructs at least one top quark pair in nearly every event that has six or more picky jets. However, \CORRAL is not strictly a top quark tagger that must distinguish events with top quarks from events without top quarks. It is designed to reconstruct top quark pairs in data samples that are predominantly made up of top quark events, as is the case for the T2tt part of this analysis. In Fig.~\ref{fig:CORRAL_eff}, the efficiency for correctly resolving the top quark pair is shown at each stage of the algorithm. These efficiencies are calculated for T2tt events with $\mStop = 600\GeV$ and $\mLSP = 50\GeV$, but they do not depend strongly on the signal mass parameters. The two hadronic top quark decays are each resolved into three distinct picky jets in 15--70\% of events, depending on the boost of the quarks. In nearly all of these events the correct six jets pass the \CORRAL jet seeding and top quark candidate selection requirements and are used to form the correct top quark pair among a number of top quark pairs found in the event. The correct pair is then chosen to be used in the analysis in 30--80\% of events.

\begin{figure*}[h!tb]
\centering
   \includegraphics[width=0.49\textwidth]{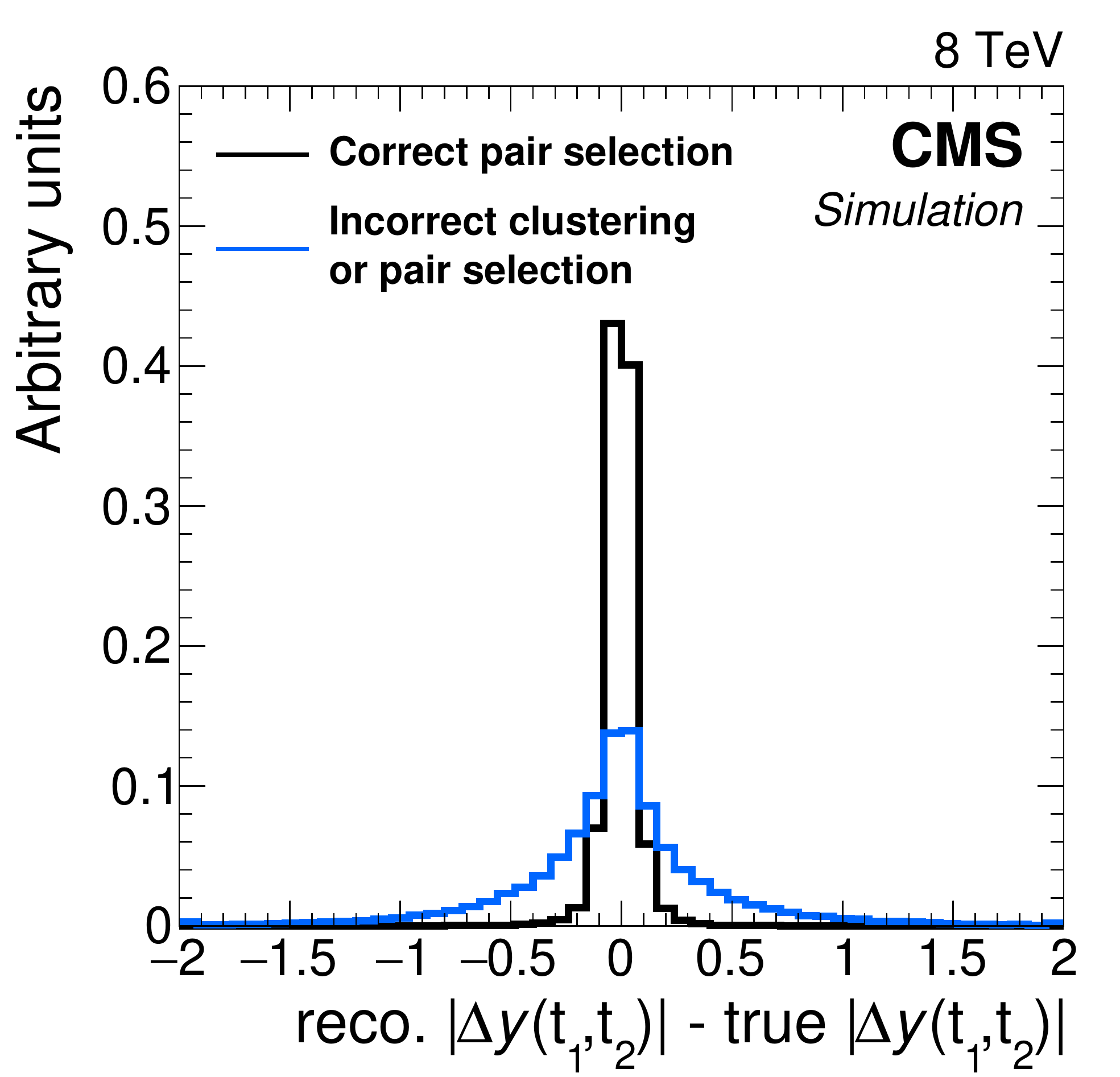}
   \includegraphics[width=0.49\textwidth]{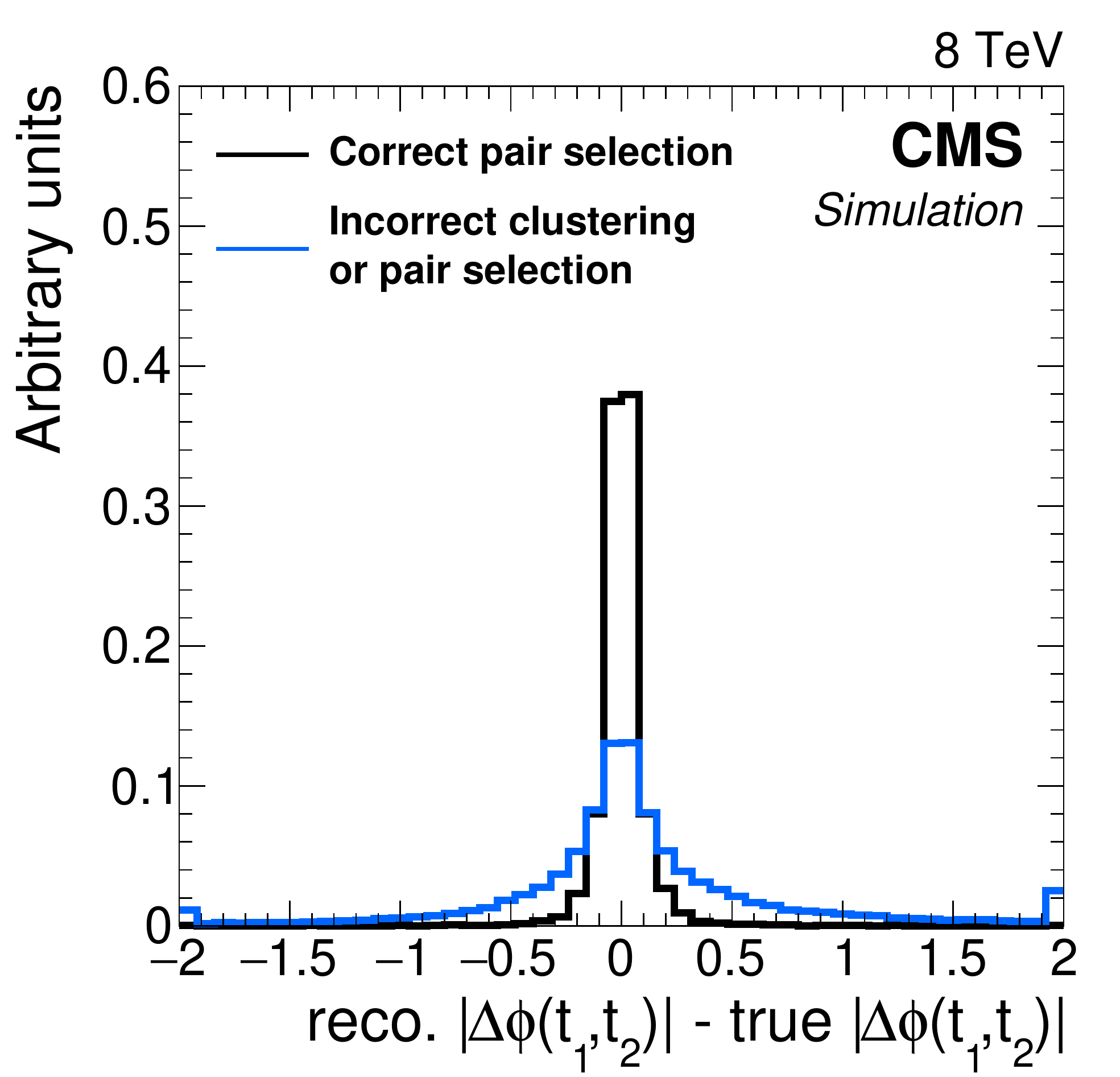}
   \includegraphics[width=0.49\textwidth]{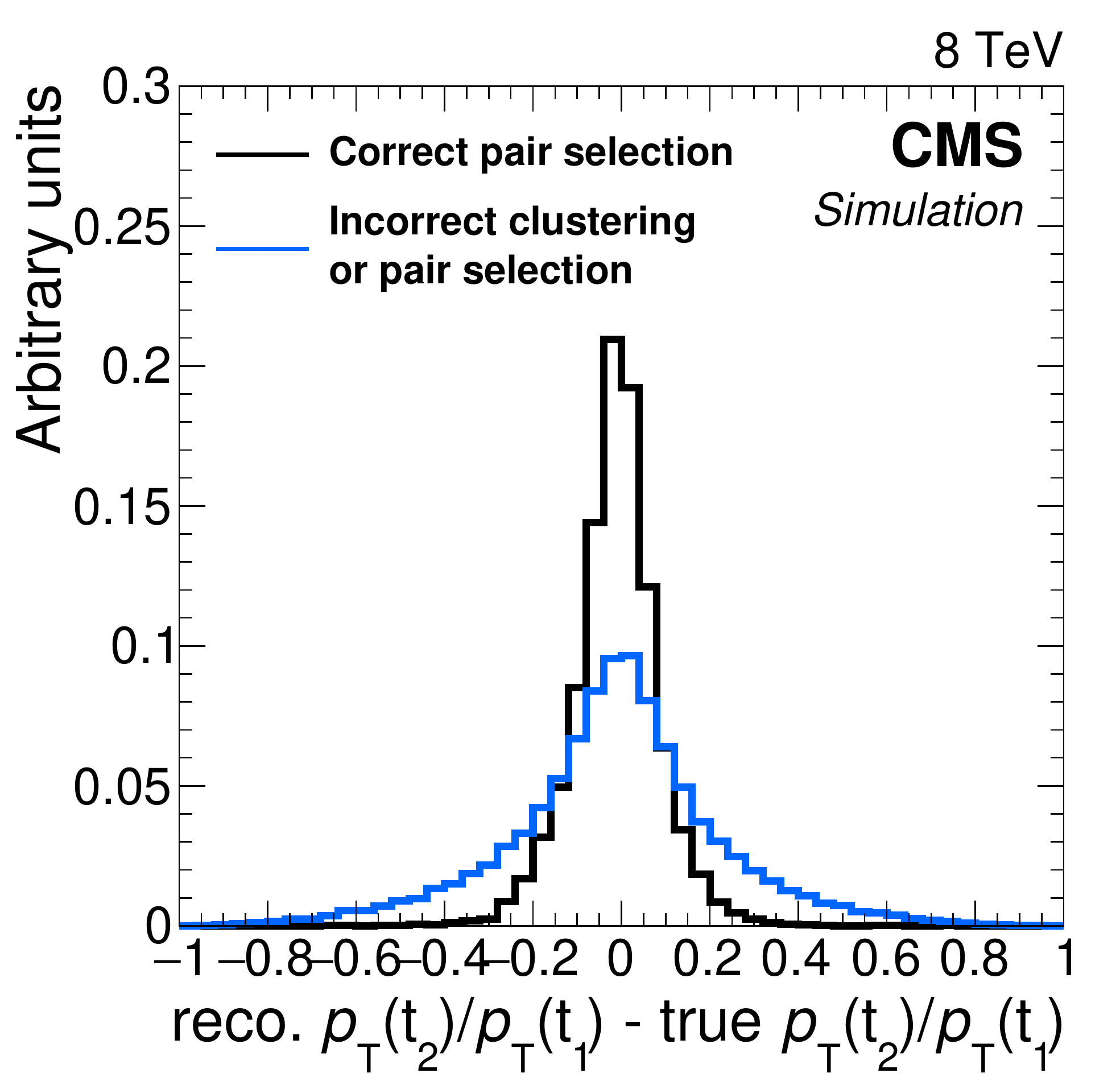}
  \caption{\label{fig:CORRAL_recoproperties} Properties of the reconstructed top quark pair used in the analysis are compared to their true properties in all-hadronic T2tt events with $\mStop = 600\GeV$ and $\mLSP = 50\GeV$. The label ``Correct pair selection'' corresponds to events in which the two top quark decays are each resolved into three distinct picky jets and these jets are used to reconstruct the two top quarks. The label ``Incorrect clustering or pair selection'' is used for all other events.  The top two figures show comparisons of the angular separation between the two top quarks in rapidity, $y\equiv-(1/2)\ln[(E+p_z)/(E-p_z)]$, and azimuthal angle $\phi$. The bottom figure compares the relative $\pt$ of the two top quarks. In all cases, $\PQt_1$ refers to the top quark with the highest $\pt$.}
\end{figure*}

Properties of the reconstructed top quark pairs used in the analysis are compared to true top quark pair quantities in Fig.~\ref{fig:CORRAL_recoproperties} for signal events with at least one reconstructed top quark pair. The events in which the true top quark pair is chosen are categorized separately in the figure. In the fully resolved and selected case the reconstructed separation in $\phi$ between the two top quarks agrees with the true separation within 0.1 in over 80\% of events. Even in the case of the reconstructed top quark pair not being fully resolved or selected, there is reasonable agreement because the top quark pair is constructed with five of the six correct jets in the majority of these events.

\begin{figure}[h!tb]
\centering
    \includegraphics[width=0.49\textwidth]{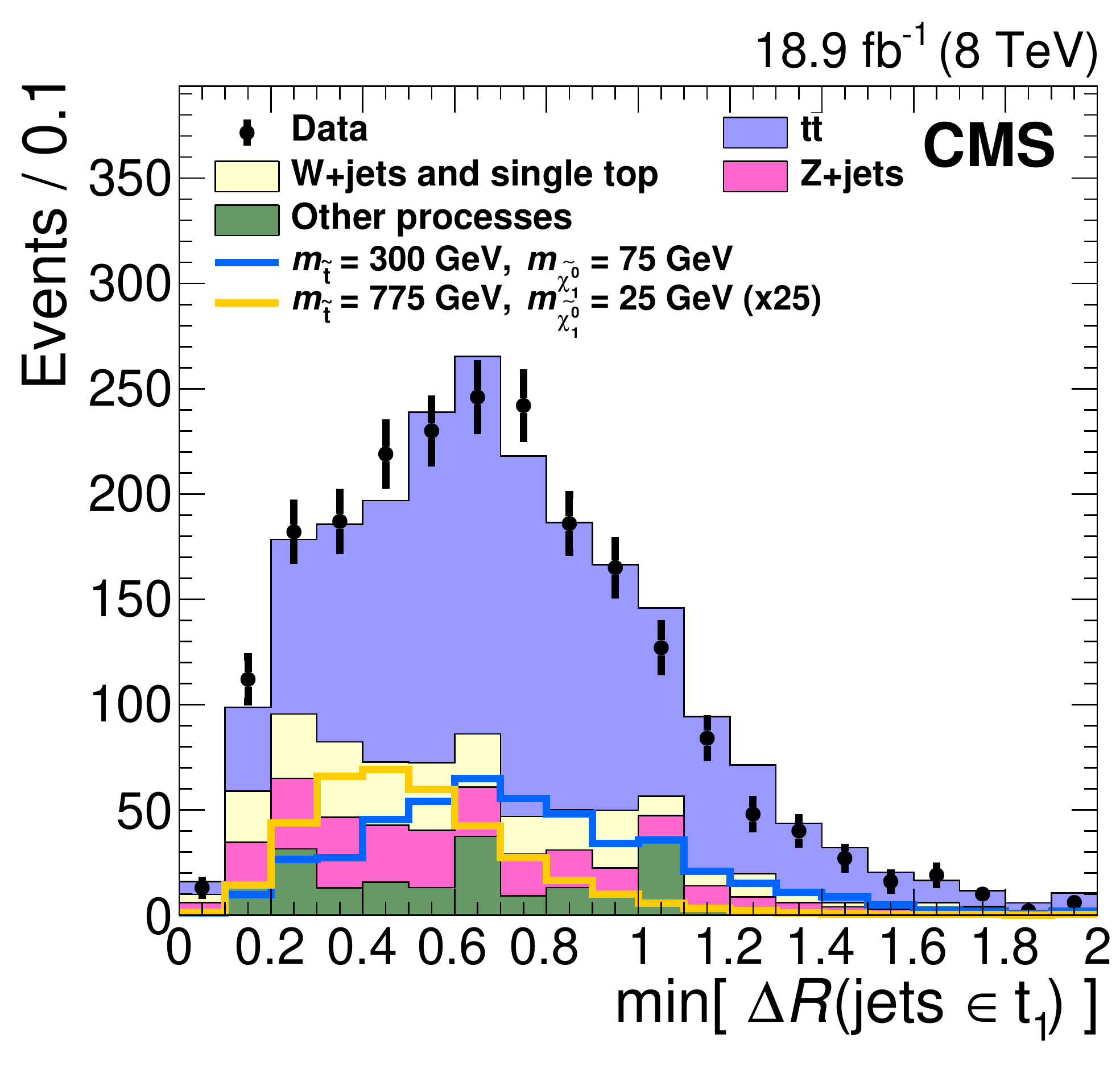}
    \includegraphics[width=0.49\textwidth]{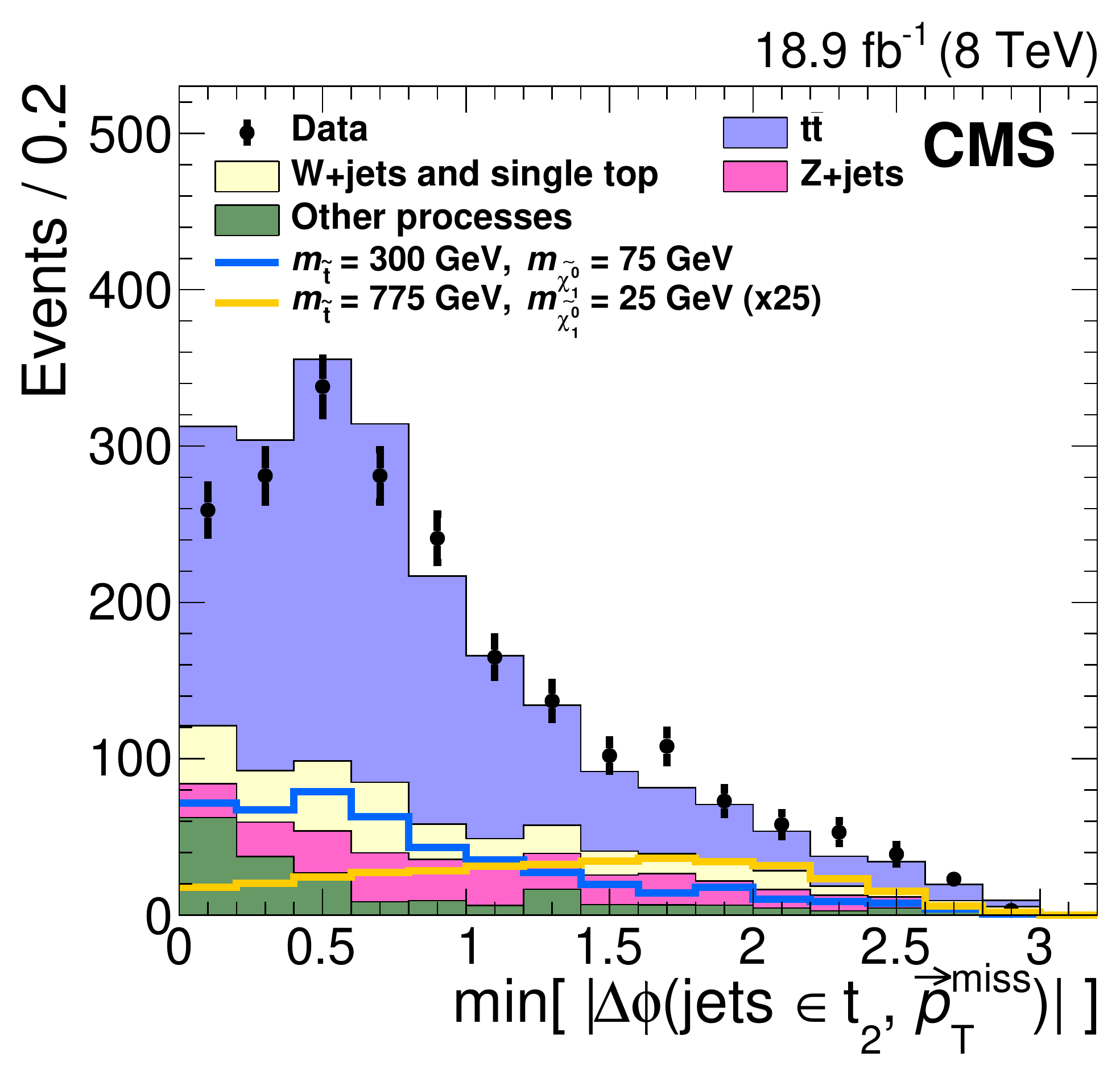}
  \caption{\label{fig:CORRAL}Distributions of properties of reconstructed top quark pairs for data together with signal and background MC data samples after the baseline selection for two choices of $\mStop$ and $\mLSP$. For the case $\mStop=775\GeV$, $\mLSP=25\GeV$ the expected signal is multiplied by a factor of 25. The \cmsLeft\ plot shows the minimum separation in the $\eta$-$\phi$ plane between any two jets in the leading reconstructed top quark, defined as the one with the highest discriminator value, while the \cmsRight\ plot shows the separation in $\phi$ between \ptvecmiss  and the jet in the sub-leading reconstructed top quark for which this separation is the smallest. Both variables are inputs to the T2tt search region BDT discriminators, which are described in Section~\ref{sec:searchregions}.}
\end{figure}

The signal discrimination that is achieved by exploiting differences in the kinematics of the reconstructed top quark pairs in simulated signal samples and those in simulated SM background samples is illustrated in Fig.~\ref{fig:CORRAL}. The \cmsLeft\ plot shows the minimum separation in the $\eta$-$\phi$ plane between any two jets in the reconstructed top quark candidate with the highest discriminator value,  labelled $\PQt_1$. The separation tends to be smaller in T2tt signal events because the top quarks with the highest discriminator value are more likely to be boosted. Similarly, the \cmsRight\ plot shows the distribution for the separation in $\phi$ between the jet direction and \ptvecmiss for the jet with the smallest such separation from the sub-leading reconstructed top quark, labelled $\PQt_2$. The distribution for the semileptonic $\ttbar$ background, involving $\ttbar$ events in which one $\PW$ boson decays leptonically, is shifted to low values of $\Delta\phi$ because the $\PQt_2$ top quark candidates in $\ttbar$ events typically use the b jet from the leptonically decaying top quark, which is correlated in angle with the $\ptvecmiss$ from the leptonically decaying $\PW$ boson.

\section{Rejection of isolated leptons}
\label{sec:leptonvetoes}
The main backgrounds for this analysis arise from events with lost or misidentified leptons. Sensitivity to signal is therefore improved by identifying and rejecting events with charged leptons originating from prompt $\PW$ boson decays as efficiently as possible. On the other hand, signal events often contain charged leptons that arise from decays of heavy flavour hadrons or charged hadrons that have been misidentified as charged leptons. It is advantageous to retain these events in order to achieve high signal efficiency.    In events with $\ETm > 175\GeV$ and five or more jets, the standard CMS lepton identification algorithms operating at their tightest working points~\cite{Khachatryan:2015hwa,Chatrchyan:2012xi} can identify semileptonic $\ttbar$ events with efficiencies of 54\% and 60\% for final states involving electrons and muons, respectively. This analysis makes use of MVA techniques to achieve higher efficiencies for the identification and rejection of semileptonic $\ttbar$ events, while retaining high signal efficiency.

The MVAs used here combine a number of moderately discriminating quantities into a single metric that can be used for electron and muon identification. Electrons and muons must have $\pt > 5$\GeV, $\abs{\eta}<2.4$, and are required to satisfy the conditions for the loose working point of the standard CMS identification algorithms, for which the efficiencies for electrons and muons in the tracker acceptance are above 90\%. The discriminating variables used in the training of the muon identification BDT are the $\pt$ of the muon, its track impact parameter information, relative isolation in terms of charged and neutral particles, and the properties of the jet nearest to the muon. Isolation in terms of charged and neutral hadrons is defined by means of separate sums of the $\pt$ of charged and neutral PF particles, respectively, in a region near the lepton, divided by the lepton $\pt$. The properties of the nearest jet that are used include the separation from the lepton in the $\eta$-$\phi$ plane, the momentum of the lepton relative to the jet axis, and the CSV b tagging discriminator value for the jet. For electron identification, the variables include all of those used for the muon, plus several electron-specific variables that are used in the standard CMS electron identification MVA~\cite{Khachatryan:2015hwa}.

\begin{figure}[!htb]
\centering
    \includegraphics[width=0.49\textwidth]{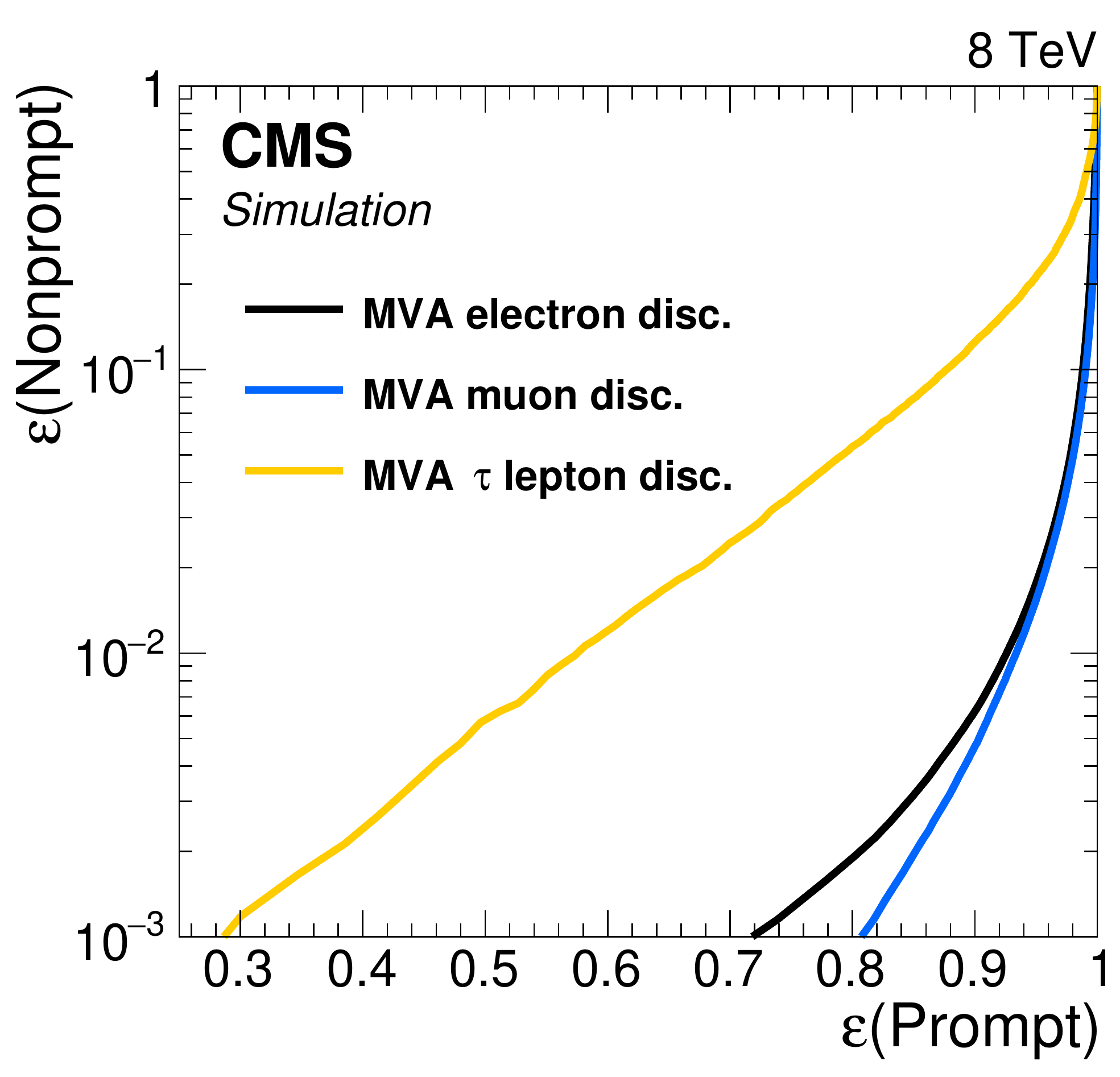}
    \includegraphics[width=0.49\textwidth]{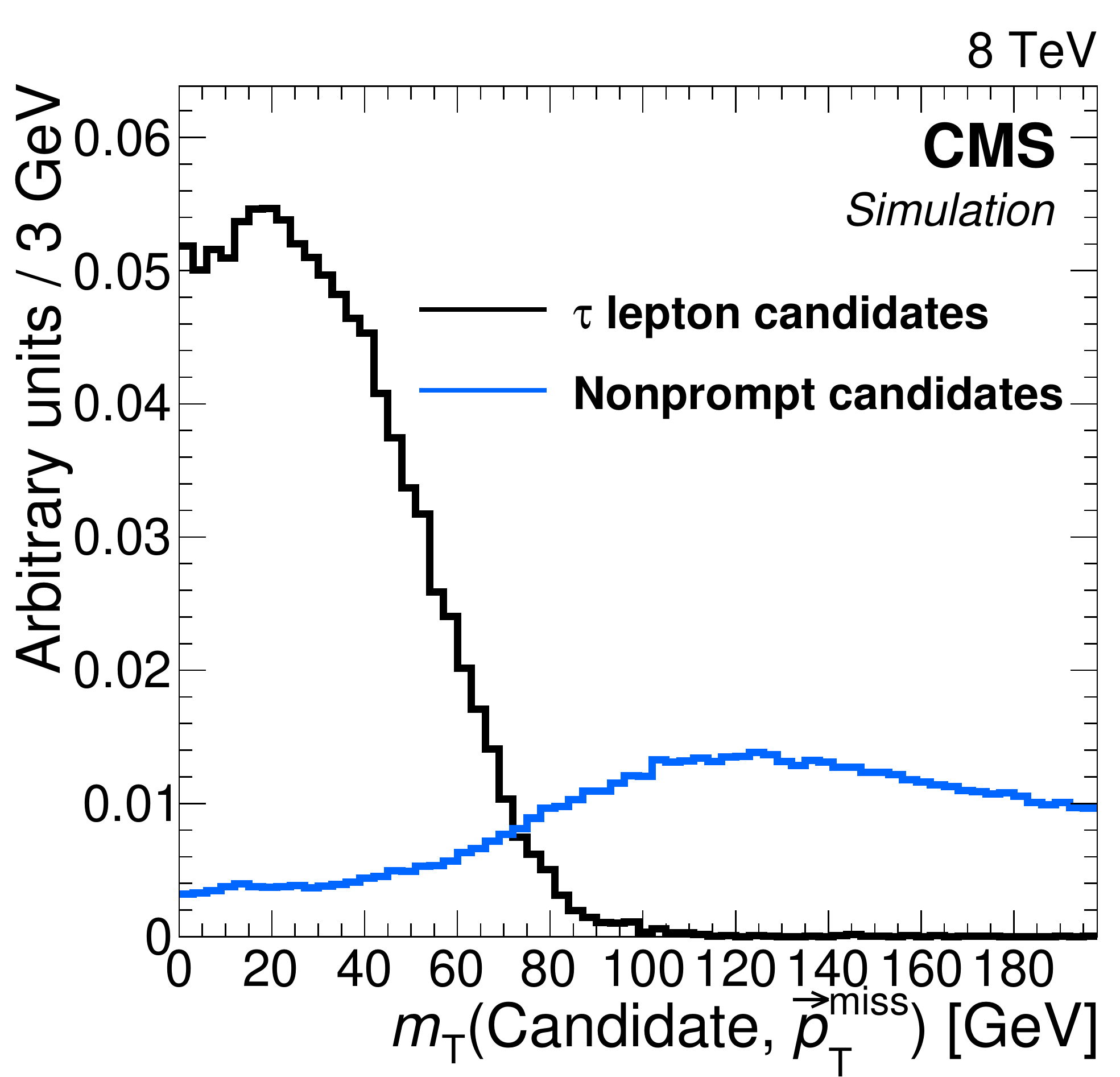}
  \caption{\label{fig:leptons}\cmsLeft: Comparison of BDT discriminator selection efficiencies for non-prompt and prompt leptons. Prompt leptons are those matched to lepton candidates in semileptonic $\ttbar$ events  whereas non-prompt leptons are those that are matched to lepton candidates in all-hadronic $\ttbar$ in the case of electrons and muons, or all-hadronic T2tt signal events in the case of $\Pgt$ leptons. It follows that the non-prompt category includes misidentified charged hadrons and leptons from decays of hadrons. \cmsRight: The \mt calculated from \ptvecmiss and the momentum of the visible $\Pgt$ lepton decay products, for $\Pgt$ lepton candidates matched to $\Pgt$ lepton  decays in semileptonic $\ttbar$ events, and all $\Pgt$ lepton candidates in a simulated all-hadronic T2tt signal sample ($\mStop = 620\GeV$, $\mLSP = 40\GeV$).  }
\end{figure}

{\tolerance=800
The BDTs are trained using simulated event samples with electrons or muons. In particular, single-lepton $\ttbar$ events are the source of prompt leptons, while electrons or muons in all-hadronic $\ttbar$ events are used for non-prompt leptons. The non-prompt lepton selection efficiency in signal events is similar to that in $\ttbar$ events. The \cmsLeft\ plot in Fig.~\ref{fig:leptons} shows the selection efficiency, by lepton type, for non-prompt leptons as a function of that for prompt leptons in the BDT training samples. The curves are obtained by varying the cutoff on the corresponding BDT discriminator value above which events are accepted. In this analysis, the discriminator values that are chosen have efficiencies of 98\% for events with electrons and muons from $\PW$ boson decays that pass the preselection requirements, while incorrectly selecting no more than 5\% of all-hadronic $\ttbar$ events. The latter gives some indication of the expected loss of all-hadronic top squark signal events. Upon including reconstruction and acceptance inefficiencies, these requirements eliminate 80\% of single-electron and single-muon $\ttbar$ events with $\ETm > 175\GeV$  and five or more jets.
\par}

A similar approach is used to identify hadronically decaying tau leptons originating from semileptonic $\ttbar$ decays. The $\Pgt$ identification algorithm focuses on decays involving a single charged hadron in conjunction with neutral hadrons because the majority of hadronic $\Pgt$ decays are to final states of this type, which are often referred to as `one-prong' decays. No attempt is made to specifically reconstruct the sub-dominant `three-prong' decays. A $\Pgt$ candidate is thus defined by a track and a nearby electromagnetic cluster produced by the photons from $\pi^0\to\gamma\gamma$ decay, if present, in order to include more of the visible energy from the $\Pgt$ lepton decay. Since every charged particle with $\pt>5\GeV$ and $\abs{\eta}<2.4$ could be considered to be a $\Pgt$ candidate, we reduce the pool of candidates by using \mt calculated from \ptvecmiss and the momentum of each candidate. As seen in the \cmsRight\ plot in Fig.~\ref{fig:leptons}, the \mt distribution for genuine $\Pgt$ candidates has an endpoint at the mass of the $\PW$ boson for semileptonic $\ttbar$ events, reflecting the fact that the neutrinos associated with $\PW$ boson and $\Pgt$ lepton decays are the largest source of \ETm in these events. Fully hadronic signal events with large $\ETm$ do not have this constraint, and so each $\Pgt$ candidate is required to have $\mt<68\GeV$.

The variables used in a BDT discriminator for the identification of the $\Pgt$ candidate are the track $\pt$, $\abs{\eta}$, and distance of closest approach to the primary vertex, as well as the isolation quantities and general properties of the jet in which the $\Pgt$ candidate is contained.  The isolation variables include the separate sums of the transverse momenta of charged and neutral PF particles, in cones of radii 0.1, 0.2, 0.3, and 0.4 centered on the candidate, and the distance between the candidate and the nearest track. The jet variables used are the separation in the $\eta$-$\phi$ plane between the track and the jet axis, and the b tagging discriminator value for the jet. This BDT is trained with hadronically decaying $\Pgt$ candidates originating from semileptonic $\ttbar$ decays in MC simulation for prompt candidates, while all $\Pgt$ candidates in all-hadronic T2tt events with $\mStop = 620\GeV$ and $\mLSP = 40\GeV$ are used for the non-prompt candidates. The samples produced with these T2tt mass parameters are not included in the final array of T2tt samples used in the later stages of this analysis.  The T2bW baseline selection is applied to all events  in order to have training samples whose kinematic selection criteria are consistent with those used to select the data samples used for the search. The \mt cutoff value and the BDT discriminator value are chosen to keep losses below 10\% in the all-hadronic signal samples targeted by this analysis. The efficiency for correctly selecting the background of semileptonic $\ttbar$ events with hadronically decaying tau leptons is 65\%. This  efficiency is defined relative to events for which the $\Pgt$ lepton decay products include at least one reconstructed charged particle with $\pt > 5\GeV$.

The efficiencies for selecting leptons in simulation are corrected to match those measured in data after applying the T2bW baseline selection criteria. The multiplicative correction factors applied to the simulated electron and muon selection efficiencies for this purpose are  $0.95 \pm 0.03$ and $1.01 \pm 0.03$, respectively. The corrections to the simulated $\Pgt$ selection efficiency are $1.30 \pm 0.10$ for $\Pgt$ candidates with $\pt < 10\GeV$ and $0.98 \pm 0.04$ for all other candidates.

\section{Search regions}
\label{sec:searchregions}
As discussed above, this analysis makes use of MVA  techniques based on BDTs to achieve sensitivity to direct production of top squark pairs in the all-hadronic final states of the T2tt and T2bW simplified models in the presence of three main classes of much more copiously produced SM backgrounds.  The signal space of the T2tt simplified model is parameterised by the masses of the top squark and the neutralino. The T2bW simplified model also includes an intermediate chargino, and is therefore parameterised by three masses. For each model, a large set of simulated event samples is prepared, corresponding to a grid of mass points in two dimensions for T2tt, and in three dimensions for T2bW. A large set of moderately to strongly discriminating variables, discussed in more detail below, serves as input to each BDT to yield a single discriminator value ranging between $-1.0$ and $+1.0$ for each event considered. Events with values closer to 1 ($-1$) are more like signal (background).

\begin{figure*}[hbt!]
\centering
  \includegraphics[width=0.49\textwidth]{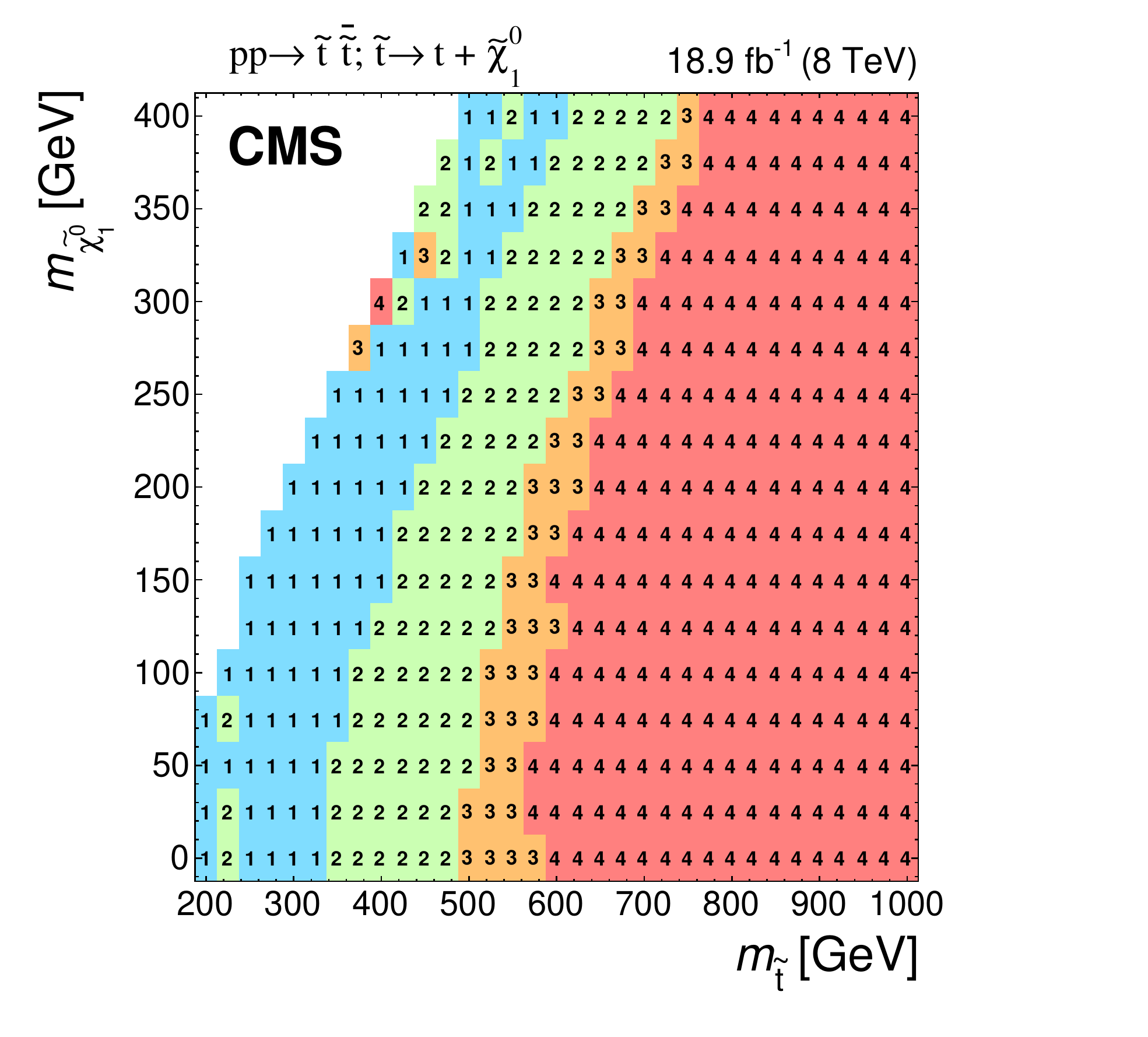}
   \includegraphics[width=0.49\textwidth]{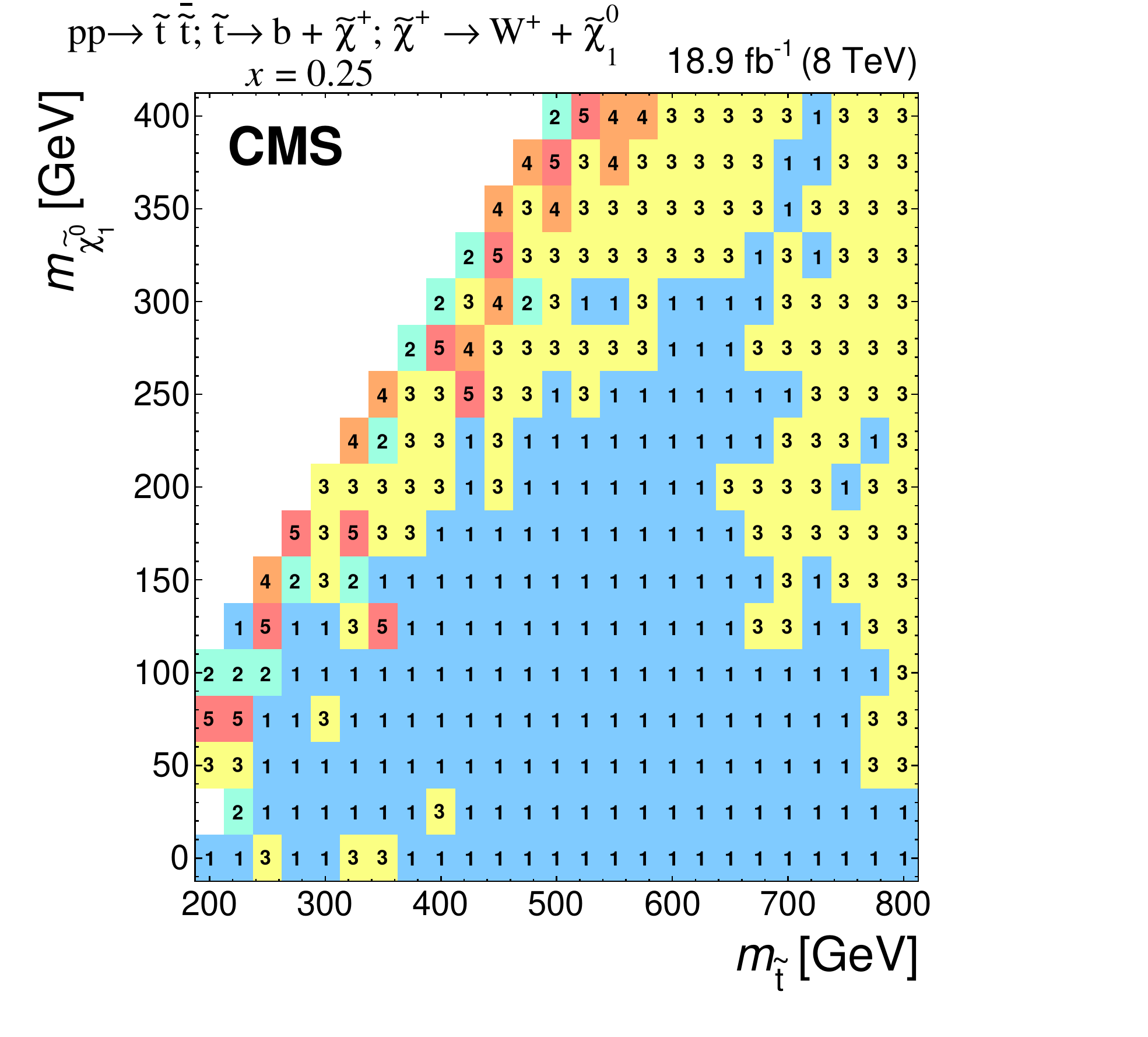}
   \includegraphics[width=0.49\textwidth]{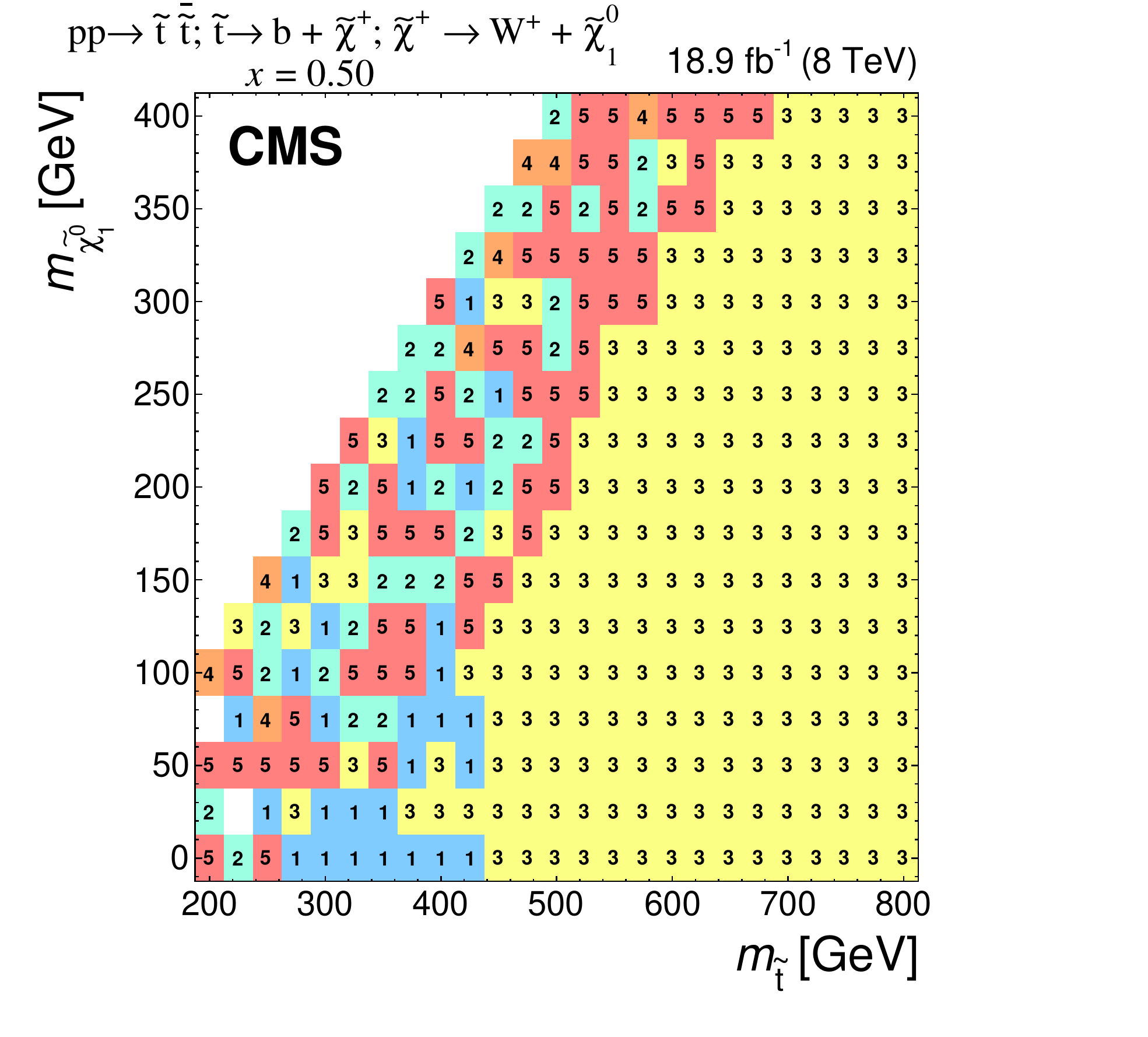}
   \includegraphics[width=0.49\textwidth]{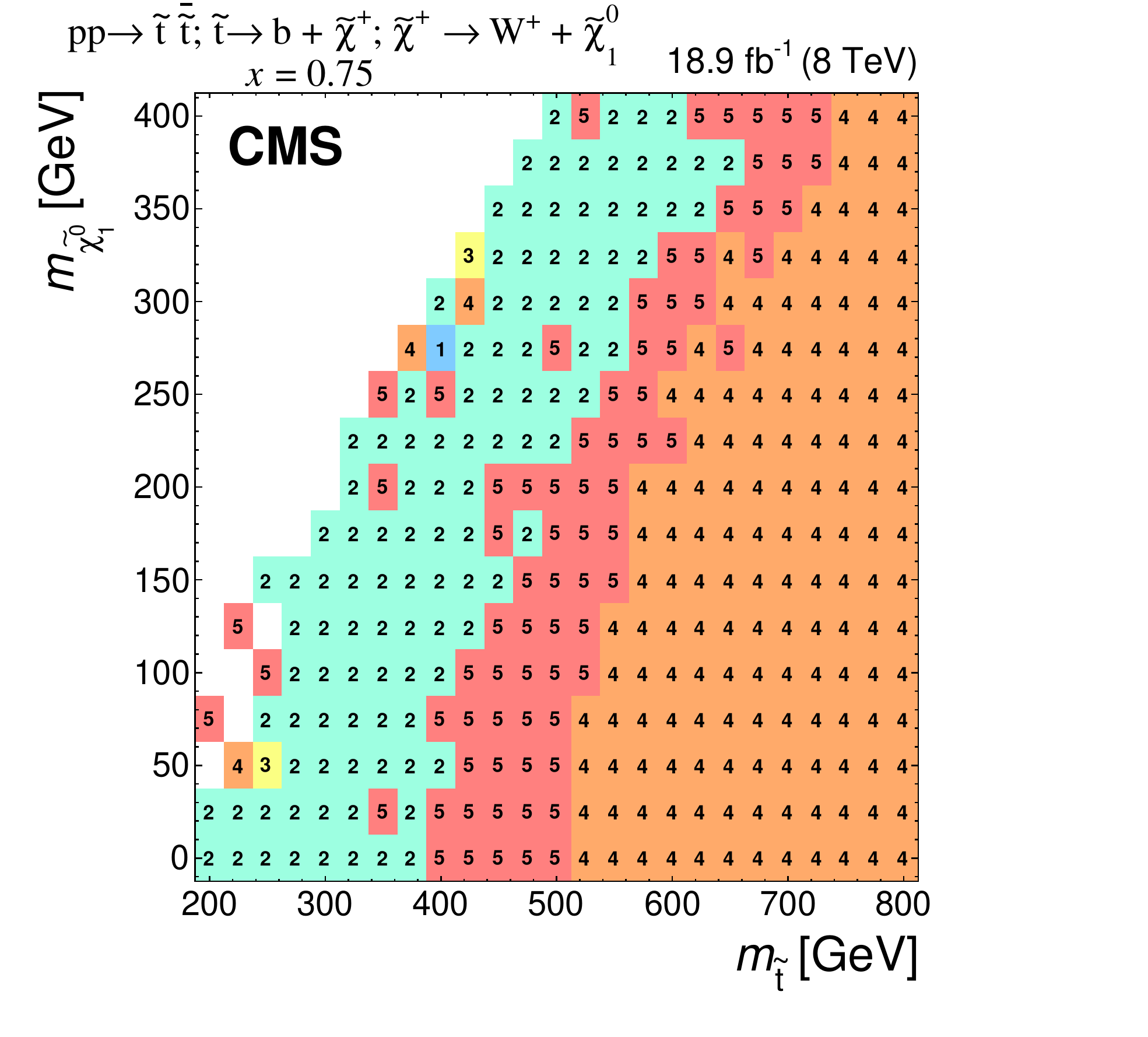}
  \caption{\label{fig:BestRegion} Search regions providing the most stringent limits in the $\mStop$-$\mLSP$ plane in the T2tt signal topology (top left) and the T2bW signal topologies for mass splitting parameter values  $x$ = 0.25, 0.50, 0.75. The T2tt\_LM, T2tt\_MM, T2tt\_HM, and T2tt\_VHM search regions are numbered 1, 2, 3, and 4, respectively. The T2bW\_LX, T2bW\_LM, T2bW\_MXHM, T2bW\_VHM, and T2bW\_HXHM search regions are numbered 1, 2, 3, 4, and 5 respectively.  In some regions, particularly with $\mLSP$ similar to $\mStop$, the different search regions can have similar sensitivity, which can lead to the fluctuations in choice of search regions in neighboring bins that is seen in some areas.}
\end{figure*}

Since there are potentially significant differences in the kinematic characteristics of signal samples at different points in the mass grids described above, it is not known a priori what is the minimum number of distinct BDTs that are needed to achieve the near optimal coverage of the signal spaces. To this end, a minimum number of BDTs that provides  sufficient coverage of each signal space is selected from a larger superset that includes BDTs that are each uniquely trained on grid points separated by $\approx$100\GeV in top squark mass and $\approx$50\GeV in neutralino mass for both signal types. For T2bW, there are also 3 different values of chargino mass that are considered, corresponding to $x=0.25$, 0.5, and 0.75.
Sensitivity to signal is probed by varying  discriminator thresholds from 0.5 to 1.0 in steps of 0.01. Ultimately it is determined that four BDTs for T2tt and five for T2bW are adequate to cover the largest possible parameter space with near optimal  signal sensitivity. Each BDT tends to cover a specific portion of signal space, referred to as a search region.  The optimisation of the overall search does not depend strongly on the specific signal points that are used to train individual BDTs. Moreover, adding more regions is not found to increase the sensitivity of the analysis. Table~\ref{tab:searchregions} lists the search regions for both signal types, the mass parameter points used to train each BDT, and the optimal BDT discriminator cutoffs that are used to define the final samples. Figure~\ref{fig:BestRegion} displays the most sensitive search regions in T2tt and selected T2bW mass planes. The colour plotted in any given partition of the plane corresponds to the search region BDT with the strongest expected limit on the signal production cross section.

For the T2tt search a total of 24 variables are used. They can be divided into variables that do or do not rely upon top quark pair reconstruction by the \CORRAL algorithm. The latter include $\ETm$, jet multiplicity, and $\mt$ calculated with $\ptvecmiss$ and the $\ptvec$ of the b-tagged picky jet that is closest to $\ptvecmiss$ in $\phi$. Of these, the most important variables for $\ttbar$ suppression are $\ETm$ and $\mt$. The $\mt$ distribution is peaked near the top quark mass for semileptonic $\ttbar$ events because nearly all of the $\ETm$ originates from the leptonic W decay, and the corresponding lepton is usually soft. On the other hand, there is no peak in the distribution for fully hadronic signal events. One variable suppresses SM background by exploiting the higher probability for jets in SM events, particularly Z+jets and W+jets,  to originate from gluons. It is the product of the quark-gluon likelihood values~\cite{CMS-PAS-JME-13-002} that are computed for each jet in the event. Two additional variables, the $\eta$ of the peak in jet activity and the $\Delta\eta$ between two peaks in jet activity, provide a measure of the centrality of the event activity. They are obtained by a kernel density estimate (KDE)~\cite{rosenblatt1956,parzen1962} of the one dimensional jet $\pt$ density. The KDE uses the jet $\eta$ as input with a jet $\pt$ weighted gaussian kernel function and a bandwidth parameter optimized on an event by event basis such that two peaks in the KDE are found. Another variable counts the number of unique combinations of jets that can form reconstructed top quark pairs. The remaining seventeen variables are all built with information pertaining to the candidate top quark pair obtained from \CORRAL. The invariant mass of the top quark pair and the relative $\pt$ of the two reconstructed top quarks are used to take into account correlations between the two top quark candidates that generally differ for  signal and background. The degree of boost or collimation of each top quark candidate is measured with three variables,  including the minimum cone size in the $\eta$-$\phi$ plane that contains all of the reconstructed particles from the top quark decay. Two variables use the \CORRAL discriminator value for each of the two top quarks as a measure of the quality of the reconstruction. Two other variables measure the angular correlation with \ptvecmiss for the lower-quality member of the top quark pair. The last eight variables are the $\pt$ values for the six jets in the top quark pair and two CSV b jet discriminator values that each correspond to the highest b tagging discriminator value obtained for the three jets that make up each of the two top quark candidates. While the properties of the reconstructed top quark pairs differ between signal events with two hadronic top decays and all SM background events with one or no hadronic top decays, the variables measuring the quality of the reconstruction are particularly useful for the suppression of Z+jets and W+jets since no reconstructed top quark candidates originate from hadronic top decays. A similar situation occurs  for the variables utilizing b jet discriminator values since these processes typically have fewer jets that originate from b quarks than signal processes. As explained in Section~\ref{sec:toptagger}, the kinematics of the reconstructed top quarks, such as their angular correlation with $\ptvecmiss$, are used for $\ttbar$ suppression.

There are 14 variables used to train the BDTs that target the T2bW final state, half of which are the same or very similar to those used for the T2tt final state. Four of these are commonly used to distinguish SM background from SUSY signals. They are $\ETm$, jet multiplicity, multiplicity of jets passing the CSV b tagger medium working point, and the azimuthal separation of the third-leading jet from $\ptvecmiss$. Variables that are sensitive to correlations between b jets and the rest of the event are the invariant mass formed with the two highest $\pt$ b-tagged jets; $\mt$ formed with $\ptvecmiss$ and the nearest b-tagged jet; and the standard deviation of the separation in pseudorapidity between the b-tagged jet with the highest $\pt$ and all other jets in the event. Three additional variables make use of quark-gluon likelihood values for the jets in the event, and a further set of three make use of jet kinematics. Of the last the most important is the scalar sum over $\pt$ of jets whose transverse momenta are within $\pi/2$ of the direction of $\ptvecmiss$, (i.e. $\Delta\phi(\ptJet,\ptvecmiss)<\pi/2$) divided by the corresponding sum for all jets that do not meet this criterion. This variable is particularly useful for suppression of Z+jets and W+jets since the jets and $\ptvecmiss$ in these events are typically opposite in $\phi$. This is not the case for signal events, for which the direction of $\ptvecmiss$ and hadronic activity is less correlated. For the calculation of the final variable, jets are first grouped into unique pairs by requiring the smallest separation distances in $\eta-\phi$ space. Of these, the invariant mass of the pair with the highest vector sum $\pt$ is found in simulation to have a high probability to correspond to the decay of a $\PW$ boson and is used to suppress $\Z$+jets events with $\Z\to \nu\PAGn$.

\begin{table*}[hbt]
\topcaption{\label{tab:searchregions}Search regions for the T2tt and T2bW channels. The table lists the SUSY particle masses used for the training of the BDTs, the cutoff on the BDT output, and the efficiency for the signal to pass the BDT selection relative to the baseline selection. The event counts of the T2bW discriminator training samples are limited and so four nearby mass points were used. They are the four combinations of the two $\PSQt$ and two $\PSGczDo$ masses listed. The signal efficiency in each row of the table is then that of the best case of the four, which in every case is the point with the largest $\mStop$ and smallest $\mLSP$ values of those indicated.   }
\centering
\begin{tabular}{lccccc}
\hline
    Search region &   $\mStop$ [GeV] & $\mLSP$ [GeV] & $x$ & Cutoff & Signal efficiency [\%] \\
\hline
    T2tt\_LM  &  300 & 25  & \NA & 0.79             & 8\\
    T2tt\_MM  &  425 & 75  & \NA & 0.83             & 16\\
    T2tt\_HM  &  550 & 25  & \NA & 0.92             & 25\\
    T2tt\_VHM &  675 & 250 & \NA & 0.95             & 19\\
    T2bW\_LX    & 550 \& 575 & 175 \& 200 & 0.25 & 0.94 & 25\\
    T2bW\_LM    & 350 \& 375 & 75 \& 100  & 0.75 & 0.73 & 10\\
    T2bW\_MXHM  & 550 \& 575 & 125 \& 150 & 0.50 & 0.92 & 14\\
    T2bW\_HXHM  & 400 \& 425 & 25 \& 50   & 0.75 & 0.82 & 10\\
    T2bW\_VHM   & 550 \& 575 & 25 \& 50   & 0.75 & 0.93 & 12\\
\hline
\end{tabular}
\end{table*}

\section{Estimation of SM backgrounds}
\label{sec:background}
We divide the important SM backgrounds into three classes. The first class, referred to as EW backgrounds, includes semileptonic and dileptonic decays of $\ttbar$, $\PW$+jets, single top, and $\Z$+jets with $\Z\to \nu\PAGn$. The second class of backgrounds originates from high-\ETm QCD multijet processes, and the third arises from associated production of $\ttbar\Z$ with $\Z\to \nu\PAGn$ and both top quarks decaying to hadrons. The latter produces a final state that is extremely similar to that of the signal but is fortunately very rare. The diboson contributions to search regions are studied in simulation and found to be negligible.

The estimation of the EW and QCD multijet backgrounds is based on MC samples in which the events have been reweighted by scale factors with values that are generally within a few percent of unity. As discussed in Section \ref{sec:mc-reweighting}, the scale factors are extracted from data-MC comparisons in control regions. The reweighting of the events assures that the simulation samples match data samples with regard to distributions of quantities that are relevant to the selection of events in the signal regions. However, it is important to note that the reweighted MC samples are not used directly to estimate backgrounds in the signal region. Rather, the search region yields and uncertainties are estimated by comparing the reweighted MC samples to data in background-specific control regions that differ from the search regions only in that they are obtained with selection criteria that simultaneously increase the purity of a single background and reduce any potential signal contamination.  In the case of the EW backgrounds the control regions are selected by requiring one or more isolated leptons, while for the QCD multijet background it is selected by requiring $\ptvecmiss$ to be aligned with one of the leading jets.

The $\ttbar\Z$ background is estimated directly from a sample of next-to-leading-order (NLO) MC simulation events generated with \MCATNLO. This procedure is motivated by the fact that $\ttbar\Z$ has a much lower cross section than other SM processes, making it impossible to define control regions that are both kinematically similar to the search regions and sufficiently well-populated to enable the extraction of scale factors.

\subsection{EW and QCD background estimates with MC reweighting}
\label{sec:mc-reweighting}

This analysis uses MC samples as the basis for the estimation of SM backgrounds in signal regions. These simulations have been extensively tested and tuned in CMS since the start of LHC data taking in 2009. As a result, they accurately reproduce effects related to the detailed geometry and material content of the apparatus, as well as those related to physics processes such as initial-state and final-state radiation. Nevertheless, the MC samples are not assumed to be perfect, discrepancies being observed with data in some kinematic regions.  Comparisons between data and MC simulation are therefore performed to derive scale factors in order to reduce the observed discrepancies.

The scale factors fall into two conceptually different categories. The first category involves  effects associated with detector modelling and object reconstruction that are manifested as discrepancies in jet and \ETm energy scales and resolutions, lepton and b jet reconstruction efficiencies, and trigger efficiencies. The second category corresponds to discrepancies associated with theoretical modelling of the physics processes as represented by differential cross sections in collision events. The scale factors in this category are estimated separately for each SM background process. The main sources of discrepancy here are finite order approximations in matrix element calculations and phenomenological models for parton showering and hadronisation. Scale factors are parameterised as a function of generator-level quantities controlling post-simulation event characteristics relevant to the final selection criteria used in the analysis. The scale factors are derived by comparing distributions of variables after full reconstruction that are particularly sensitive to these generator-level quantities, as seen in comparisons of MC with data.  D'Agostini unfolding with up to four iterations~\cite{DAgostini:1994z}, implemented with RooUnfold~\cite{Adye:2011gm}, is used to determine the correct normalization of  the generator-level quantities such that the distributions agree after full reconstruction. The scale factors are defined as the ratio of the corrected values of generator-level quantities to their original values. The MC events are reweighted by these scale factors, thereby eliminating any observed discrepancies with data. The scale factors are generally found to be close to unity as a result of the high quality of the MC simulation. The inclusive kinematic scale factors lead to no more than 10\% shifts in any regions of the distributions of \HT and number of jets that are relevant to this analysis.

\subsubsection{Detector modelling and object reconstruction effects}
The detector modelling and object reconstruction scale factors are grouped into the following categories: lepton identification efficiency, jet flavour, jet \pt, and \ptvecmiss.

For the lepton identification efficiency,  the event yields of simulated data passing the lepton vetoes in the search regions are corrected by scale factors as described in Section~\ref{sec:leptonvetoes}. The associated uncertainties in the search region predictions are denoted as ``MVA lepton sel. scale factors'' in Tables~\ref{tab:results_T2tt} and~\ref{tab:results_T2bW}. Similarly, in the control regions defined by the presence of a single lepton as described in Section~\ref{sec:eventselection}, scale factors are applied to the simulated electron and muon reconstruction, identification, and trigger efficiencies. These scale factors are measured by applying a ``tag-and-probe'' technique to the pairs of leptons coming from Z boson decays~\cite{Khachatryan:2015hwa,Chatrchyan:2012xi,Khachatryan:2010xn}.

Identification of jet type via b tagging is important for the \CORRAL top reconstruction algorithm and the signal discriminator used in the T2tt search. Both use the CSV b tagging algorithm output values directly rather than setting a particular cutoff value as is done for standard CMS loose, medium, and tight working points~\cite{Chatrchyan:2012jua}. It is therefore important that the CSV discriminator output distributions in simulated event samples match those  seen in corresponding data samples. To this end, the CSV discriminator output of each picky jet is corrected so that the CSV output distributions for simulated $\ttbar$ and $\Z$+jets event samples match those observed in the inclusive $\ttbar$ and $\Z$+jets control samples, respectively. Similarly, the quark-gluon likelihood distribution for jets is corrected to match data. The jet energy scale is corrected as described in Section~\ref{sec:eventselection}, and the simulated  picky jet $\pt$ spectrum is corrected as described in Section~\ref{sec:toptagger}.

The rejection of SM backgrounds in this analysis is very much dependent on the measurement of \ptvecmiss and its resolution, which is not modelled perfectly in simulation. Corrections are therefore applied to MC simulated samples of EW and QCD multijet processes in order to obtain good agreement with data in search region variables that depend on the correlation of event activity with \ptvecmiss. There are three separate corrections~\cite{CMS:vgm} applied for EW processes that are derived from a control sample of $\Z$+jets events with $\zll$ where, by conservation of energy and momentum, the reconstructed $\Z$ boson provides an accurate measure of the energy associated with all other activity in the event as measured in the transverse plane. Sources of genuine \ETm such as neutrinos in these events are rare and have a negligible effect on the derived corrections. The corrections are based upon comparisons of data to simulation in the inclusive Z+jets control sample in which \ptvecmiss is decomposed into components parallel and perpendicular to the direction of the $\Z$ boson $\ptvec$. The components and their resolutions are then investigated as a function of a variety of quantities to look for systematic trends and biases that can then be corrected. In this way, an \ETm scale correction of order 1\% is obtained as a function of both the boson $\pt$ and the distribution of hadronic energy in the event relative to the energy of the boson. The second and third corrections involve an increase in the jet resolution by 9\% and a smearing of the \ptvecmiss in both the directions parallel to the boson and perpendicular to it by approximately 4.5 $\GeV$. The measured resolutions of the components of \ptvecmiss along and perpendicular to the boson direction as obtained in simulation match those found in the data control regions after these corrections are applied.

For the EW backgrounds the \ptvecmiss corrections are parameterised in such a way that the corrected MC samples are consistent with data in \ptvecmiss-related quantities, such as the reconstructed $\PW$ boson $\mt$. In contrast, for the discrimination between QCD multijet events and SUSY signal events, the angular correlations between \ptvecmiss and the $\ptvec$ of leading jets in the event are the most important variables. Corrections are therefore obtained expressly for this background process with the inclusive QCD multijet control sample. The corrected simulation samples provide a good match to the angular correlations between \ptvecmiss and the leading jets in data.

\subsubsection{Corrections to the theoretical modelling of EW background processes}
\label{sec:kinrew}
The kinematic distributions of simulated EW processes are validated and corrected with three control samples having charged leptons in the final state: the high purity $\ttbar$, the inclusive Z+jets, and the inclusive W+jets control samples. Based on the physically reasonable assumption that the kinematics of the rest of the event should be largely independent of the boson decay(s) in these processes, the control samples are used in conjunction with corresponding MC samples to extract scale factors described below that are parameterised by generator-level quantities. They are then applied to MC samples in the search regions to estimate background contributions.

The scale factors are extracted as functions of the \pt of the boson in the case of $\PW$+jets and $\Z$+jets or of the momenta of the top quarks in the case of $\ttbar$. They also depend on the multiplicity and flavour of radiated jets as well as \HT. Because the control samples have finite sizes, the scale factors are organised into subsets that are derived and used sequentially. That is, prior to each derivation step, the scale factors extracted in the previous derivation steps are applied. For example, scale factors for correcting the $\ttbar$ jet multiplicity and top quark spectra are obtained and applied prior to calculating those used to correct the production of $\Z$ bosons in conjunction with heavy-flavour jets, since as much as 60\% of the events in the $\Z$ control sample are $\ttbar$ events.

There is no suitable control region to accurately measure corrections to the theoretical modelling of the single top process. However, a precise modelling of this process is not important as its contribution in the search regions is much smaller than that of $\ttbar$. A 50\% systematic uncertainty on the single top yield, estimated with simulation, is therefore used. It appears under the label ``Single top kinematics'' in Tables~\ref{tab:results_T2tt} and~\ref{tab:results_T2bW}.

\subsubsection{Estimation of EW background}
The corrections to the MC event samples based on scale factors, as discussed above, result in an agreement between MC and data distributions that is typically within 10\% for all control samples, including samples that were not used to extract the scale factors. This level of agreement is also found for distributions of many kinematic variables for which no corrections were explicitly applied.  There are a few regions in which kinematic distributions disagree at the level of 20\%, but these disagreements have been found to have a negligible impact on the search region predictions. A bootstrapping procedure is used~\cite{tEFR82a} to take into account statistical uncertainties in the derived scale factors for distributions of kinematic quantities and their correlations. The corresponding statistical uncertainty in the search region predictions is labelled ``Kinematics reweighting'' in Tables~\ref{tab:results_T2tt} and~\ref{tab:results_T2bW}. While the corrected MC and data distributions are found to agree in many control regions, the corrected MC is not used to directly estimate the background in the search regions. Instead, corrections specific to each search region are derived in addition to the more general scale factors previously described.

After correcting MC simulation samples for detector, reconstruction, and kinematic discrepancies, a closure correction and its uncertainty are measured, where closure is defined as the largest residual data-MC difference seen in a number of kinematic distributions. To this end, data-MC comparisons are performed in a variety of leptonic control regions for which the kinematic distributions under study are as similar as possible to those in the search regions as seen for MC samples that pass the signal selection criteria. The leptonic control samples used for the closure tests are obtained by applying the full set of baseline requirements, with the exception of the lepton vetoes. The control samples used to correct the $\ttbar$, $\PW$+jets and single top processes, referred to as the ``$1\ell$ closure samples,'' are subsets of the inclusive $\ttbar$ control sample, in which exactly one charged lepton has been identified. The charged lepton is removed from the list of physics objects in the event, leading to an additional component of \ptvecmiss that simulates the case in which the W boson decay has a large invisible component, which is common for events passing the search region selection.  As a result, many events with low intrinsic \ETm  pass the search region selection criteria, thereby enhancing the data statistics and significantly reducing the closure uncertainty. For similar reasons, this procedure also reduces potential contamination by semileptonic signal events to negligible levels. Likewise, ``$2\ell$ closure samples'' are subsets of the inclusive Z+jets control sample and are used to correct the $\Z$+jets process. The charged leptons are removed from the event, altering the \ptvecmiss to simulate the case in which the $Z$ boson decays to neutrinos.

\begin{figure*}[htb]
\centering
\includegraphics[width=0.40\linewidth]{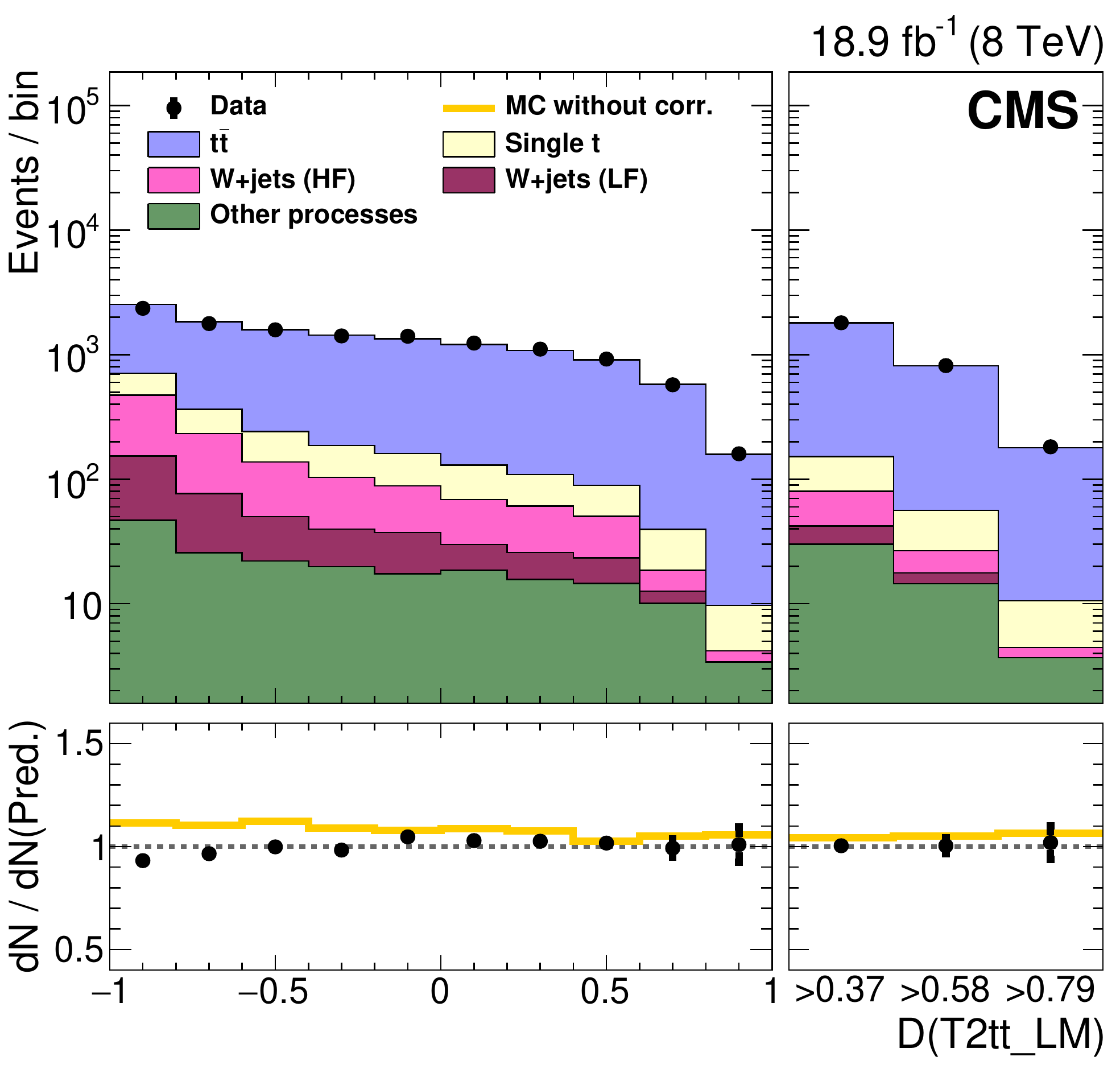}
\includegraphics[width=0.40\linewidth]{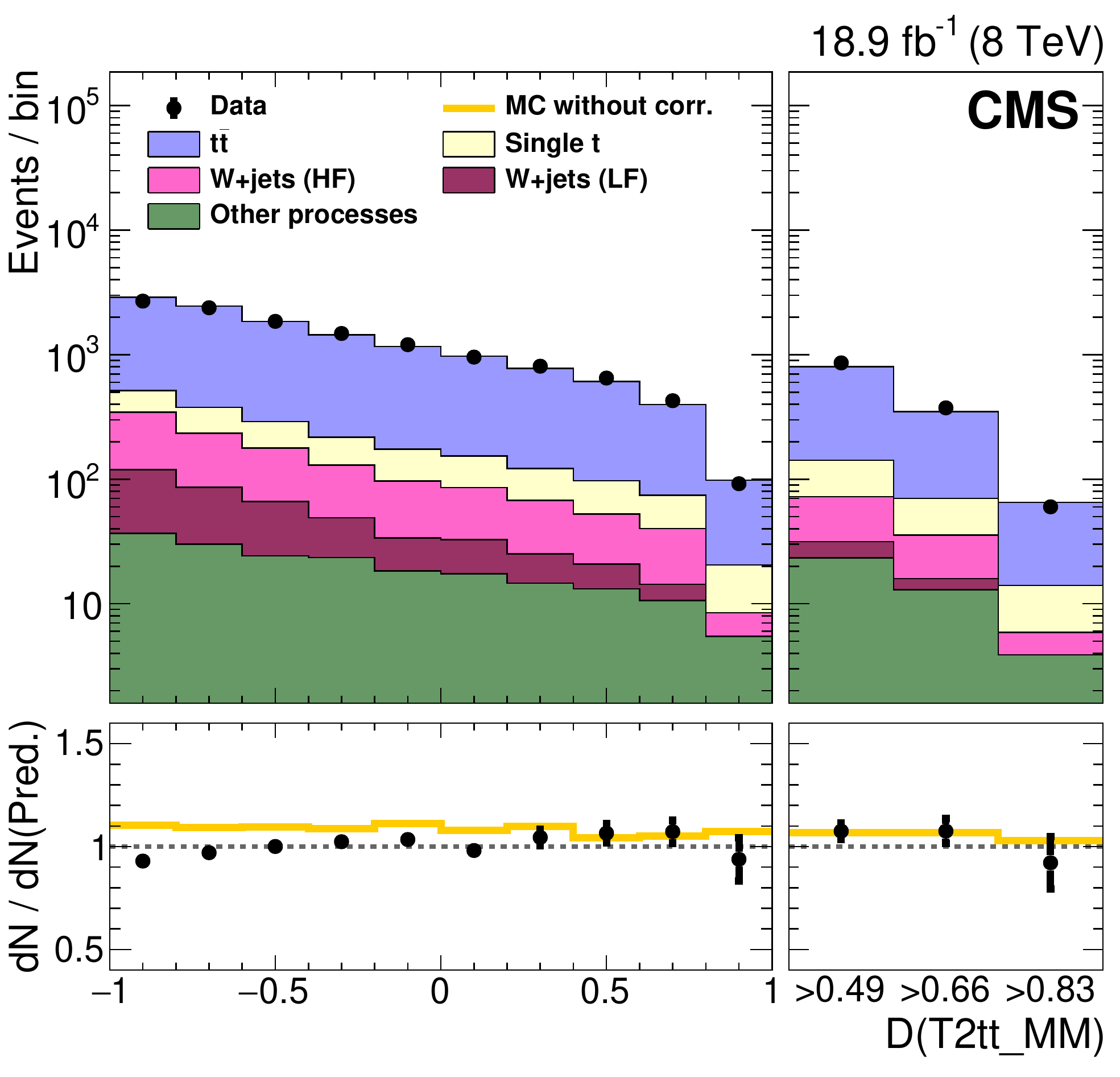}
\includegraphics[width=0.40\linewidth]{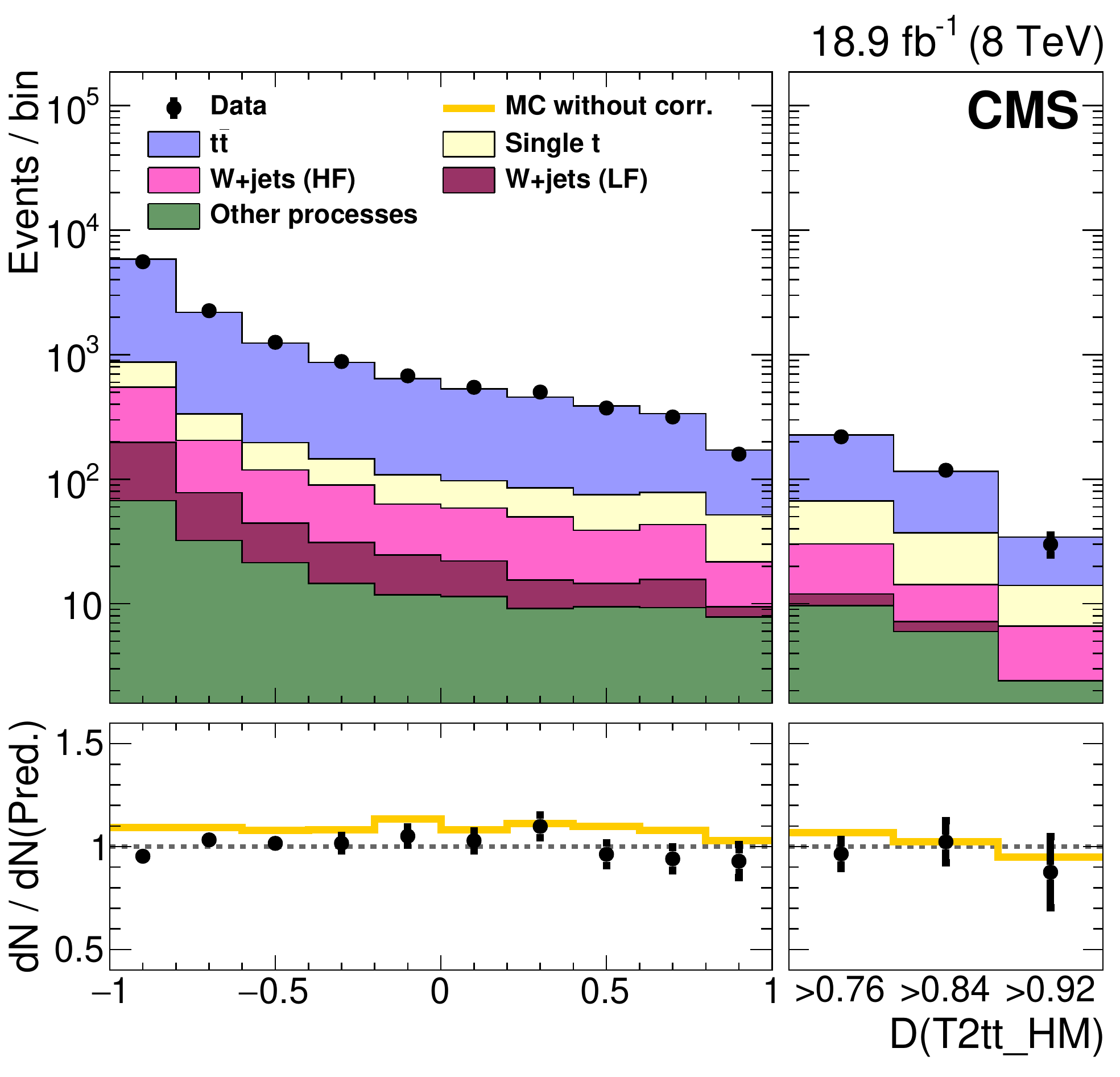}
\includegraphics[width=0.40\linewidth]{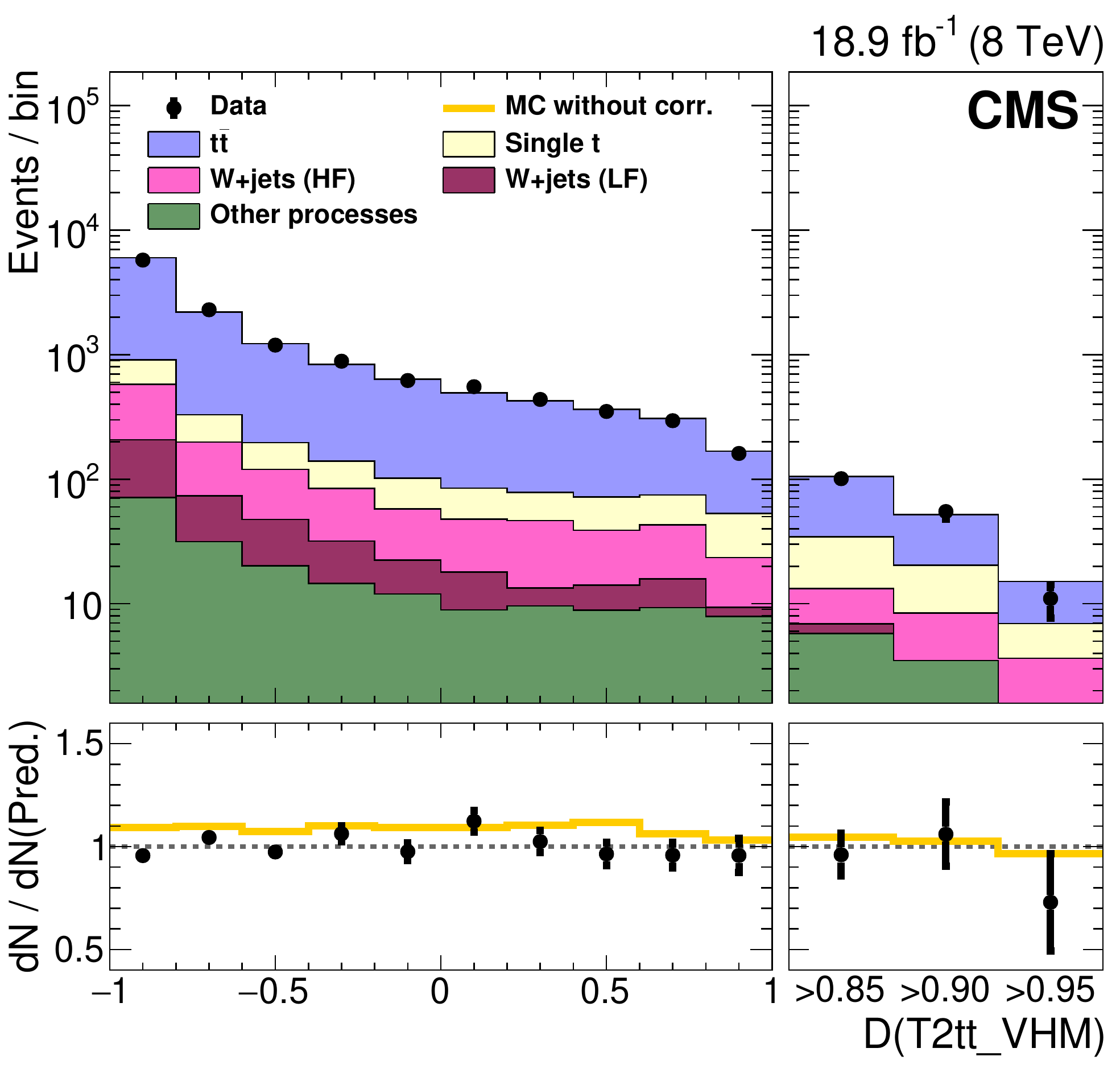}
\caption{\label{fig:mcrw-t2tt-closure-1l} Comparisons of BDT discriminator (D) outputs for data and corrected MC simulation for the $1\ell$ closure samples, with leptons removed, for the four T2tt validation regions.   The three bins at the far right in each plot are used to validate the MC performance in the signal region and its two extensions.  The points with error bars represent the event yields in data. The histogram labelled ``MC without corr." in the bottom pane of each figure plots the ratio whose numerator is the total MC event count before corrections and whose denominator is the event count for the corrected MC shown in the upper pane. The other histograms indicate the contributions of the various background processes. The ``LF'' and ``HF'' labels denote the subsets of the W+jets process in which the boson is produced in association with light and heavy flavour (b) quark jets, respectively.}
\end{figure*}

\begin{figure*}[htbp]
\centering
\includegraphics[width=0.40\linewidth]{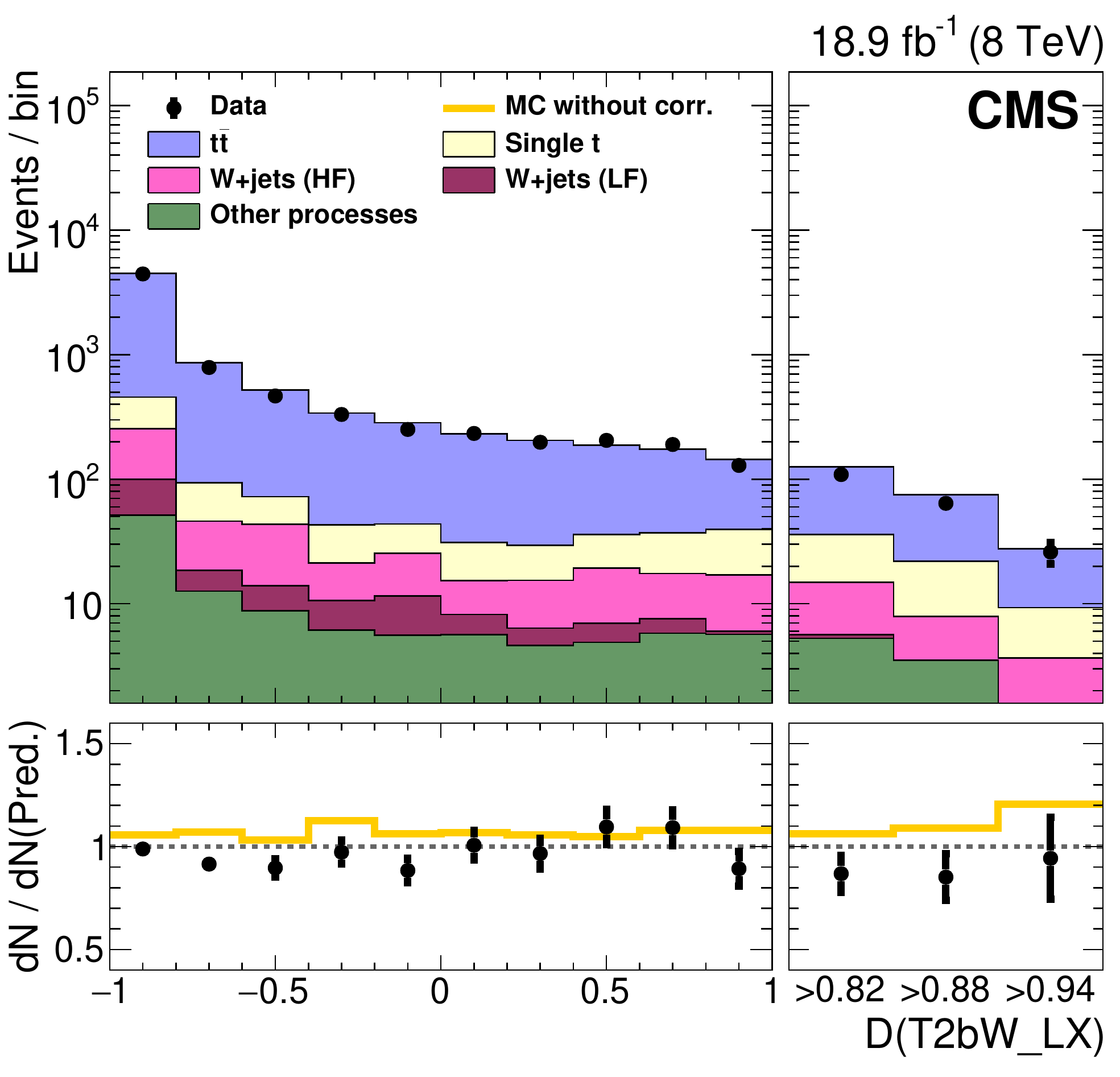}
\includegraphics[width=0.40\linewidth]{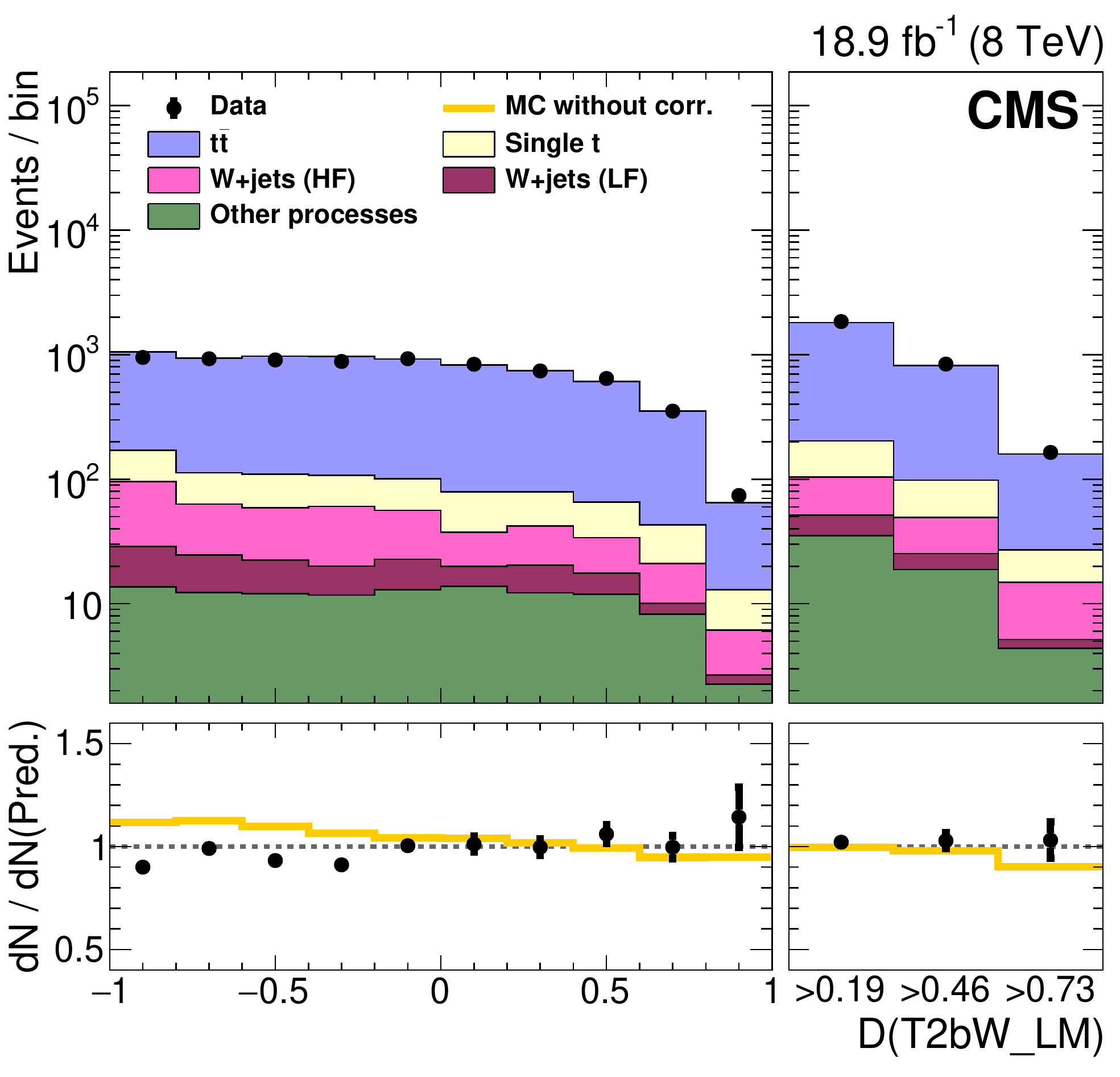}
\includegraphics[width=0.40\linewidth]{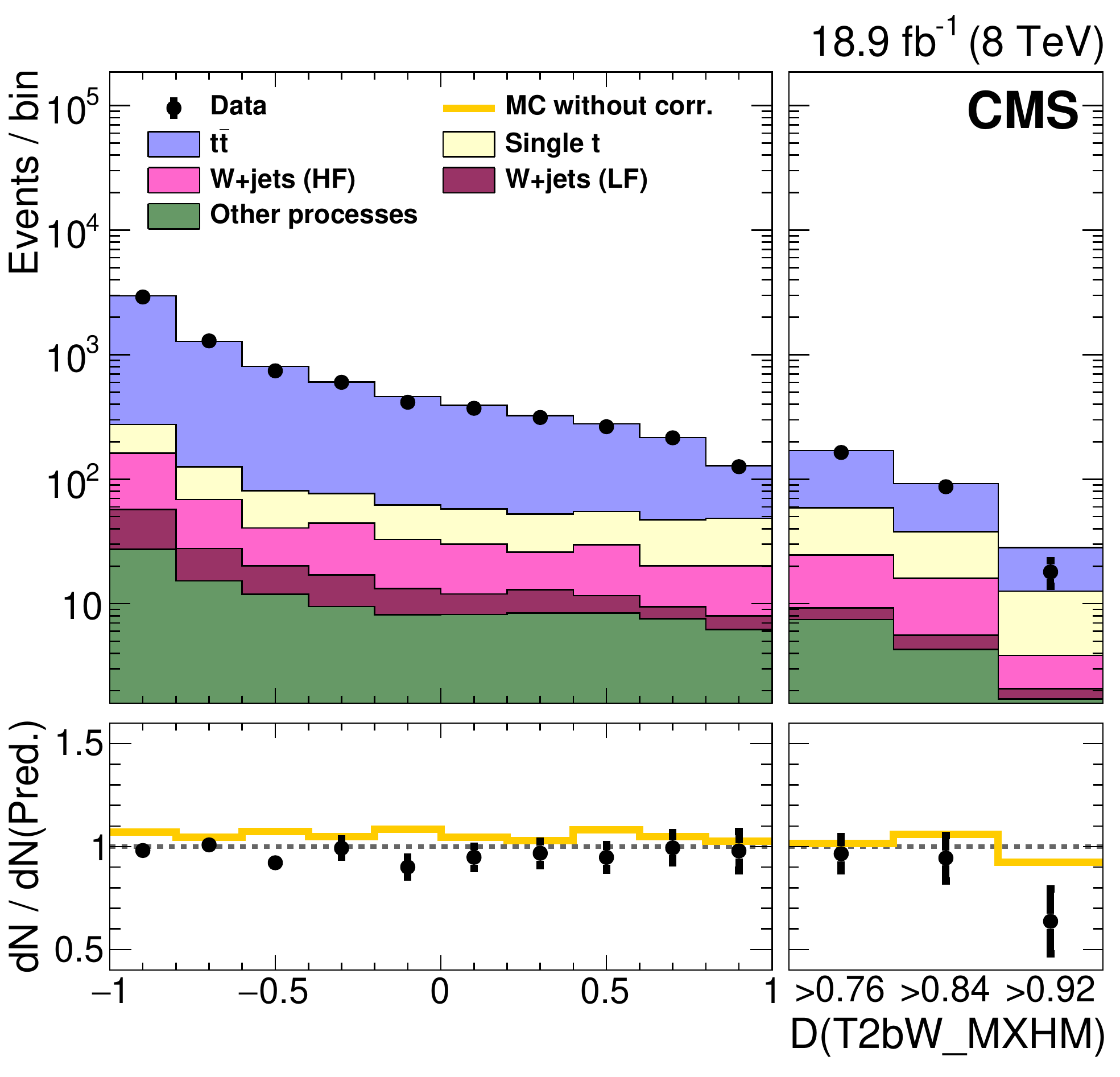}
\includegraphics[width=0.40\linewidth]{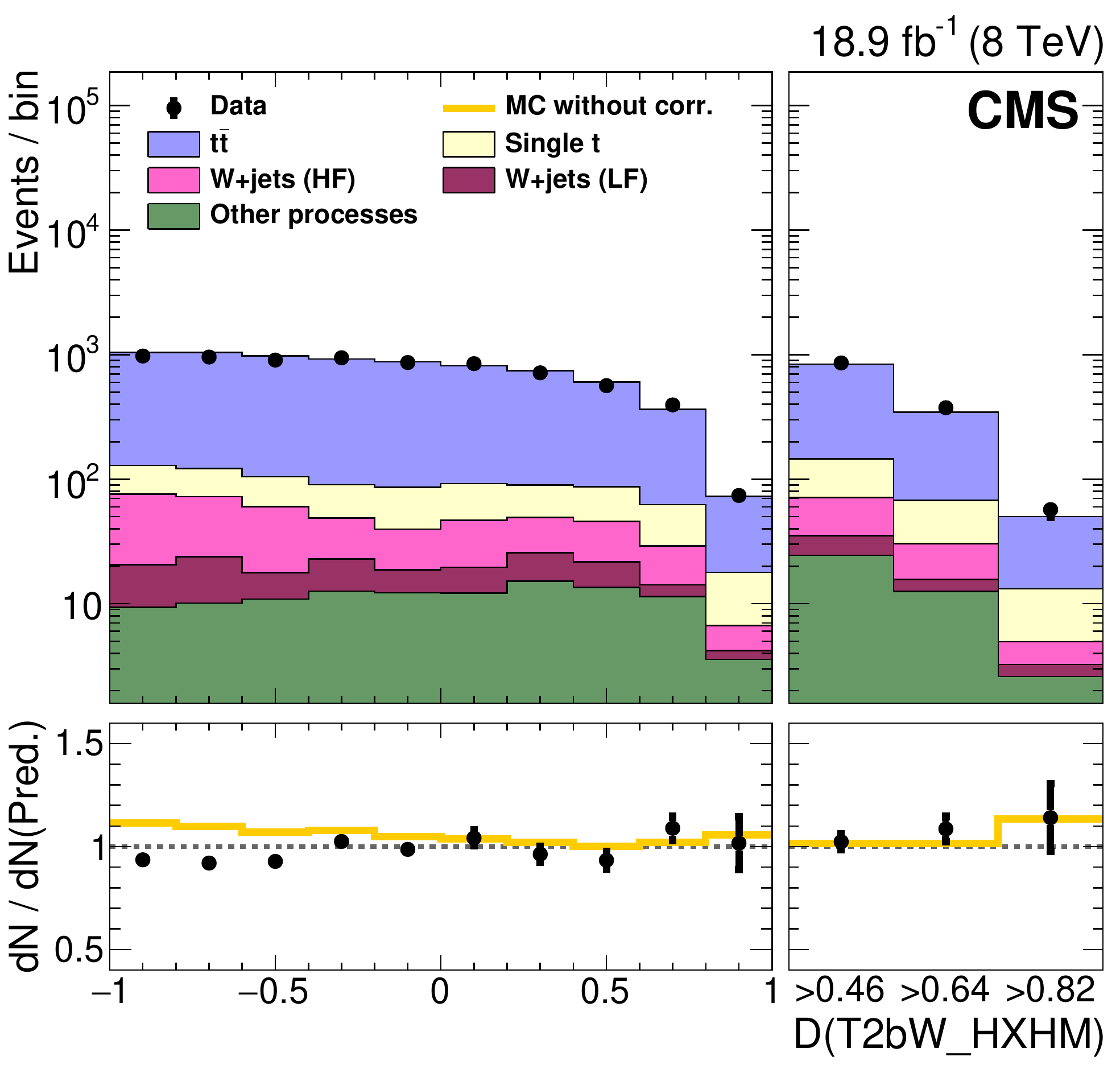}
\includegraphics[width=0.40\linewidth]{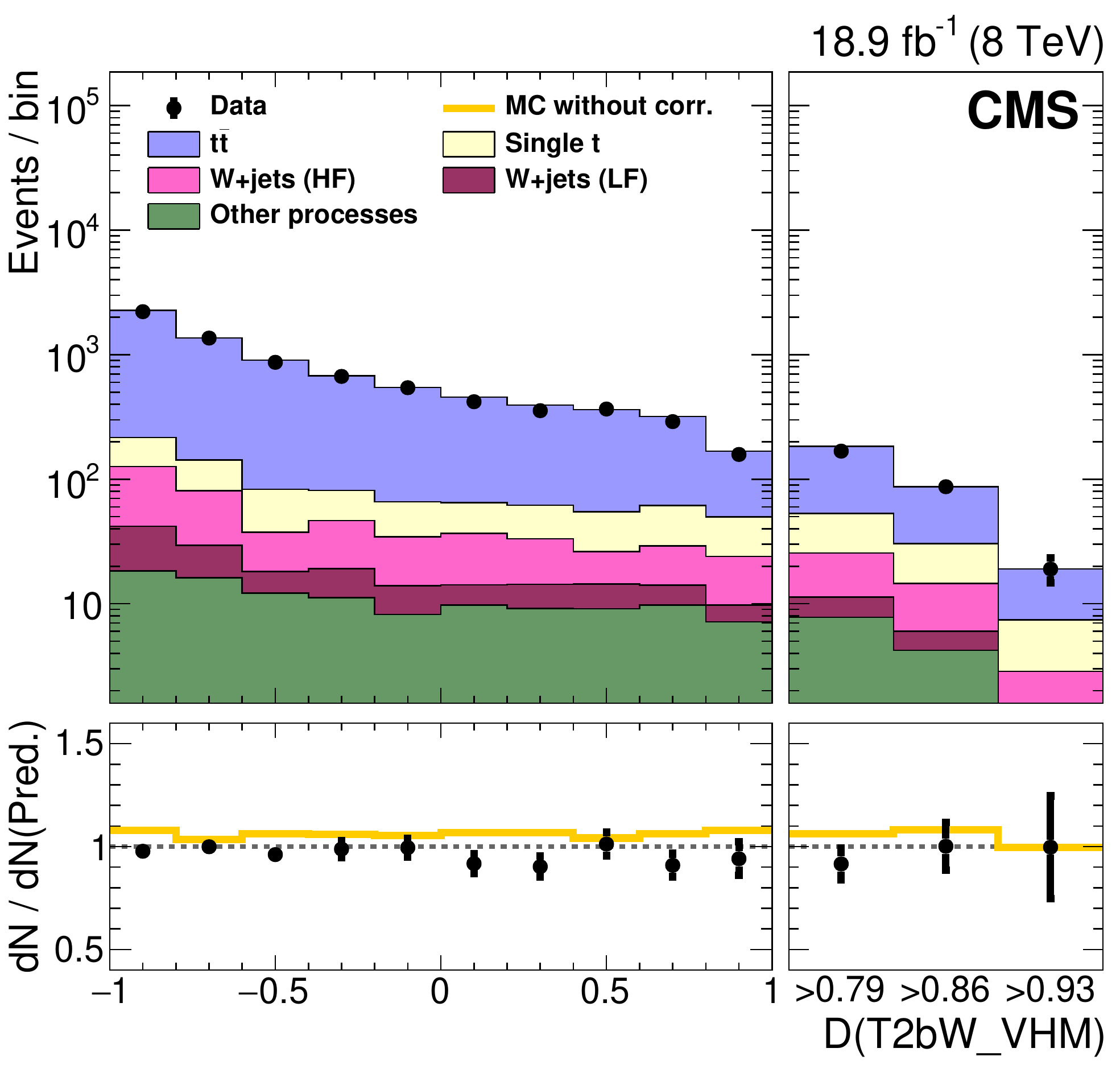}
\caption{\label{fig:mcrw-t2bw-closure-1l} Comparisons of BDT discriminator (D) outputs for data and corrected MC simulation for the $1\ell$ closure samples, with leptons removed, for the five T2bW validation regions.  The three bins at the far right in each plot are used to validate the MC performance in the signal region and its two extensions.  The points with error bars represent the event yields in data. The histogram labelled ``MC without corr." in the bottom pane of each figure plots the ratio whose numerator is the total MC event count before corrections and whose denominator is the event count for the corrected MC shown in the upper pane.   The other histograms indicate the contributions of the various background processes. The ``LF'' and ``HF'' labels denote the subsets of the W+jets process in which the boson is produced in association with light and heavy flavour (b) quark jets, respectively.}
\end{figure*}

Comparisons of the BDT discriminator outputs for data and corrected MC simulation for the $1\ell$ closure samples, after removal of the single identified charged lepton in each event, are shown in Figs. \ref{fig:mcrw-t2tt-closure-1l} and \ref{fig:mcrw-t2bw-closure-1l}, with the first ten bins in each plot covering the full BDT discriminator range. The closure is quantified by comparing the predicted event counts in MC simulation to those found in data in a `validation region', defined as the region containing the events with a single lepton that pass all of the final signal selection criteria after the lepton is removed, and in two control regions that extend the final search region to lower BDT discriminator values. The latter are defined by doubling and tripling the difference between unity and the discriminator cutoff value used for the final search region. These two additional regions are needed because the search region is statistically limited in some cases.  The results for the signal region and the two extended regions are shown in the last three bins in Figs. \ref{fig:mcrw-t2tt-closure-1l} and \ref{fig:mcrw-t2bw-closure-1l}, for the four T2tt and five T2bW BDT discriminators, respectively. The differences seen in the event counts for data and MC simulation in the extended regions are in general statistically compatible with the difference seen in the search region. Therefore, the data over simulation ratio in the first extended region is used as a correction for any potential residual bias in the event counts obtained with MC samples in which the events pass all of the signal region selection criteria, now including the lepton veto requirements.   The uncertainty in the correction is taken to be the statistical uncertainty in the data over simulation ratio in the last bin, which we have referred to as the validation region. This choice assures that the uncertainty covers any potential unknown differences between the search region and the first extended search region. For the four separate T2tt search regions, the largest correction is $1.08\pm0.13$ in the medium-mass region, with the closure uncertainties ranging from $\pm0.08$ in the low-mass region to $\pm0.24$ in the very-high-mass region. For the five separate T2bW search regions, the largest correction is $0.85\pm0.20$, and the uncertainties in the corrections range from $\pm0.09$ to $\pm0.25$. This uncertainty in the search region predictions is denoted as ``Closure ($1\ell$)'' in Tables~\ref{tab:results_T2tt} and~\ref{tab:results_T2bW}.

\begin{figure*}[htb]
\centering
\includegraphics[width=0.40\linewidth]{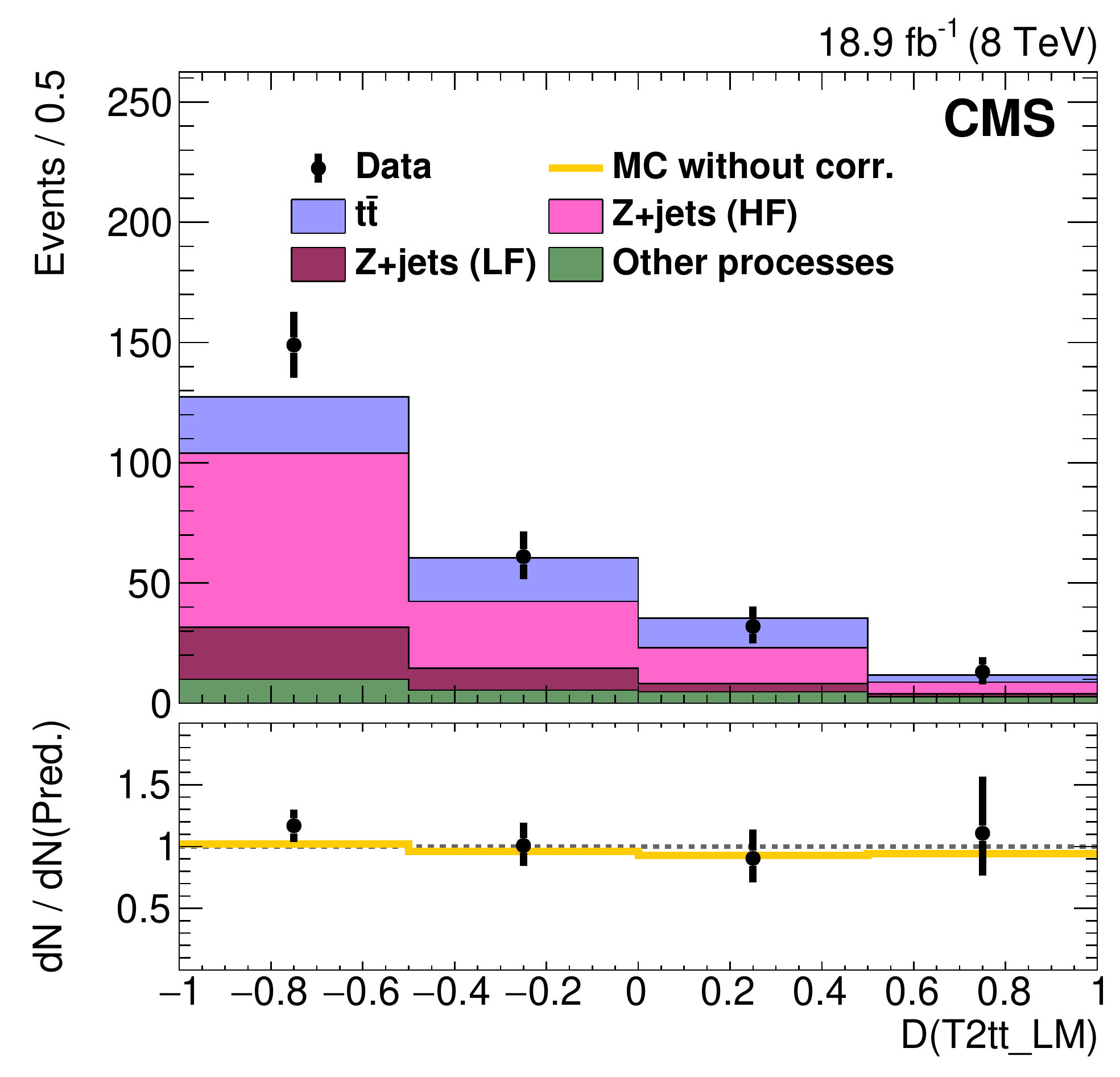}
\includegraphics[width=0.40\linewidth]{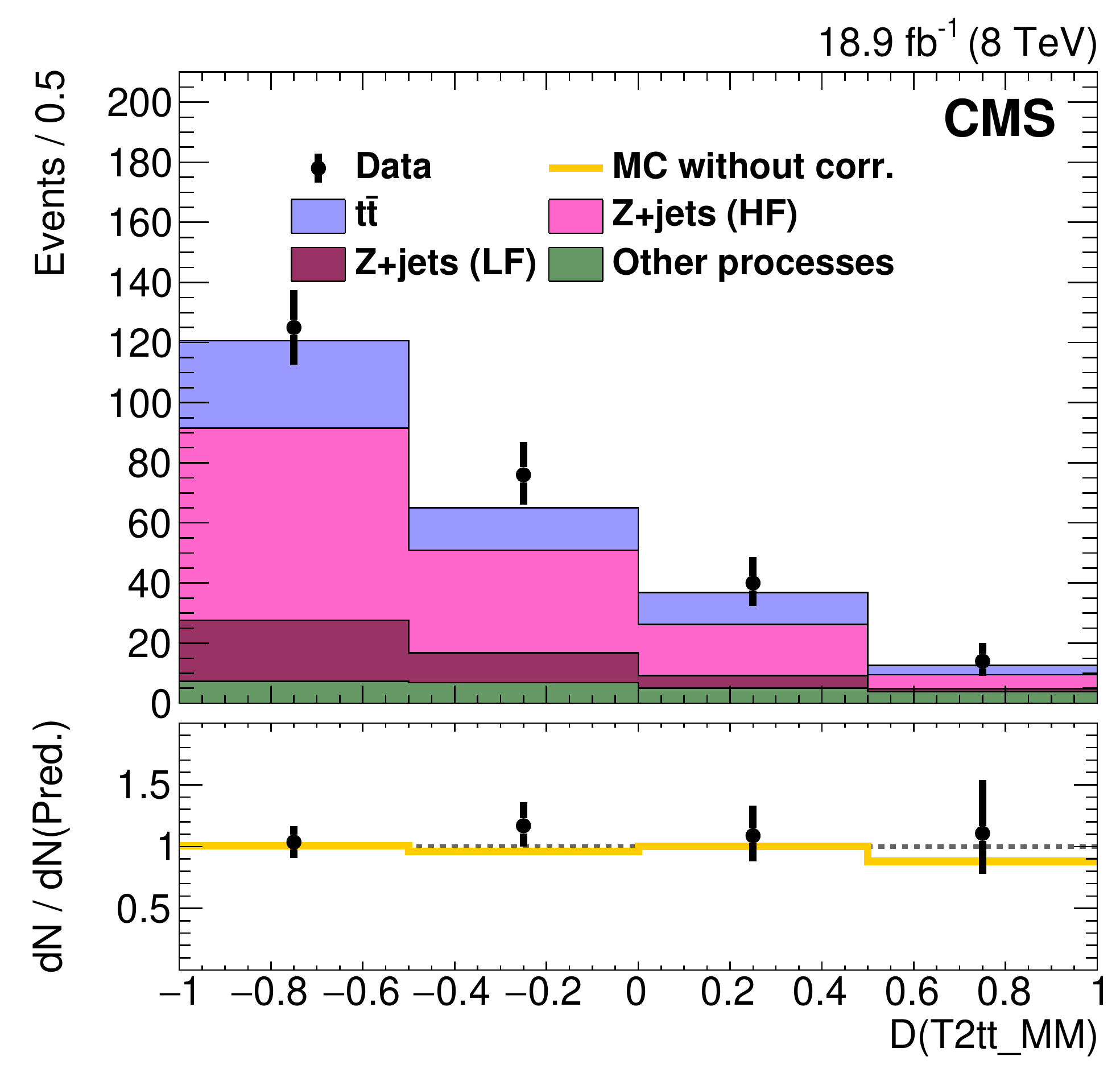}
\includegraphics[width=0.40\linewidth]{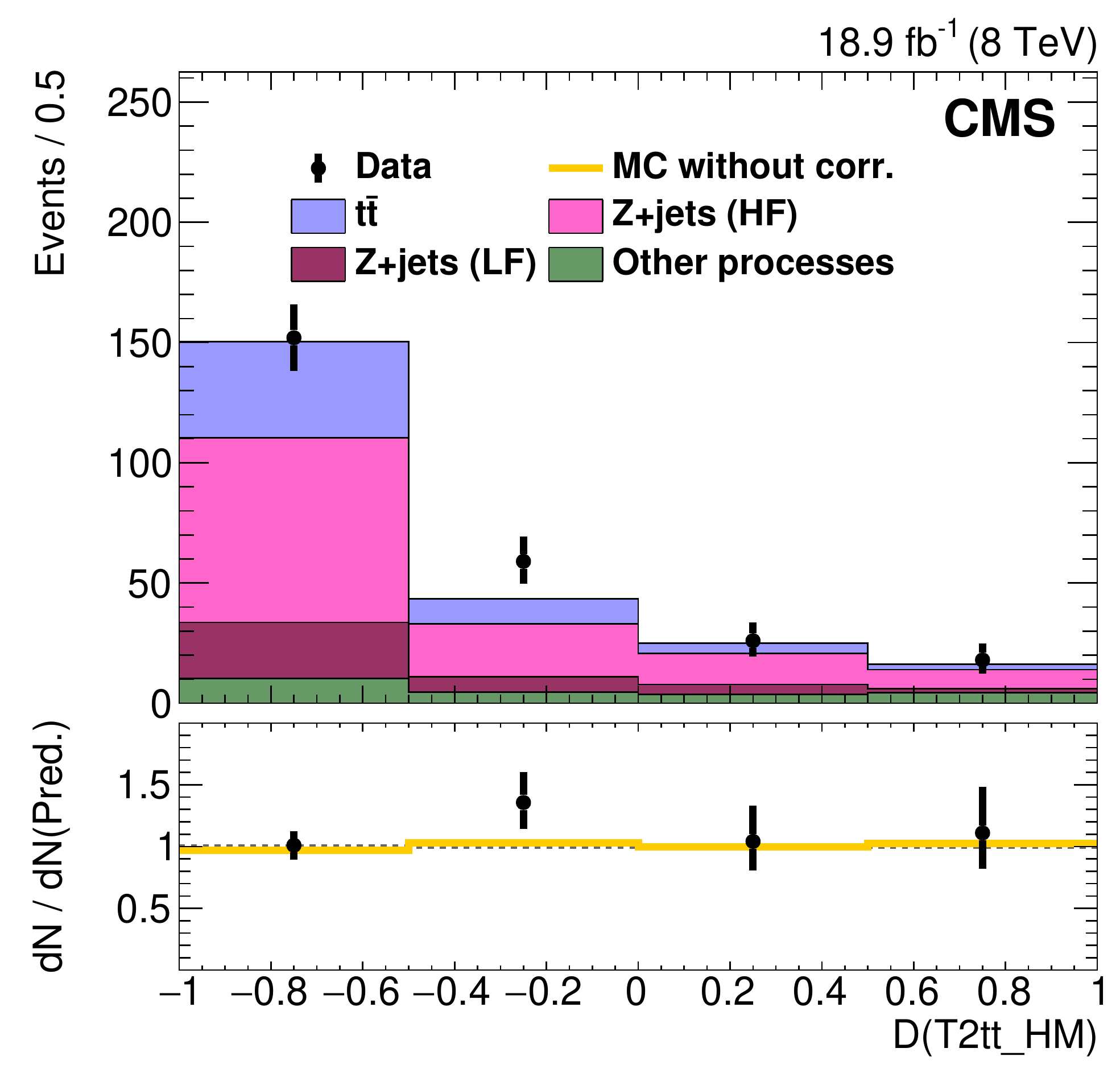}
\includegraphics[width=0.40\linewidth]{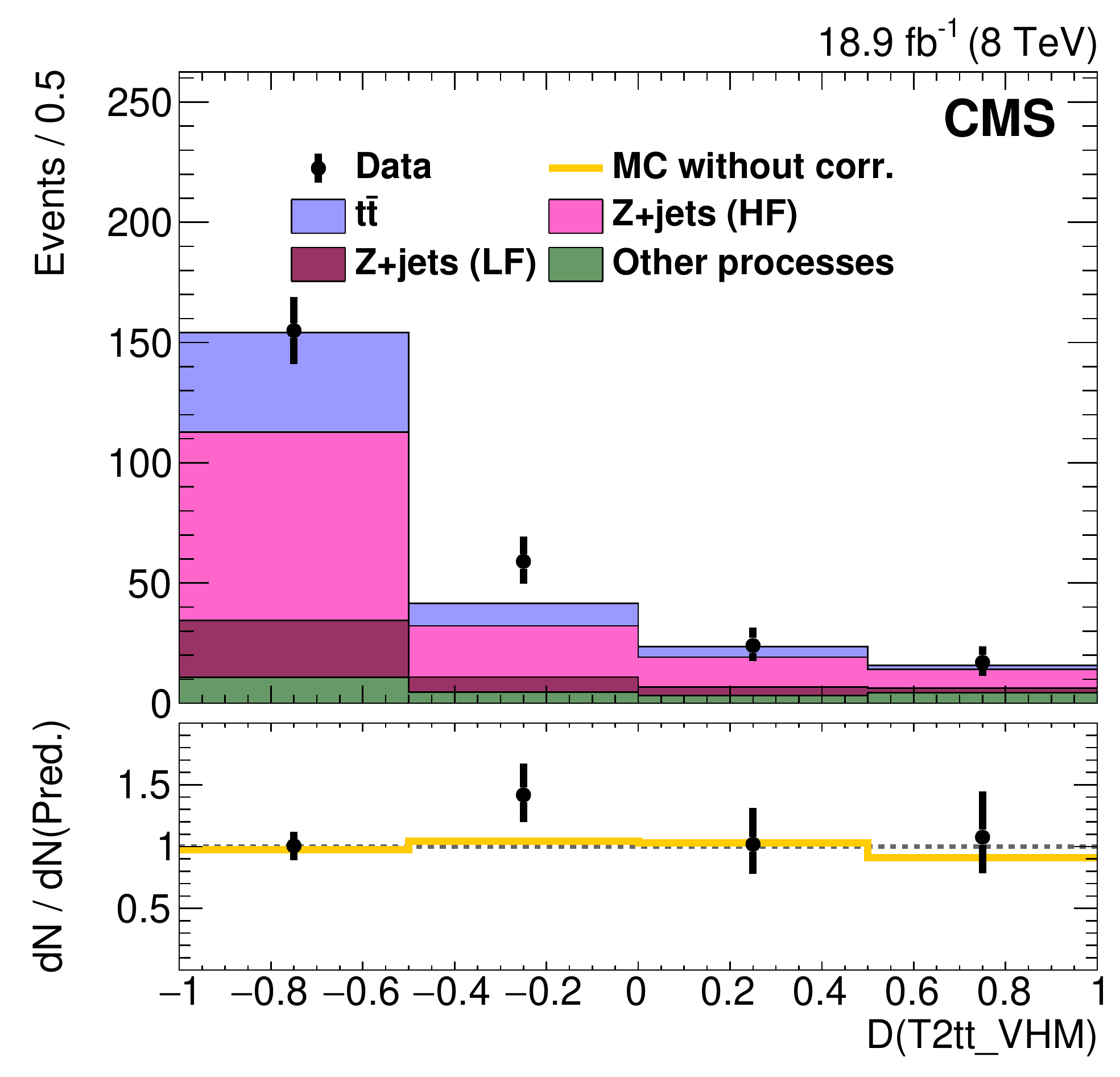}
\caption{\label{fig:mcrw-t2tt-closure-Zll} Comparisons of BDT discriminator (D) outputs for data and corrected MC simulation for the $2\ell$ closure samples, with leptons removed. All four T2tt validation regions are plotted.  The points with error bars represent the event yields in data. The histogram labelled ``MC without corr." in the bottom pane of each figure plots the ratio whose numerator is the total MC event count before corrections and whose denominator is the event count for the corrected MC shown in the upper pane.    The other histograms provide the contributions of the various background processes. The ``LF'' and ``HF'' labels denote the subsets of the Z+jets process in which the boson is produced in association with light and heavy flavour (b) quark jets, respectively.}
\end{figure*}

\begin{figure*}[htbp]
\centering
\includegraphics[width=0.40\linewidth]{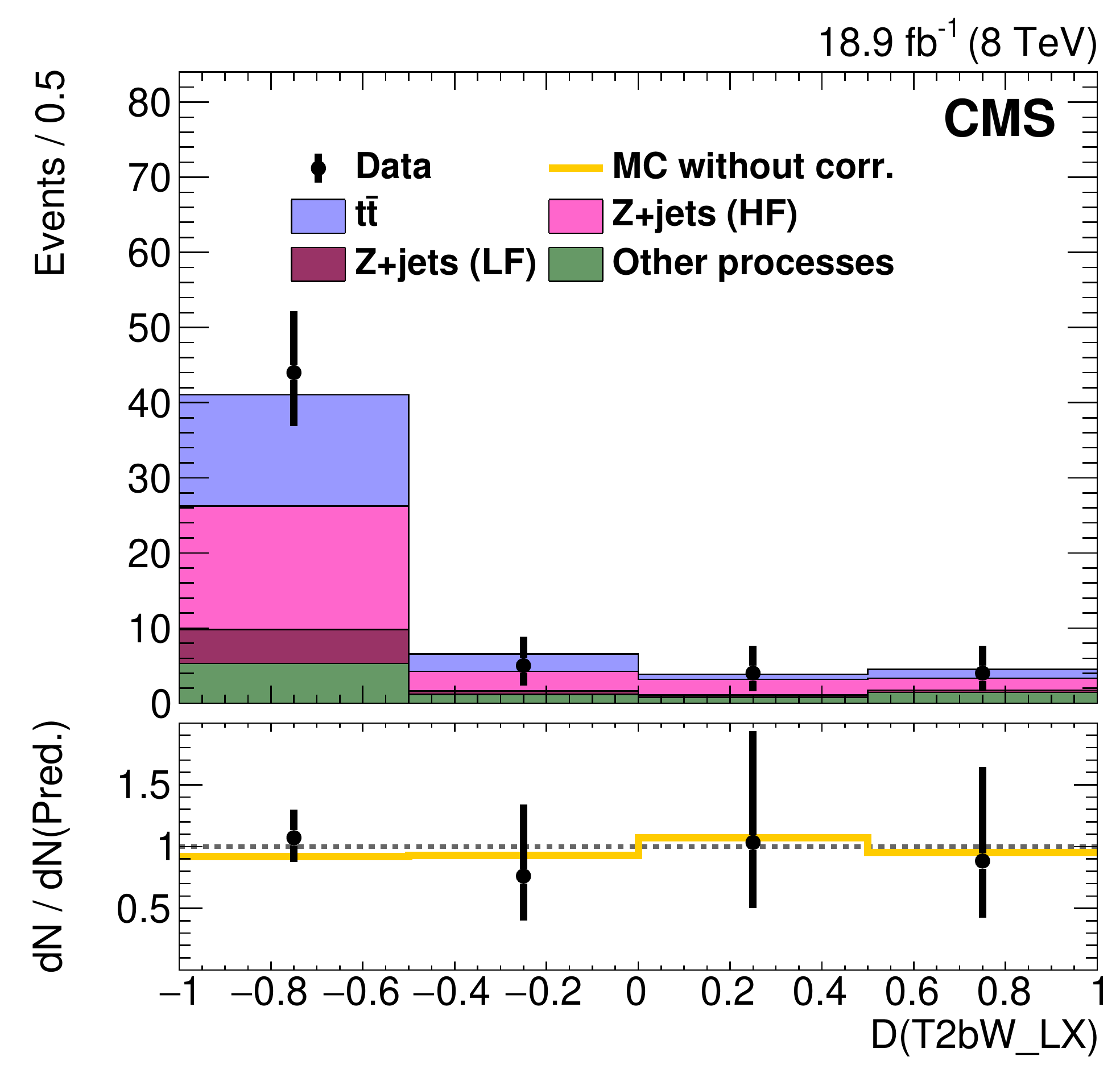}
\includegraphics[width=0.40\linewidth]{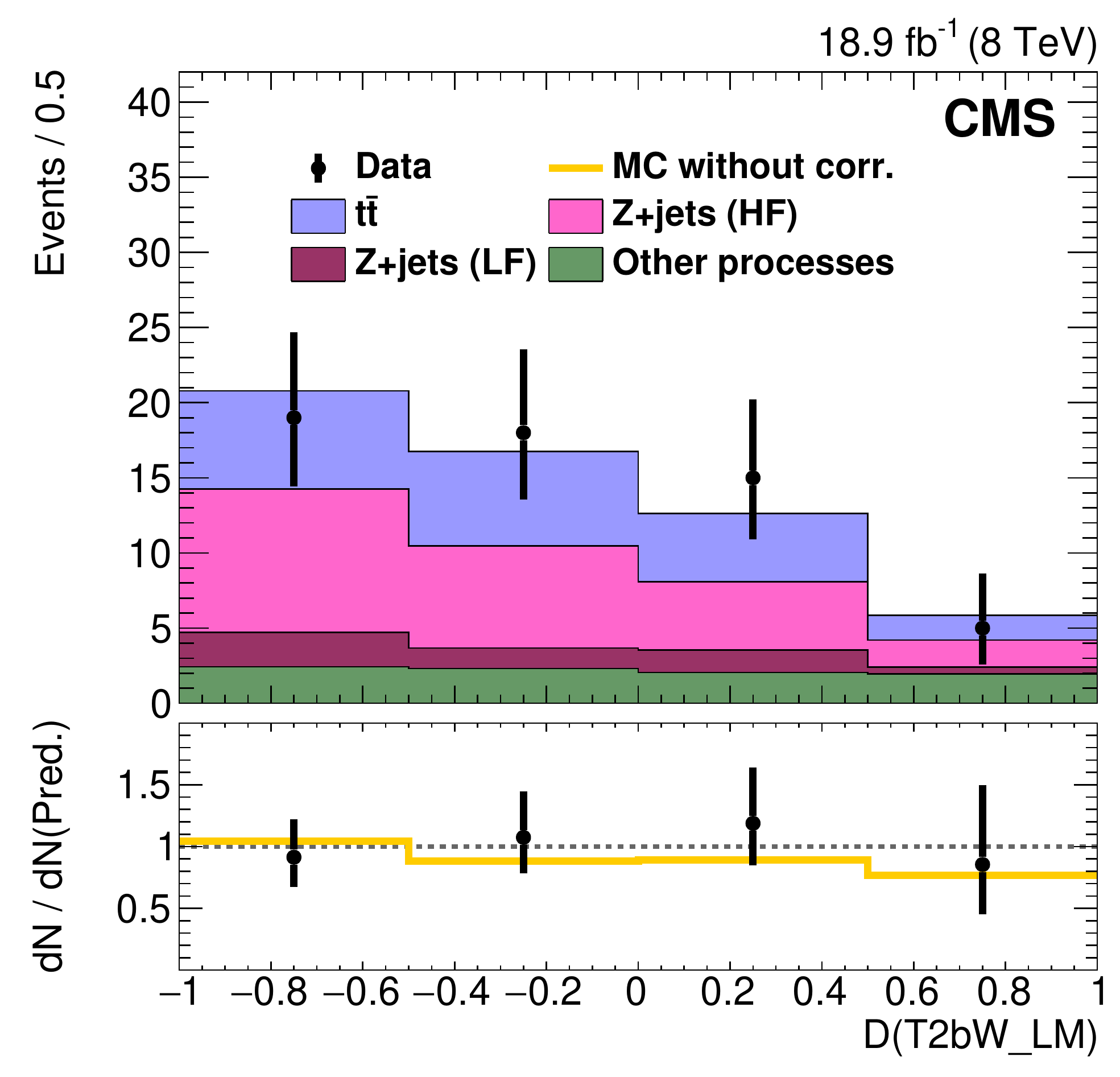}
\includegraphics[width=0.40\linewidth]{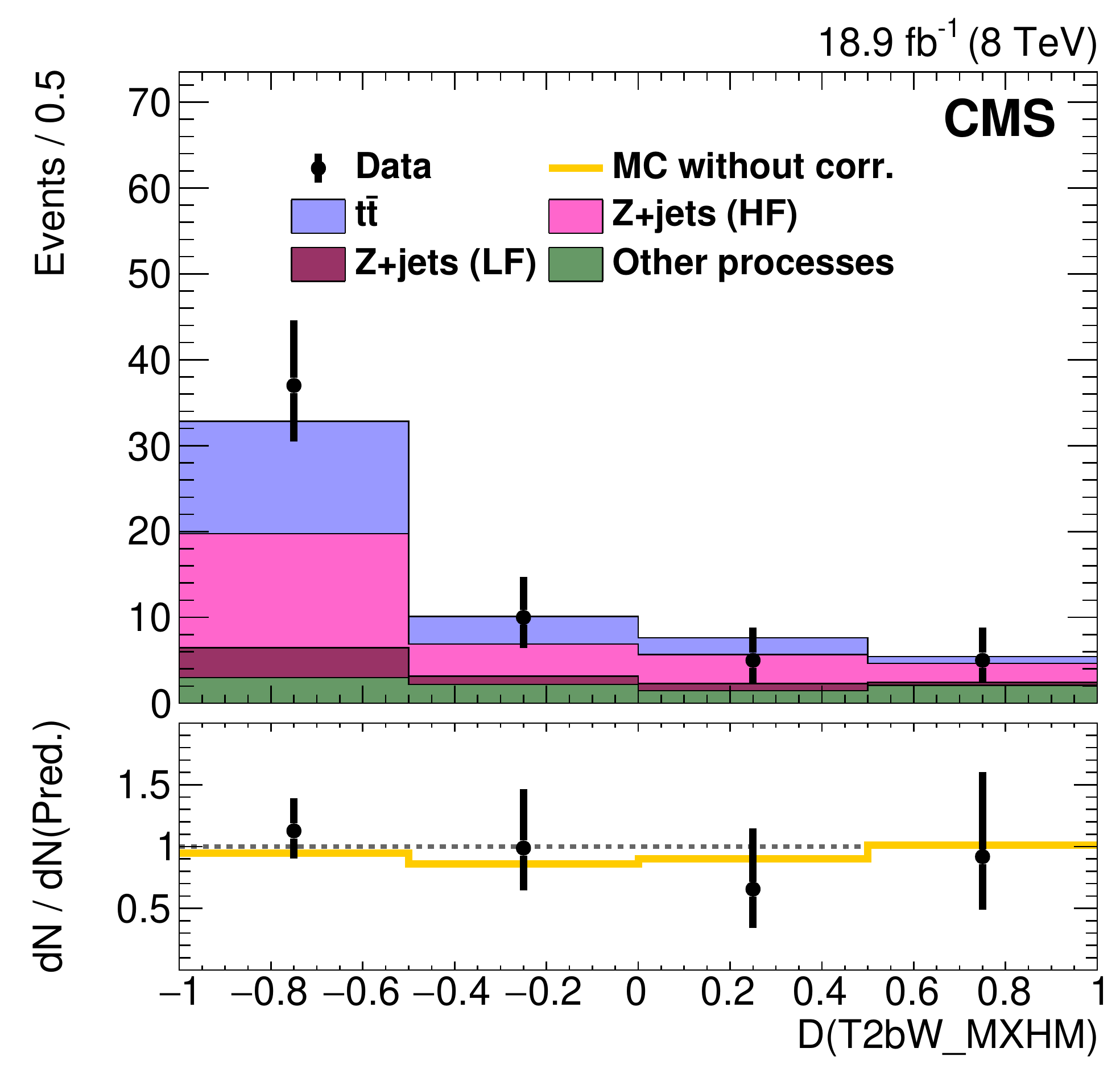}
\includegraphics[width=0.40\linewidth]{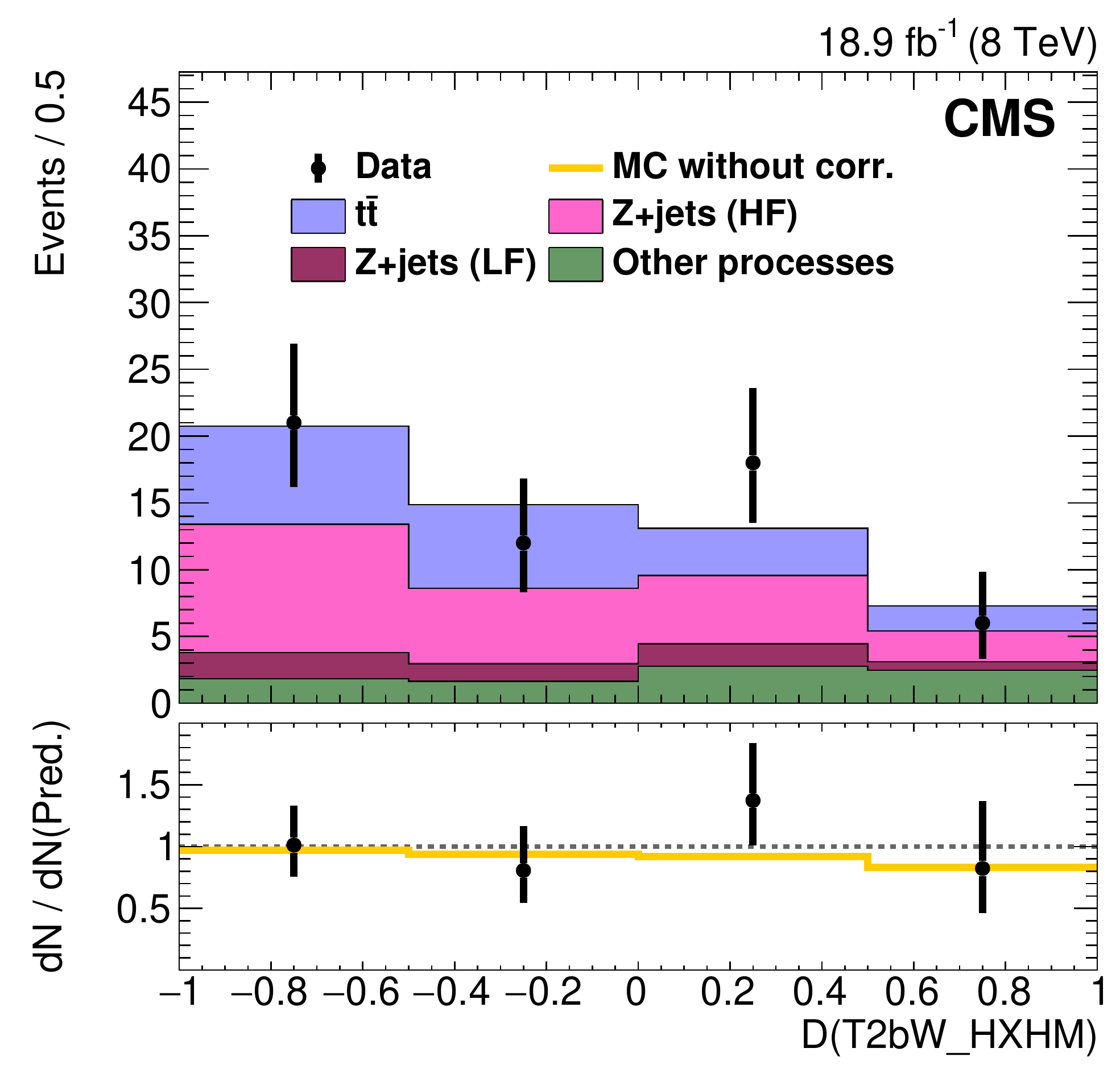}
\includegraphics[width=0.40\linewidth]{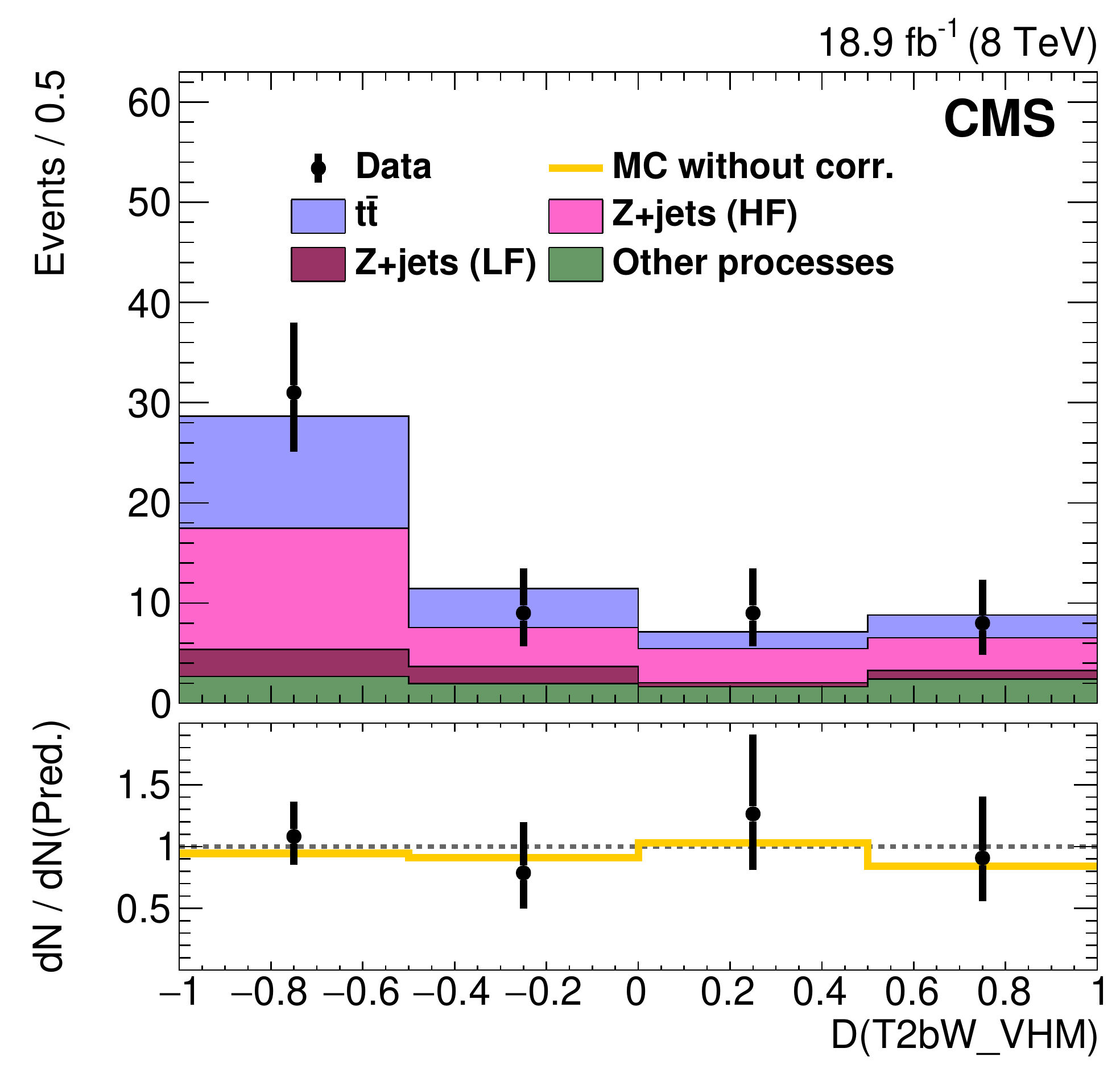}
\caption{\label{fig:mcrw-t2bw-closure-Zll} Comparisons of BDT discriminator (D) outputs for data and corrected MC simulation for the $2\ell$ closure samples, with leptons removed. All five T2bW validation regions are plotted.  The points with error bars represent the event yields in data. The histogram labelled ``MC without corr." in the bottom pane of each figure plots the ratio whose numerator is the total MC event count before corrections and whose denominator is the event count for the corrected MC shown in the upper pane.    The other histograms provide the contributions of the various background processes. The ``LF'' and ``HF'' labels denote the subsets of the Z+jets process in which the boson is produced in association with light and heavy flavour (b) quark jets, respectively.}
\end{figure*}

The simulated data are similarly compared to data in the $2\ell$ closure samples in Figs. \ref{fig:mcrw-t2tt-closure-Zll} and \ref{fig:mcrw-t2bw-closure-Zll}. No statistically significant lack of closure is observed for any of the T2tt and T2bW search regions. However, the small sample size makes it impossible to probe comparisons near to the search regions. An uncertainty is therefore obtained by measuring the largest data-MC discrepancy for each individual MVA input variable in the kinematic phase space of the search regions. This is defined for each input variable and search region as the ratio of event yields in data relative to MC simulation after reweighting both distributions. The weights that are used come from MC simulated distributions of the input variables after applying the MVA discriminator cutoff that is used for the search region. The distributions are normalised to unit area and the normalised bin contents are the final weights.  The weights are applied to binned events in both samples before taking the data/MC ratio in the control region where we measure the uncertainty.
The uncertainty in the $\Z$+jets background prediction is then taken to be the difference with respect to unity of this ratio for the variable with the largest degree of nonclosure, defined as  $\abs{({\text{Data}}/{\text{MC}}) - 1} / \sigma$ where $\sigma$ is the statistical uncertainty in the ratio. This closure test is repeated with successively tighter MVA discriminator cutoffs to check if the extracted closure uncertainty has any potential systematic trend related to discriminator cutoff. No significant trend is observed. To be conservative, the nonclosure is measured for an MVA discriminator value greater than or equal to 0.0 ($-0.5$) for T2tt (T2bW) search regions. These cutoff values are the highest ones for which the magnitude of the statistical uncertainty is smaller than the measured level of nonclosure. The uncertainties, denoted as ``Closure ($2\ell$)'' in Tables~\ref{tab:results_T2tt} and~\ref{tab:results_T2bW},  are found to range between 16\% and 39\%.

A separate control sample, which is similar to the baseline selection but with relaxed jet and b-tag requirements, is studied as an independent check of the $\Z$+jets and $\PW$+jets processes. Discrepancies of roughly 5\% in the event counts relative to those predicted are observed for both the $\Z$+jets and $\PW$+jets processes. The full magnitude of this discrepancy is taken as an additional uncertainty in the event counts for these background processes and it is included as ``Closure (relaxed baseline)'' in Tables~\ref{tab:results_T2tt} and~\ref{tab:results_T2bW}.

While the efficiencies for selecting electrons and muons in simulation are relatively well matched to what is seen in data, the efficiency for selecting $\Pgt$ leptons is observed to be significantly higher in simulation than in data for high values of some of the T2bW search region discriminators. The discrepancy is traced to a mismodelling of $\mt$, which, as discussed in Section ~\ref{sec:leptonvetoes}, is used for a preselection requirement of the tau veto. The mismodelling of $\mt$ is due to the angular component of \ptvecmiss and is uncorrelated with its magnitude. To address this, a correction and associated uncertainty are determined by means of a control region made up of modified events that is safe from signal contamination. The control region is defined by applying the full search region selection criteria to events in which search region discriminator values are calculated with a \ETm value that is randomly selected from the distribution of \ETm values obtained for the search region in MC simulation. A $\Pgt$ lepton veto efficiency is then obtained separately in data and simulation by taking the ratio of the number of events that pass the full set of signal region selection criteria but fail the $\Pgt$ lepton veto to the total number of events that pass the selection criteria prior to applying the $\Pgt$ lepton veto. The ratio of the $\Pgt$ lepton efficiency in data to the efficiency in simulation is then used to correct the efficiency for the simulated background samples with $\Pgt$ leptons from $\PW$ boson decays in the signal region. This correction reduces the data-MC discrepancy to a level that is not statistically significant and decreases the simulated $\Pgt$ lepton efficiency by a maximum of 29\% in all cases considered, with an uncertainty of 13\%. This uncertainty is included with the other lepton selection scale factor uncertainties under the label of ``MVA lepton sel. scale factors'' in Tables~\ref{tab:results_T2tt} and~\ref{tab:results_T2bW}.

The predictions in all search regions together with a breakdown of the various contributions to their uncertainties are provided in Tables~\ref{tab:results_T2tt} and~\ref{tab:results_T2bW}. After applying all corrections described in this section to the MC simulated data, no statistically significant discrepancies with data are observed in any bin of search region discriminator value for any search region.

\begin{table*}[ph!tb]
\topcaption{\label{tab:results_T2tt}Estimated contributions and uncertainties for the SM backgrounds in the T2tt search regions. The \ttbar, $\PW$+jets, single top, $\Z$+jets, and QCD multijet background estimates make use of MC simulated samples that have been weighted by scale factors obtained from data-MC comparisons as discussed in the text. The $\ttbar\Z$ background is estimated directly from simulation, with uncertainties assigned for sources of MC mismodelling.}
\centering
{
\begin{tabular}{lcccc}
\hline
& T2tt\_LM & T2tt\_MM & T2tt\_HM & T2tt\_VHM \\ \hline
\ttbar, W+jets, and single top prediction & 19.8 & 8.53 & 3.22 & 1.11 \\
Single top fraction (\%) & 3.69 & 7.71 & 19.1 & 29.8 \\
W+jets fraction (\%) & 2.27 & $<1\%$ & $<1\%$ & $<1\%$\\[\cmsTabSkip]
MC statistical uncertainty & 1.39 & 1.09 & 0.64 & 0.37 \\
MVA lepton sel. scale factors & 2.47 & 0.82 & 0.29 & 0.13 \\
Kinematics reweighting & 0.27 & 0.20 & 0.10 & 0.04 \\
Closure ($1\ell$) & 1.61 & 1.01 & 0.55 & 0.25 \\
Closure (relaxed baseline) & 0.02 & 0.01 & 0.01 & 0.01 \\
Single top kinematics & 0.37 & 0.33 & 0.31 & 0.17 \\[\cmsTabSkip]
Total uncertainty (yield) & 3.29 & 1.74 & 0.95 & 0.50 \\
Total uncertainty (\%) & 16.6 & 20.4 & 29.5 & 44.7 \\
\hline
$\Z$+jets prediction                 & {0.69} & {2.30} & {1.92} & {0.59} \\[\cmsTabSkip]
MC statistical uncertainty                         & 0.18 & 0.32 & 0.26 & 0.14 \\
Kinematics reweighting              & 0.08 & 0.38 & 0.54 & 0.18 \\
Closure ($2\ell$)     & 0.11 & 0.74 & 0.57 & 0.15 \\
Closure (relaxed baseline)               & 0.03 & 0.12 & 0.10 & 0.03 \\[\cmsTabSkip]
{Total uncertainty (yield)}  & {0.23} & {0.90} & {0.84} & {0.28} \\
{Total uncertainty (\%)}     & {33.5} & {38.9} & {43.8} & {46.4} \\
\hline
{$\ttbar\Z$ prediction}                         & {1.34} & {2.66} & {1.62} & {0.99}  \\[\cmsTabSkip]
MC statistical uncertainty                                  & 0.11 & 0.18 & 0.15 & 0.11  \\
MC simulation                               & 0.10 & 0.42 & 0.24 & 0.26  \\
MC normalisation                            & 0.42 & 0.82 & 0.50 & 0.31  \\
Kinematic closure                           & 0.21 & 0.85 & 0.49 & 0.26  \\[\cmsTabSkip]
{Total uncertainty (yield)}          & {0.49} & {1.27} &{0.75} & {0.49}  \\
{Total uncertainty (\%)}             & {36.6} & {47.7} & {46.6} & {49.5} \\
\hline
QCD multijet prediction  &	${0.33}$          &	${{<}0.01}$         &	${{<}0.01}$       &	${{<}0.01}$       \\[\cmsTabSkip]
MC statistical uncertainty                                      &	 $\pm 0.27$     &	 $\pm 0.01$   &	 $\pm 0.01$   &	 $\pm 0.01$   \\
MVA discriminator shape                         &	$\pm0.16$       &	$\pm 0.01$     &	$\pm 0.01$    &	$\pm 0.01$    \\
$\Delta\phi$ shape upper and lower bounds       &	$+1.48,-0.33$   &	$+0.22,-0.01$  &	$+0.07,-0.01$&	$+0.01$ \\
Low luminosity bins upper bound              &	\NA          &	+0.11           &	+0.02         &	+0.02         \\[\cmsTabSkip]
{Integrated uncertainty band ($\mu$)}    &	{0.91}            &	{0.17}            &	{0.04}          &	{0.01}          \\
{Integrated uncertainty band ($\sigma$)} &	{0.58}            &	{0.07}            &	{0.02}          &	{0.01}          \\
\hline
\end{tabular}
}
\end{table*}

\begin{table*}[h!tb]
\topcaption{\label{tab:results_T2bW}Estimated contributions and uncertainties for the SM backgrounds in the T2bW search regions. The \ttbar, $\PW$+jets, single top, $\Z$+jets, and QCD multijet background estimates make use of MC simulated samples that have been weighted by scale factors obtained from data-MC comparisons as discussed in the text. The $\ttbar\Z$ background is estimated directly from simulation, with uncertainties assigned for sources of MC mismodelling.}
\centering
\cmsTableResize
{
\begin{tabular}{lccccc}
\hline
& T2bW\_LX & T2bW\_LM & T2bW\_MXHM & T2bW\_HXHM & T2bW\_VHM \\
\hline
\ttbar, W+jets, and single top prediction & 6.88 & 31.3 & 3.89 & 12.7 & 2.31 \\
Single top fraction (\%) & 21.4 & 8.54 & 31.8 & 14.8 & 28.6 \\
W+jets fraction (\%) & 13.5 & 4.53 & 6.60 & 14.6 & 4.17 \\[\cmsTabSkip]
MC statistical uncertainty & 0.73 & 1.62 & 0.49 & 1.13 & 0.37 \\
MVA lepton sel. scale factors & 1.05 & 2.30 & 0.60 & 1.68 & 0.37 \\
Kinematics reweighting & 0.17 & 0.42 & 0.11 & 0.23 & 0.10 \\
Closure ($1\ell$) & 1.60 & 2.69 & 0.65 & 1.93 & 0.58 \\
Closure (relaxed baseline) & 0.05 & 0.07 & 0.01 & 0.09 & 0.01 \\
Single top kinematics & 0.73 & 1.34 & 0.62 & 0.94 & 0.33 \\[\cmsTabSkip]
Total uncertainty (yield) & 2.18 & 4.13 & 1.19 & 2.96 & 0.85 \\
Total uncertainty (\%) & 31.8 & 13.2 & 30.5 & 23.3 & 36.7 \\
\hline
   {$\Z$+jets prediction}                   &{1.88} & {4.57} & {1.66} & {1.77} & {1.24} \\[\cmsTabSkip]
MC statistical uncertainty                                                          &0.23 & 0.46 & 0.24 & 0.26 & 0.21 \\
Kinematics reweighting                                             &0.51 & 0.62 & 0.46 & 0.36 & 0.38 \\
Closure ($2\ell$)                                    &0.73 & 1.46 & 0.50 & 0.57 & 0.31 \\
Closure (relaxed baseline)                                              &0.09 & 0.23 & 0.08 & 0.09 & 0.06 \\[\cmsTabSkip]
{Total uncertainty (yield)}                                 &{0.93} & {1.67} & {0.72} & {0.73} & {0.54} \\
{Total uncertainty (\%)}                                    &{49.3} & {36.6} & {43.6} & {41.0} & {43.4} \\
\hline
    {$\ttbar\Z$ prediction}             & {0.59} & {2.46} & {0.83} & {1.72} & {0.62}  \\[\cmsTabSkip]
MC statistical uncertainty                                                          & 0.07 & 0.15 & 0.09 & 0.14 & 0.08  \\
MC simulation                                                      & 0.02 & 0.10 & 0.10 & 0.17 & 0.02  \\
MC normalisation                                                   & 0.18 & 0.76 & 0.26 & 0.53 & 0.19  \\
Kinematic closure                                                  & 0.23 & 0.79 & 0.25 & 0.55 & 0.15  \\[\cmsTabSkip]
{Total uncertainty (yield)}                                 & {0.30} & {1.11} & {0.39} & {0.79} & {0.26} \\
{Total uncertainty (\%)}                                    & {51.2} & {45.1} & {46.3} & {46.3} & {42.2} \\
\hline
{QCD multijet prediction}                     &    ${0.51}$          &    ${0.07}$          &    ${0.10}$          &    ${{<}0.01}$         &    ${{<}0.01}$        \\[\cmsTabSkip]
MC statistical uncertainty                                                        &    $\pm0.21$      &    $\pm0.06$      &     $\pm0.08$     &    $ \pm 0.01$    &     $\pm 0.01$    \\
MVA discriminator shape                                            &    $\pm0.17$       &    $\pm0.06$       &    $\pm0.08$       &    $\pm0.01$      &    $\pm 0.01$      \\
$\Delta\phi$ shape upper and lower bounds                          &    $+0.58,-0.21$   &    $+0.54,-0.07$   &    $+0.07,-0.10$   &    $+0.01,-0.01$  &    $+0.01$  \\
Low luminosity bins upper bound                                    &    +0.01           &    +0.11           &    +0.03           &    +0.02           &    +0.01          \\[\cmsTabSkip]
{Integrated uncertainty band ($\mu$)}                       &    {0.71}            &    {0.36}            &    {0.10}            &    {0.01}            &    {0.01}           \\
{Integrated uncertainty band ($\sigma$)}                    &    {0.35}            &    {0.19}            &    {0.12}           &    {0.01}           &    {0.01}           \\
\hline
\end{tabular}
}
\end{table*}

\subsubsection{Estimation of the QCD multijet background}
\label{sec:qcd}
Kinematic distributions obtained with the inclusive QCD multijet control sample are compared to those found in QCD multijet MC simulation. The same method of deriving a series of scale factors parameterised by generator-level quantities that was used in the estimation of the EW processes is applied here, but distributions of different quantities are used. In particular, the jet $\pt$ spectrum and angular correlations among jets in the event are the quantities that provide the most power in the identification of QCD background. We also consider the distributions of quantities related to heavy-flavour production and the relative momenta of jets in the event. After all corrections are applied, good closure is obtained: discrepancies between data and simulation are less than 10\% in distributions used to determine reweighting scale factors.

The one quantity that does, however, require special consideration is \ETm.  Most of the QCD multijet background is eliminated by high-\ETm requirements. The events that are not eliminated largely originate from the extreme tails of very broad distributions associated with two mechanisms. Namely, in order to produce large \ETm, a QCD multijet event must either involve production of a heavy-flavour hadron that decays leptonically, or involve one or more jets that are poorly resolved, leading to severe underestimates of their momenta.

The simulation of these sources of \ETm, particularly for the rare cases in which the events survive all selection requirements for the search regions, is not well understood, and it is difficult to study these mechanisms directly in data. This means that the QCD multijet background cannot be estimated precisely and so a reliable upper bound is found instead. This is sufficient because the QCD multijet contribution is small compared to other backgrounds.  To this end, simulation samples having sources of large \ETm are compared with \ETm-triggered data in control regions to obtain scale factors and associated uncertainties that are used to reweight simulated events. The resulting weights are then applied to simulation samples in the signal region. Additional systematic uncertainties are applied to cover the uncertainties in the extrapolations of these corrections into the search regions.

The high \ETm QCD multijet control sample, which is defined with the requirement that \ptvecmiss be aligned with one of the jets to a degree that is consistent with expectations for either of the two sources of \ETm discussed above, is used to derive scale factors. The jet with which  \ptvecmiss is aligned is referred to as the probe jet in such events. The negative vector sum of momenta of all jets in the event, other than the probe jet, provides an alternative estimate of the probe jet momentum, since \pt is conserved, within uncertainties, in the absence of other severe mismeasurements. The recoil response, defined as the ratio of the momenta of the probe jet to that for the rest of the activity in the event, ($\pti{\text{probe}}/\pti{\text{recoil}}$), is a very good estimator for the true response of the probe jet, ($\pti{\text{probe}}/\pti{\text{true}}$), in the tails of the distribution, where mismeasurement of the probe jet momentum dominates over the mismeasurement of the recoil momentum. It is therefore used to derive separate scale factors for the jet resolution, parameterised by jet $\pt$, for each of the two sources of \ETm. These scale factors range between 0.6 and 1.8.

The central values of the QCD background predictions are taken to be the MC simulation yields in the signal regions after applying all of the  corrections defined above. The various statistical and systematic uncertainties are highly asymmetric and in many cases non-Gaussian. Therefore, in each search region an MC integration procedure is used to properly combine the uncertainties. As expected from the central limit theorem, the combination of uncertainties can be approximated by a Gaussian distribution, the parameters of which are listed in Tables~\ref{tab:results_T2tt} and~\ref{tab:results_T2bW} under the label of ``Integrated uncertainty band.''

Two shape uncertainties are assigned to the QCD multijet estimation in each search region. The first is a systematic uncertainty associated with the search region MVA discriminator distribution, denoted as ``MVA discriminator shape'' in Tables~\ref{tab:results_T2tt} and~\ref{tab:results_T2bW}. It is obtained from a comparison of the distribution in MC simulation to that in data for the high \ETm QCD multijet control sample after also requiring that events pass the baseline selection criteria, with the exception of the requirements on the angular separation between the leading jets and \ptvecmiss. Dropping these criteria leads to a significant increase in the  contribution of QCD multijet events to the final sample relative to all other backgrounds or signal. A second systematic uncertainty, labelled ``$\Delta\phi$ shape upper and lower bounds'' in Tables~\ref{tab:results_T2tt} and~\ref{tab:results_T2bW}, is obtained from the same samples by comparing the MC distribution of the angle between \ptvecmiss and the leading jets to that for data for a variety of discriminator cutoffs. The distributions are found to differ increasingly with rising b-tagged jet multiplicity. The bias is eliminated by smearing the $\phi$ values of the $\ptvec$ of b jets with a Gaussian having a standard deviation of about 0.02.  The upper bound on the QCD background is then obtained by increasing the width of the Gaussian until there is a larger number of MC events predicted to pass the selection criteria than is observed in data. The upper bounds found in this way are different for different search regions as a result of variations in statistics and contributions of other SM processes. The values of the Gaussian width that are found to cover all cases are 0.07 in the case of T2tt and 0.05 in the case of T2bW.

Finally, the QCD multijet simulated data are generated in discrete bins of $\HT$ in the case of \MADGRAPH and in bins of quark and gluon $\pt$ in the case of \PYTHIA.  The effective integrated luminosity for some of the samples in particular bins can be much smaller than the 18.9\fbinv of integrated luminosity collected in proton-proton collision data. A systematic uncertainty is therefore applied to each QCD background prediction to cover a possible underprediction that could be the result of a lack of events in these highly weighted bins. It is denoted as ``Low luminosity bins upper bound'' in Tables~\ref{tab:results_T2tt} and~\ref{tab:results_T2bW}.

\subsection{Estimation of the \texorpdfstring{$\ttbar\Z$}{ttZ} background}
\label{sec:ttz}
Standard model $\ttbar\Z$ production is a rare process ($\sigma\sim0.2\unit{pb}$) that  becomes an important background in \CORRAL-based search regions for the T2tt signal model where general $\ttbar$ backgrounds have been greatly suppressed.  There are no sufficiently populated and uncontaminated data control regions in which to perform careful studies of this rare SM process.  The simulated data are studied instead, making use of variations in the parameters that control the generation and parton showering to establish systematic uncertainties in the estimated event counts in the signal regions. In addition, the relative difference in yields between the default \MCATNLO sample, with parton showering by \HERWIG, and a separate \MADGRAPH sample, with parton showering by \PYTHIA, is used to estimate a systematic uncertainty associated with MC generators. This uncertainty, listed in Tables~\ref{tab:results_T2tt} and~\ref{tab:results_T2bW} with the label ``MC simulation,'' ranges between 3\% and 26\% depending on the search region.

The uncertainty in the $\ttbar\Z$ production cross section is estimated from a data control sample with three reconstructed charged leptons drawn from a larger event sample that has been collected with a set of dilepton triggers used for multilepton SUSY searches~\cite{Khachatryan:2014qwa}. The two charged leptons picked up by these triggers most often originate from the decay of a $\Z$ boson and are thus oppositely charged, same-flavour leptons. The third lepton can arise via the semileptonic decay of a $\PW$ coming from the decay of a top quark in $\ttbar\Z$ events. The selection of events for this control sample thus includes the requirement that two of the reconstructed leptons must be consistent with the expectations for leptons from $\Z$ boson decay in flavour, charge, and the invariant mass of the pair. In order to reduce the contamination from other SM backgrounds, events are also required to have at least three or more jets, at least six picky jets, and one or more b-jets tagged with the medium CSV working point ~\cite{Chatrchyan:2012jua} in order to increase the relative contribution of the $\ttbar\Z$ process.

With a contribution of approximately 10\%, diboson production is a leading SM process in this region after $\ttbar\Z$. Thus, a diboson-enriched control region is established that makes use of the same selection criteria described above for the $\ttbar\Z$ control region, except that the b tagging requirement is inverted to form a corresponding b-tag veto. This sample is used to normalise the overall diboson process in MC simulation to that observed in data.

The $\ttbar\Z$ and the diboson processes in the enriched control regions described above have estimated event yields that are statistically consistent with the event yields predicted by simulation samples. In view of this, the data-MC scale factors are taken to have a central value of unity, and no correction is applied. The statistical uncertainty in the $\ttbar\Z$ scale factor is 31\%. This is adopted as a systematic uncertainty in the estimated yield of this background source and is denoted as ``MC normalisation'' in Tables~\ref{tab:results_T2tt} and~\ref{tab:results_T2bW}.

A final systematic uncertainty takes into account differences observed between the kinematic distributions in MC simulation and data. To this end,  we make use of the closure uncertainties in the $\PW$+jets (including $\ttbar$ and single top) and $\Z$+jets background predictions that have been derived in the lepton control regions as necessitated by the lack of an appropriate $\ttbar\Z$ data control sample. The maximum estimated uncertainty found for either of the two processes is taken to be the uncertainty in the modelling of the kinematics for the $\ttbar\Z$ process. This uncertainty ranges between 16$\%$ and 39$\%$, depending on the signal sample, and is included under the label of ``Kinematic closure'' along with the $\ttbar\Z$ prediction and all other associated uncertainties in Tables~\ref{tab:results_T2tt} and~\ref{tab:results_T2bW}.

\section{Results and interpretation}
\label{sec:results}

\begin{figure*}[tbph]
\centering
\includegraphics[width=0.40\linewidth]{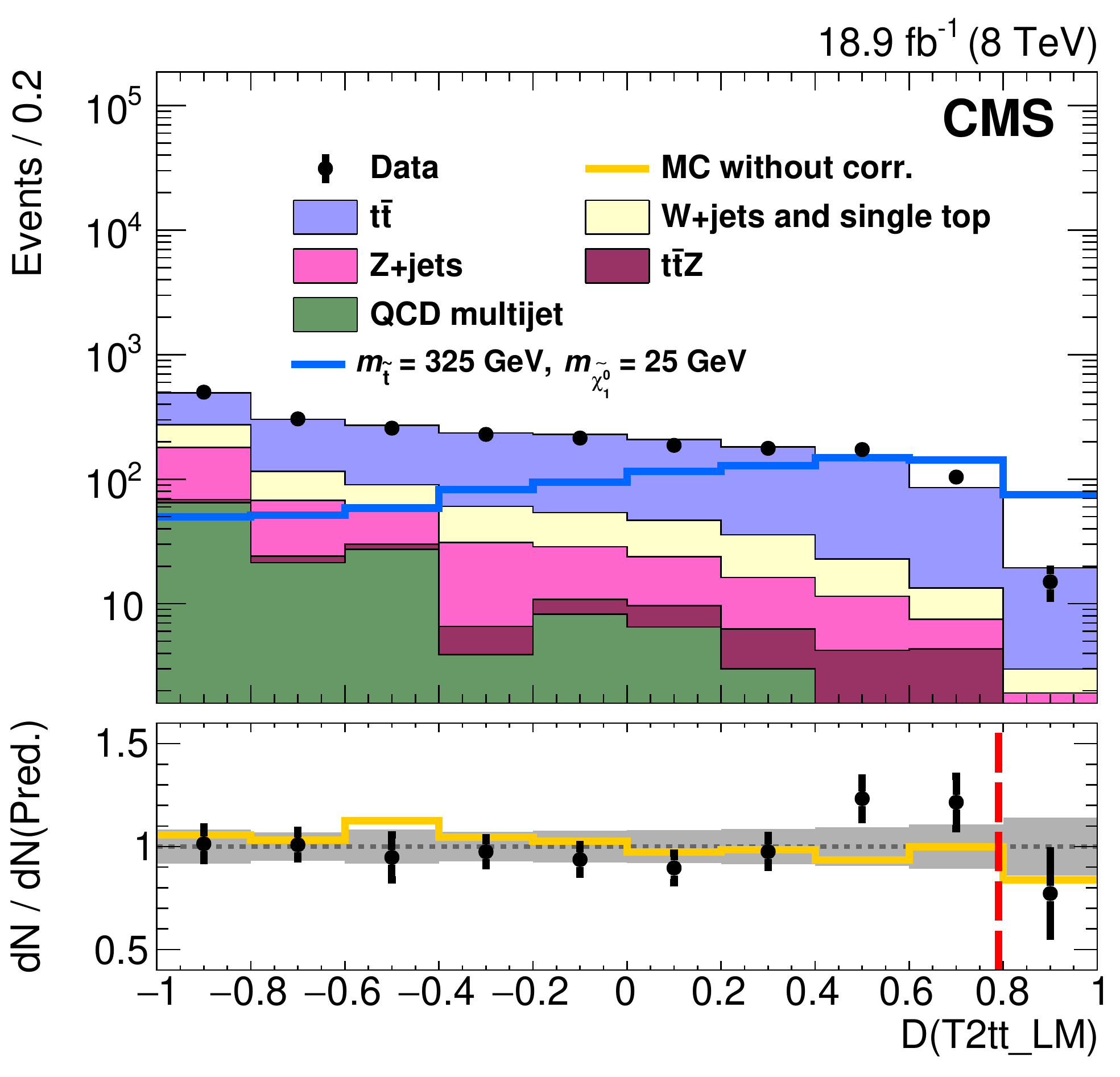}
\includegraphics[width=0.40\linewidth]{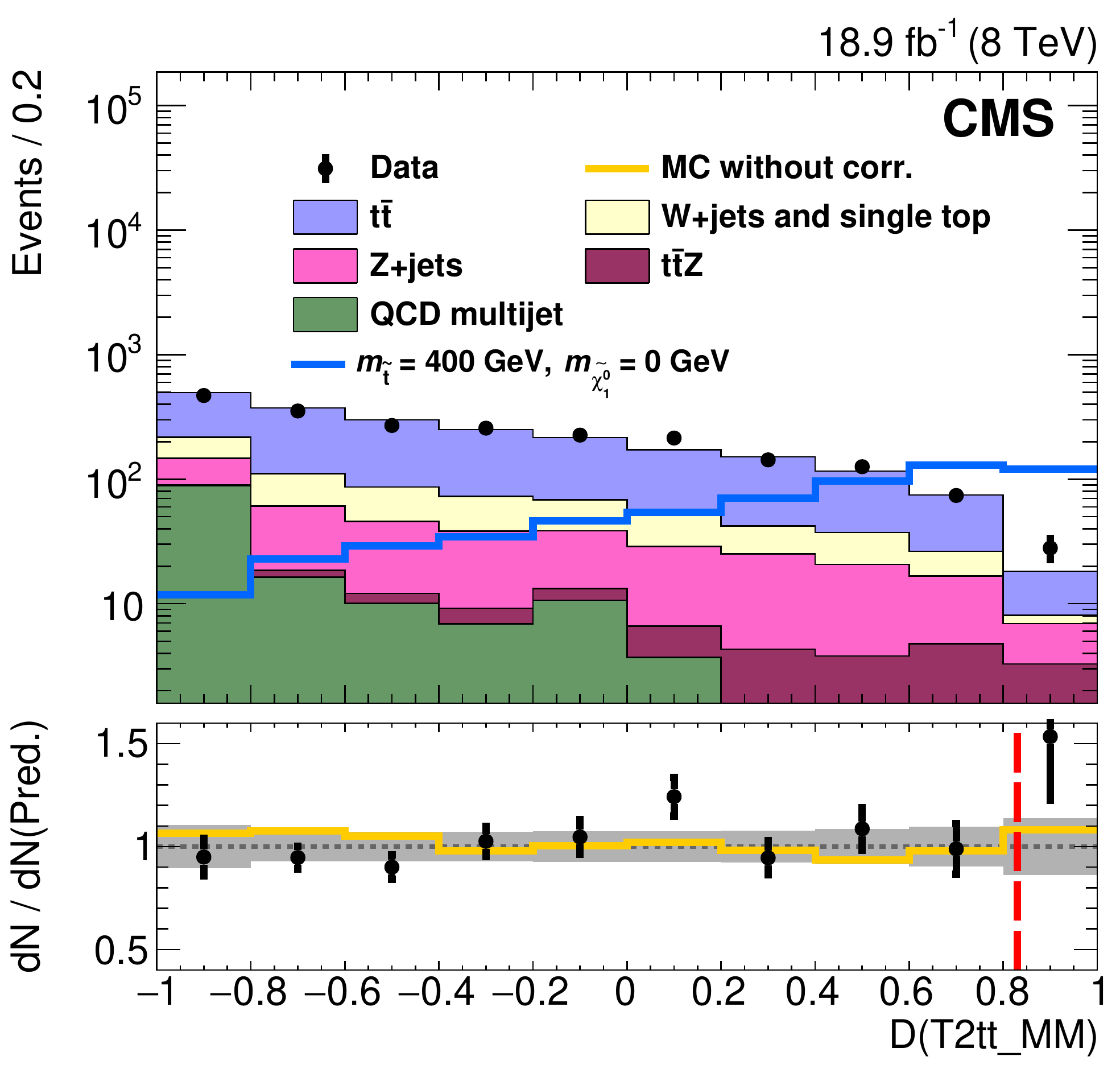}
\includegraphics[width=0.40\linewidth]{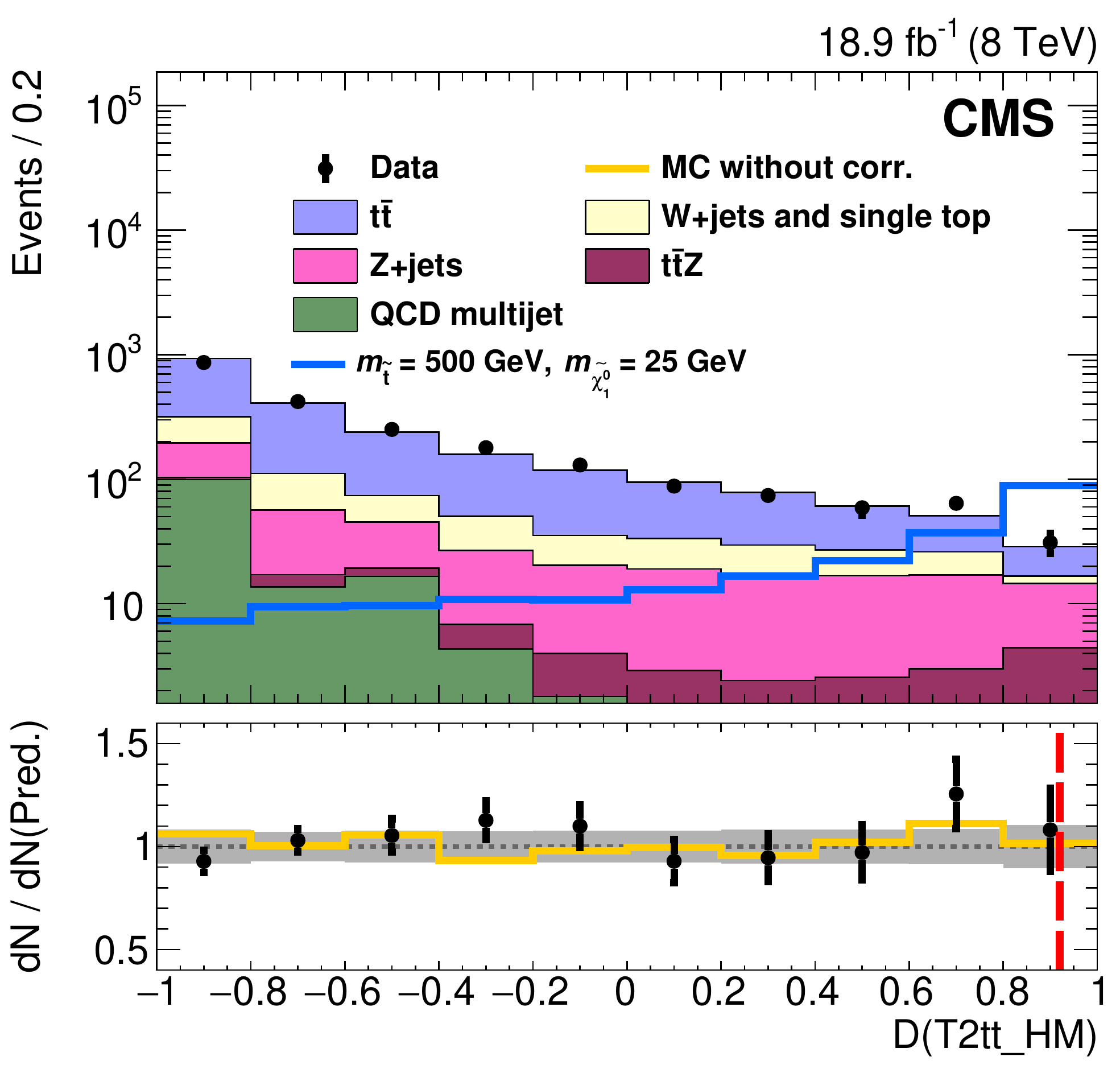}
\includegraphics[width=0.40\linewidth]{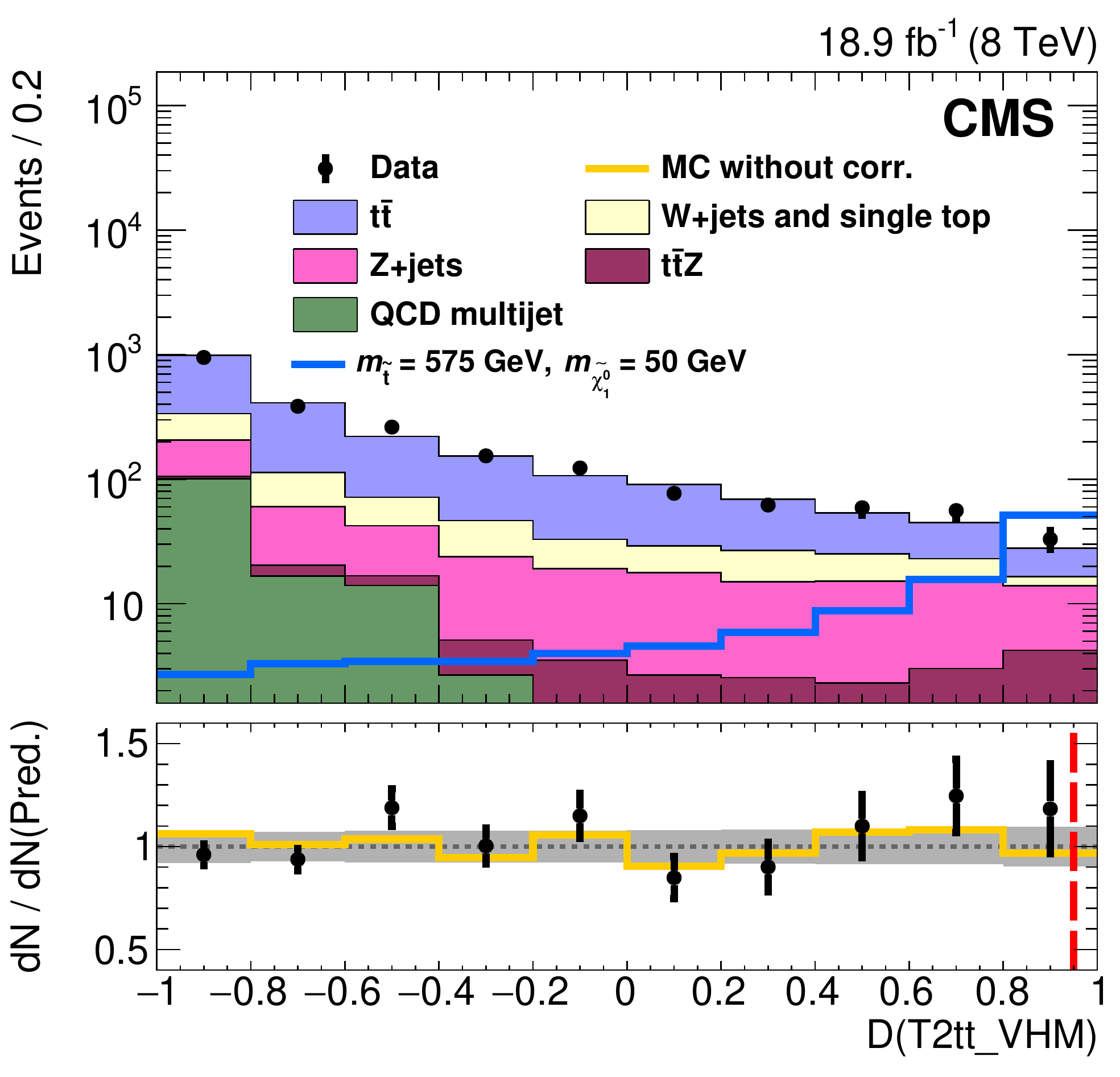}
\caption{Observed and predicted event yields for each T2tt search region discriminator (D). The bottom pane of each plot shows the ratio of observed to predicted yields where the error bars on data points only include the statistical uncertainties in the data and MC event yields. The filled bands represent the relative systematic uncertainties in the predictions.}
\label{fig:results-T2ttDisc-dist}
\end{figure*}

\begin{figure*}[tbph]
\centering
\includegraphics[width=0.40\linewidth]{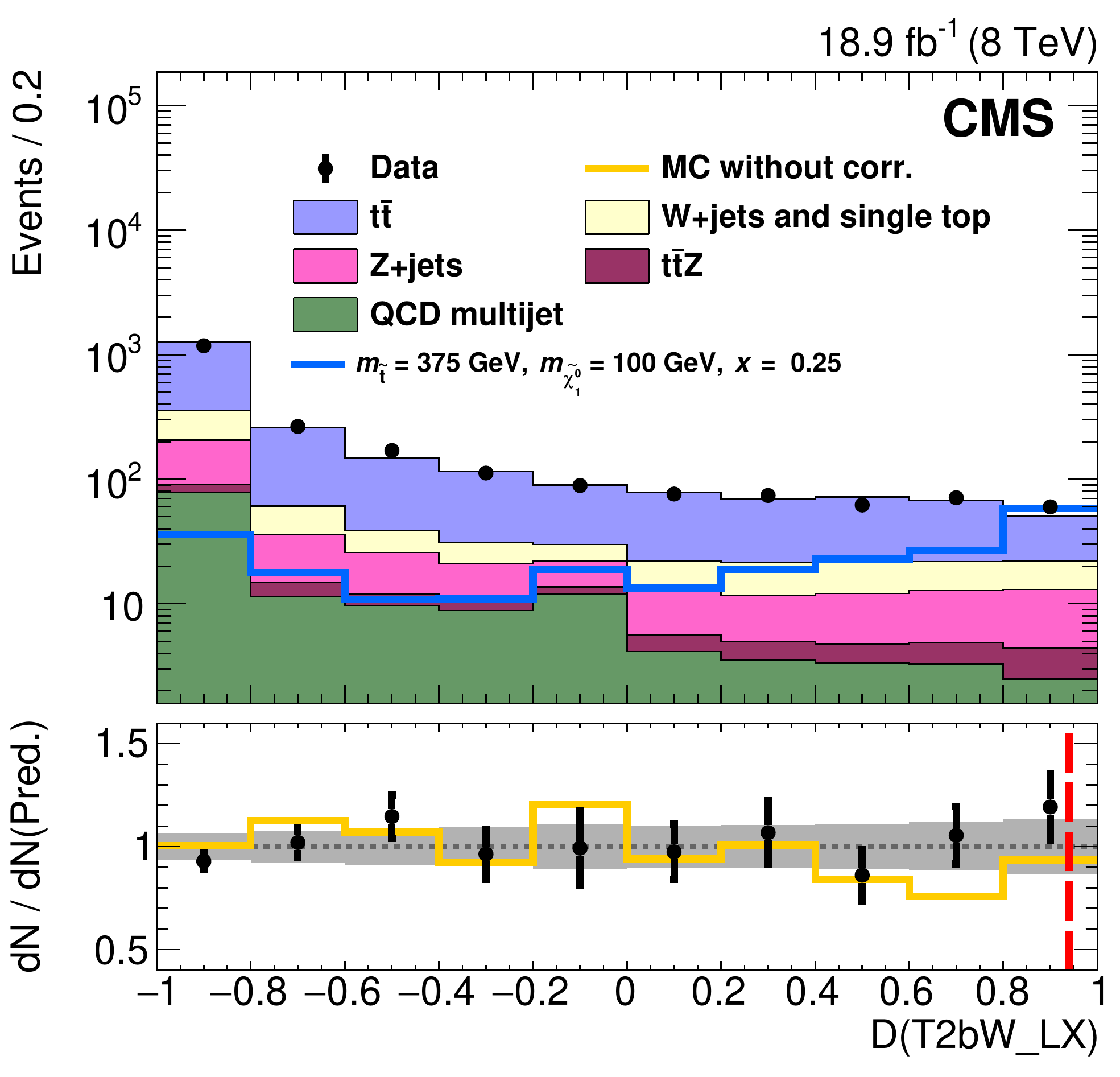}
\includegraphics[width=0.40\linewidth]{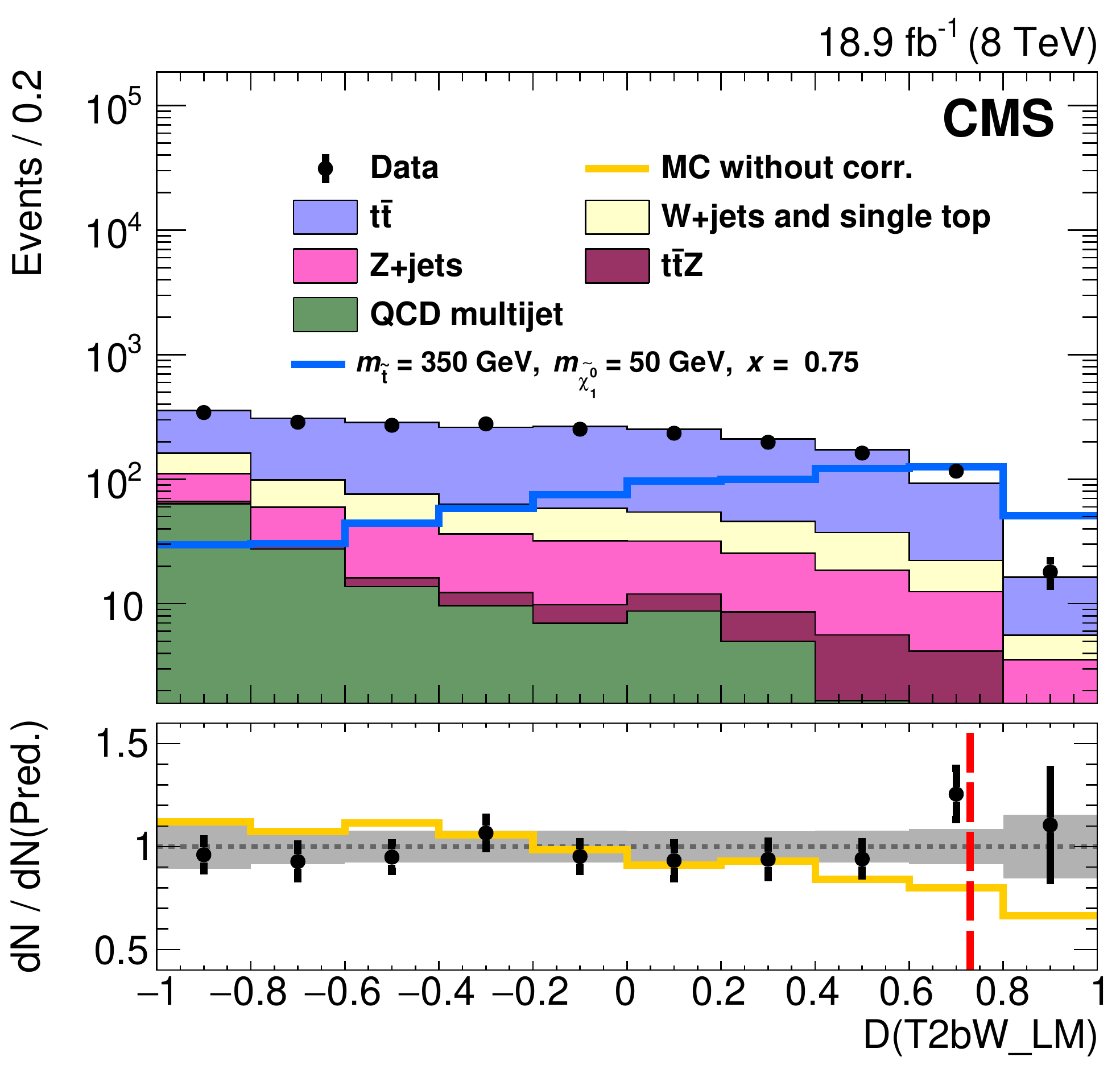}
\includegraphics[width=0.40\linewidth]{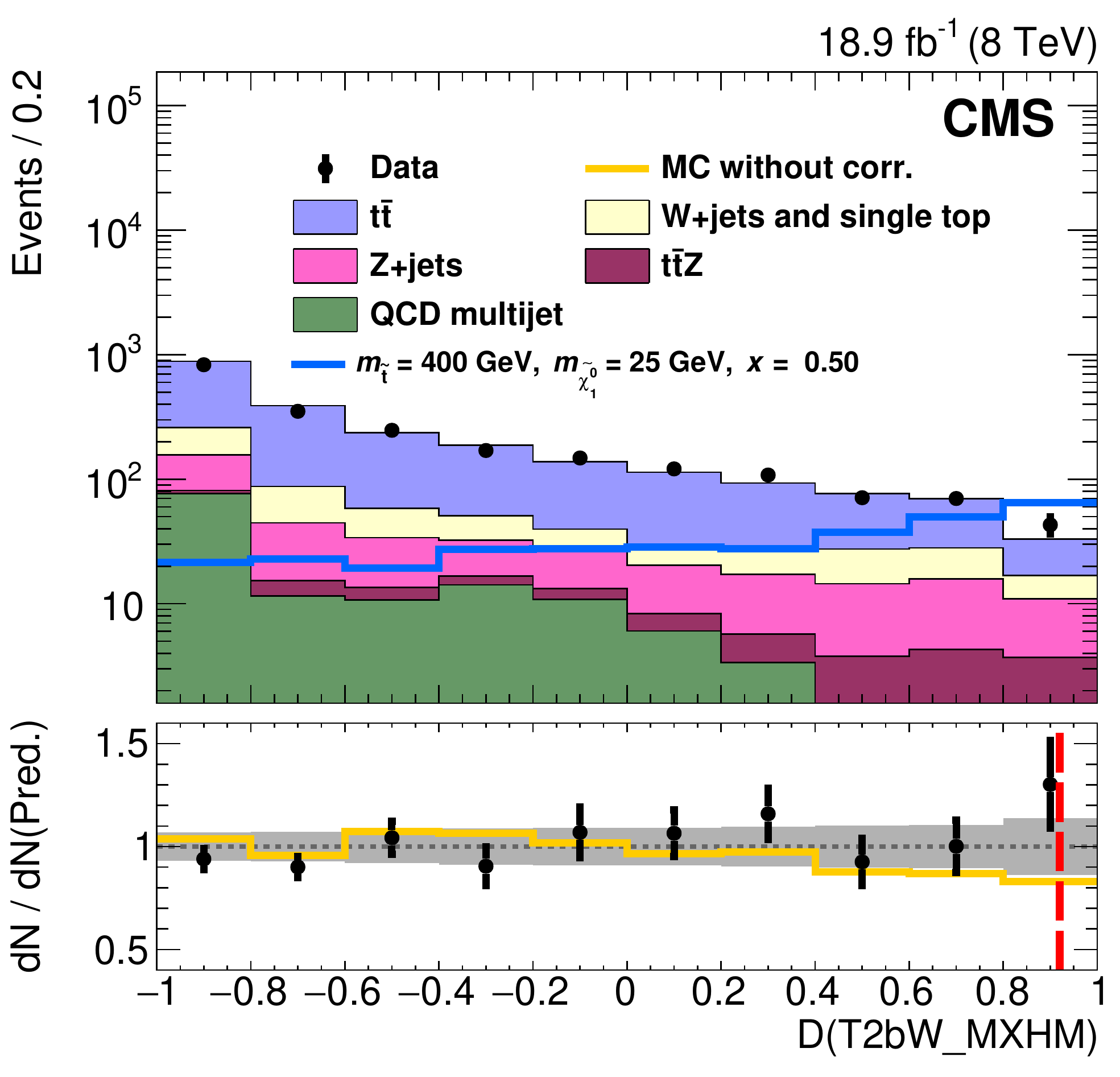}
\includegraphics[width=0.40\linewidth]{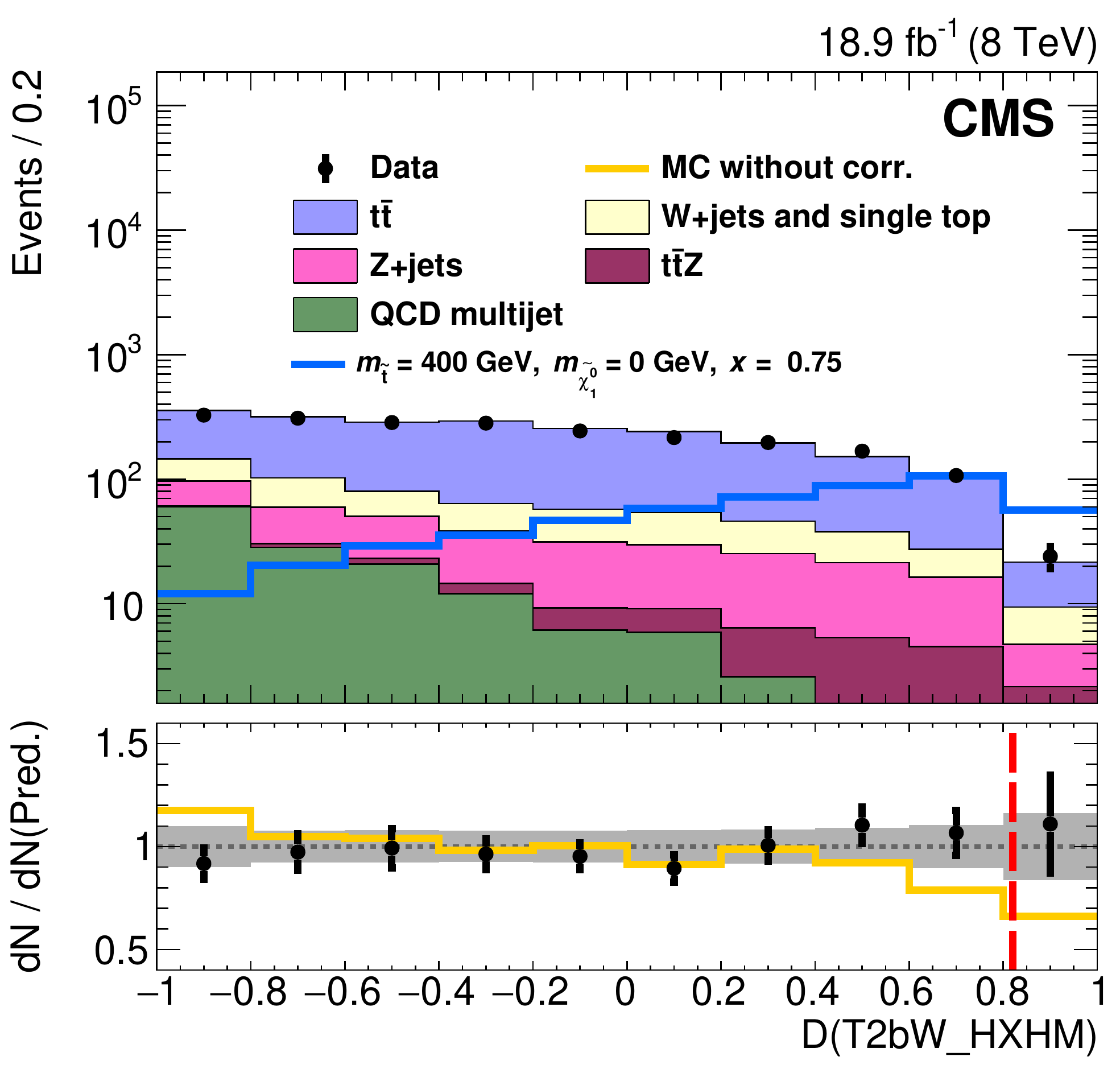}
\includegraphics[width=0.40\linewidth]{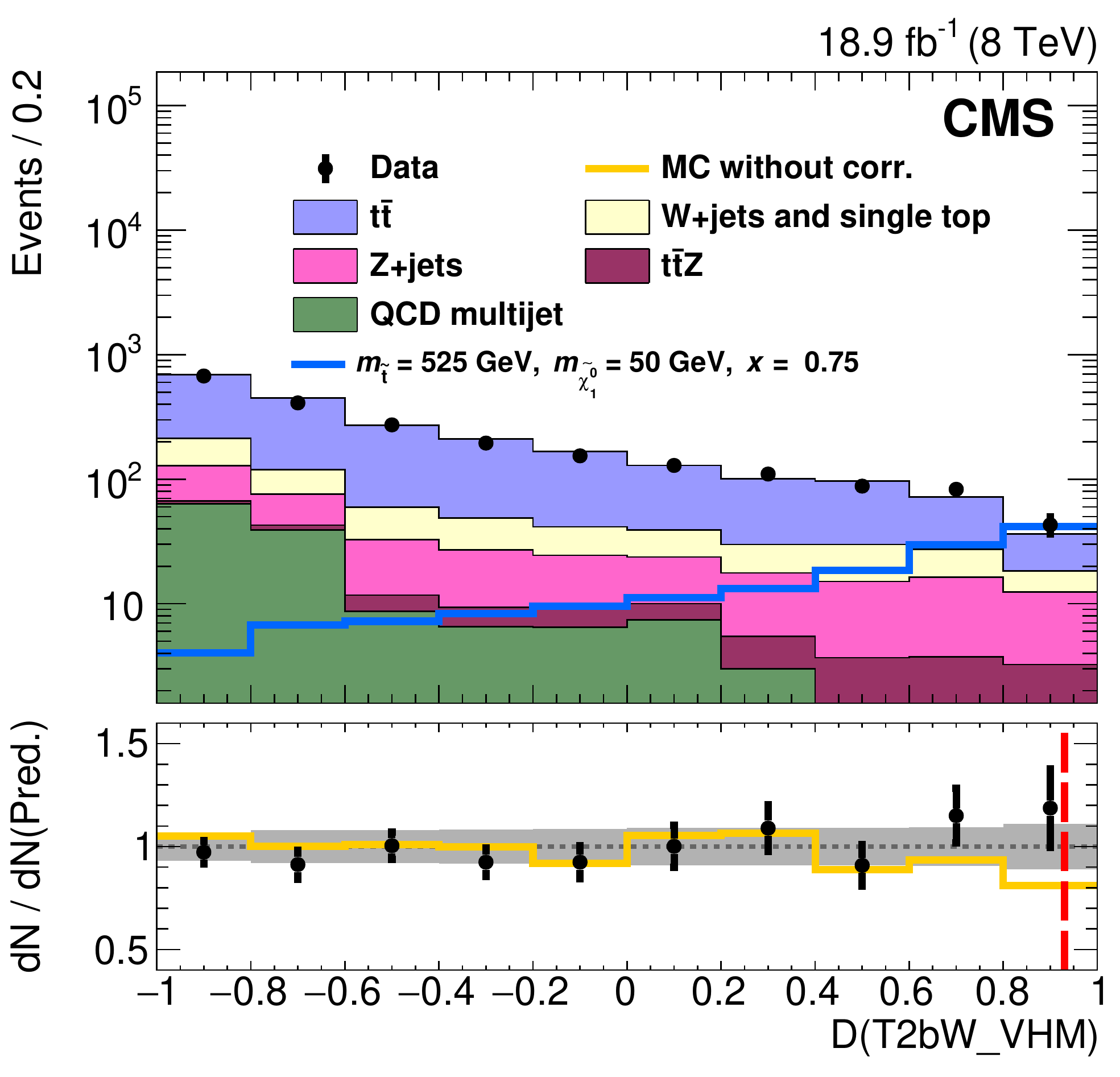}
\caption{Observed and predicted event yields for each T2bW search region discriminator (D). The bottom pane of each plot shows the ratio of observed to predicted yields where the error bars on data points only include the statistical uncertainties in the data and MC event yields. The filled bands represent the relative systematic uncertainties in the predictions.}
\label{fig:results-T2bWDisc-dist}
\end{figure*}

The predicted distributions of discriminator values for the various T2tt and T2bW searches described earlier are shown in Figs.~\ref{fig:results-T2ttDisc-dist} and~\ref{fig:results-T2bWDisc-dist}.  Event yields in data are plotted with their statistical uncertainties and compared to the SM background predictions. The latter are represented by the coloured histograms in the upper pane. Error bars on the ratios of the observed to predicted event yields in the bottom pane include only statistical uncertainties. The filled band in the lower pane of each plot represents the relative systematic uncertainty in the background predictions. A vertical dashed red line near the right edge in the lower pane of each plot marks the MVA discriminator value that is used to define the lower boundary of the search region. Note that these figures are for illustrative purposes only, and so some minor uncertainties in event yields in the more inclusive regions did not receive the detailed treatment applied to the uncertainties in the final search region yields.

\begin{figure*}[tbph]
\centering
\includegraphics[width=0.40\linewidth]{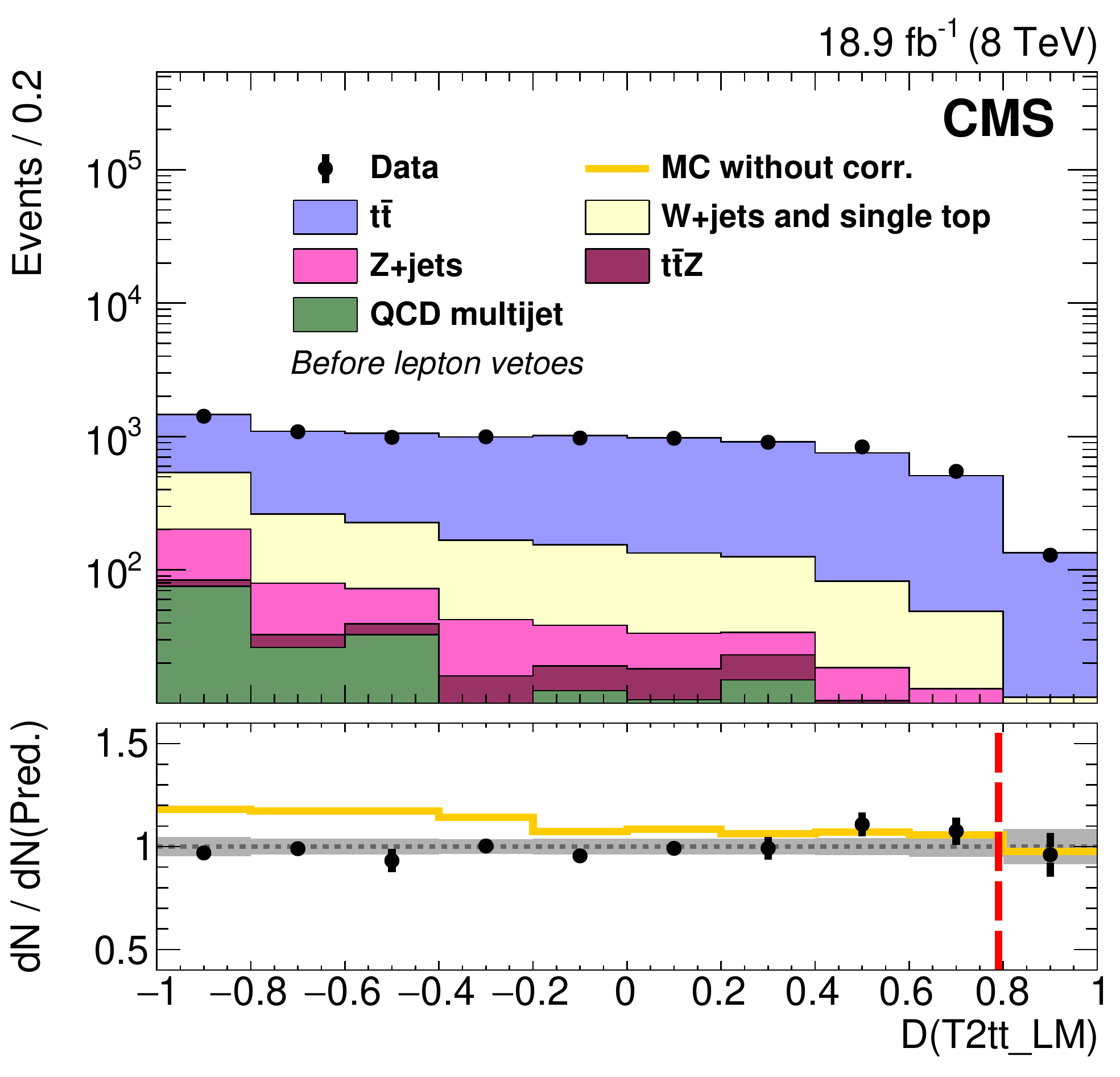}
\includegraphics[width=0.40\linewidth]{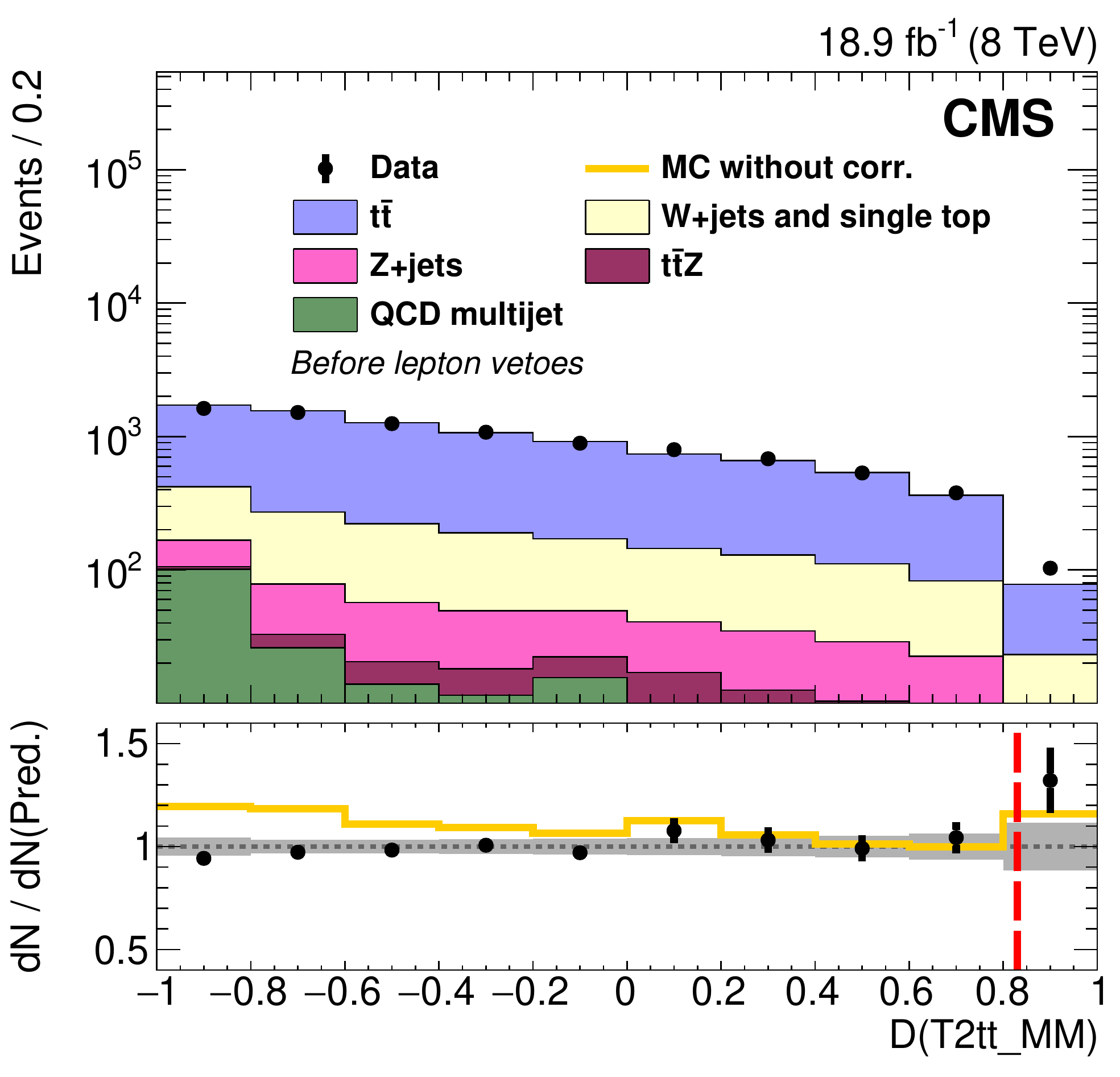}
\includegraphics[width=0.40\linewidth]{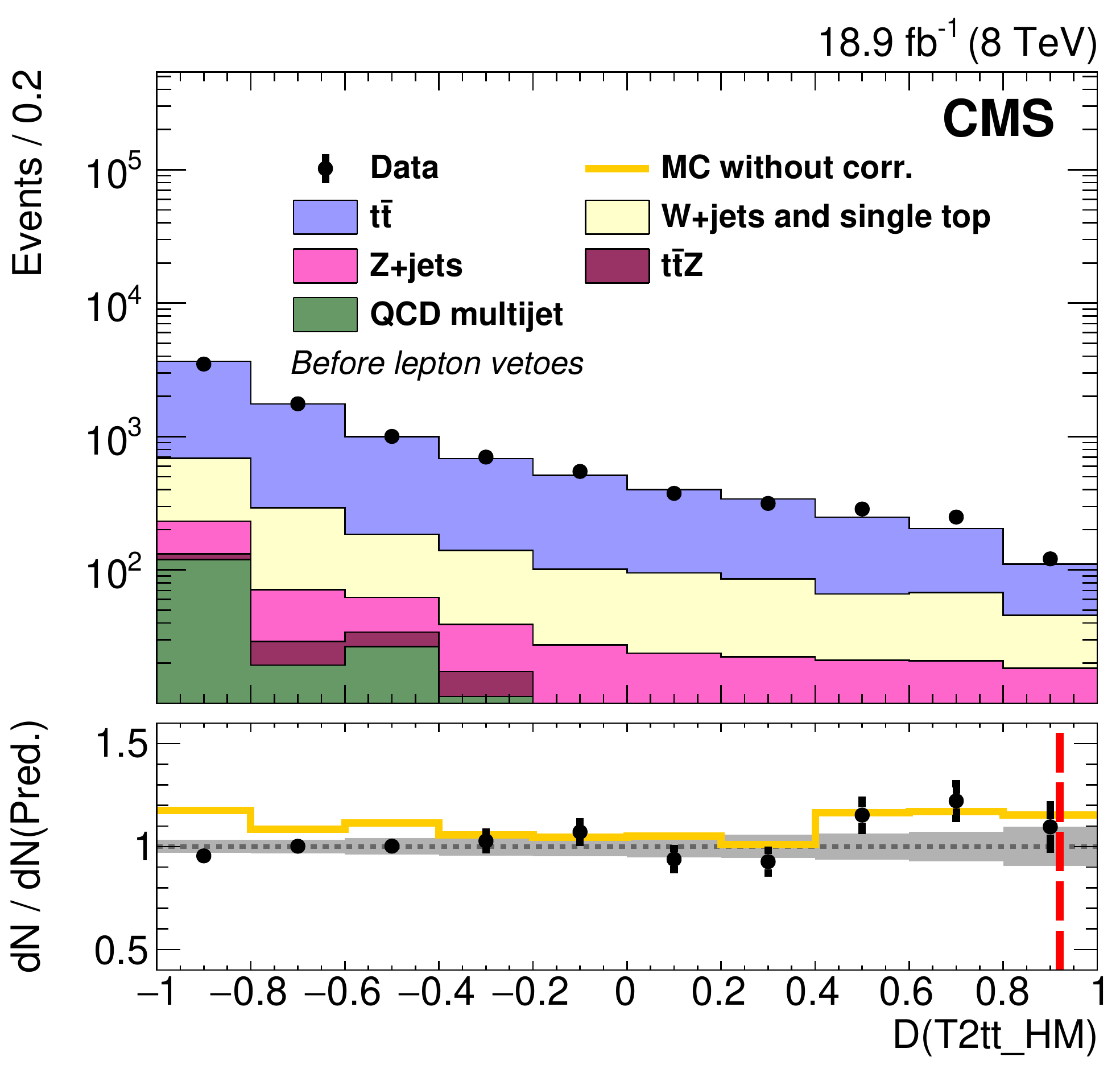}
\includegraphics[width=0.40\linewidth]{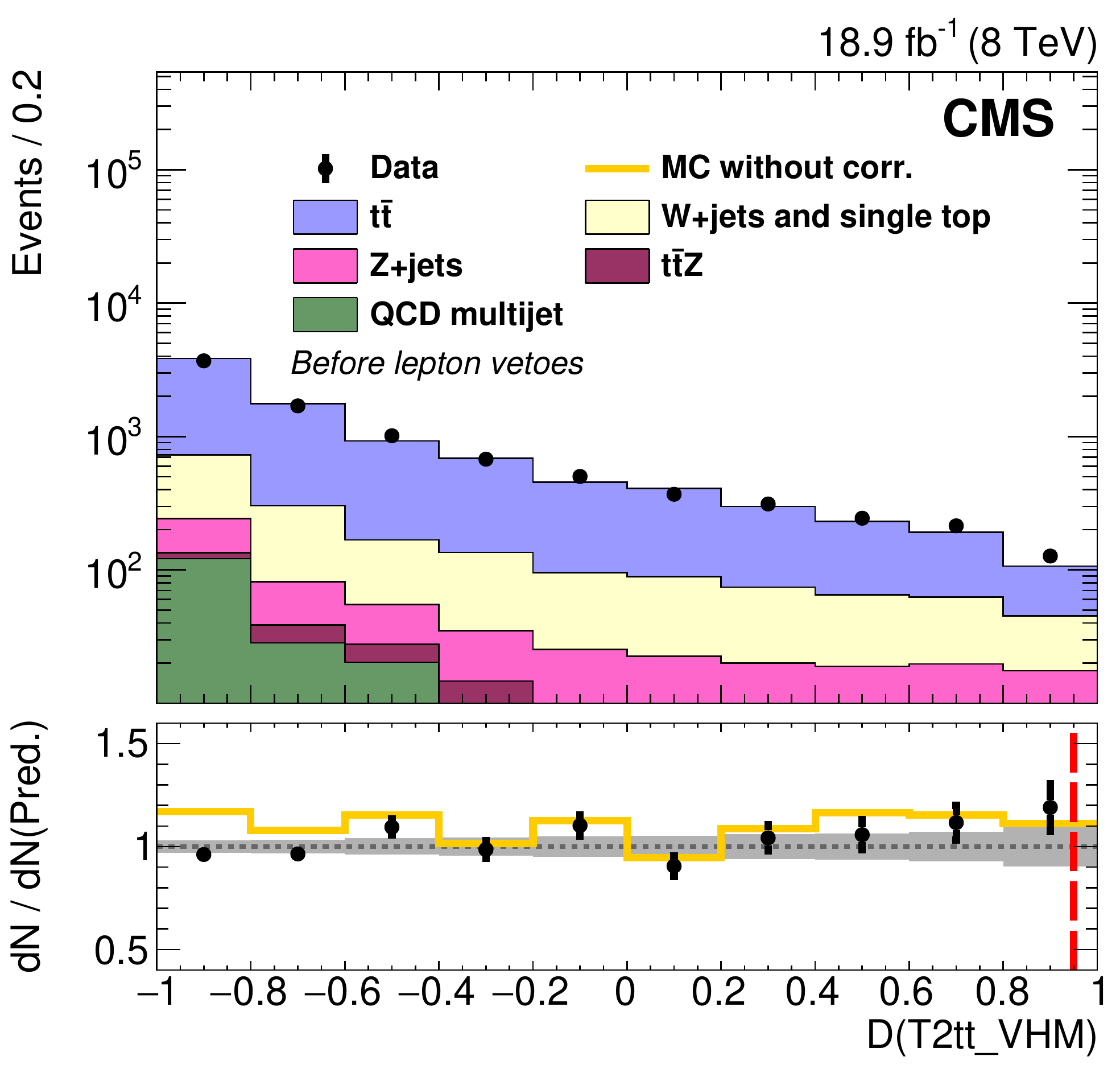}
\caption{Observed and predicted event yields for each T2tt search region discriminator (D) before lepton vetoes are applied, which are used for the cross-checks discussed in the text. The bottom pane of each plot shows the ratio of observed to predicted yields where the error bars on data points only include the statistical uncertainties in the data and MC event yields. The filled bands represent the relative systematic uncertainties in the predictions.}
\label{fig:results-T2ttDisc-incl}
\end{figure*}

\begin{figure*}[tbph]
\centering
\includegraphics[width=0.40\linewidth]{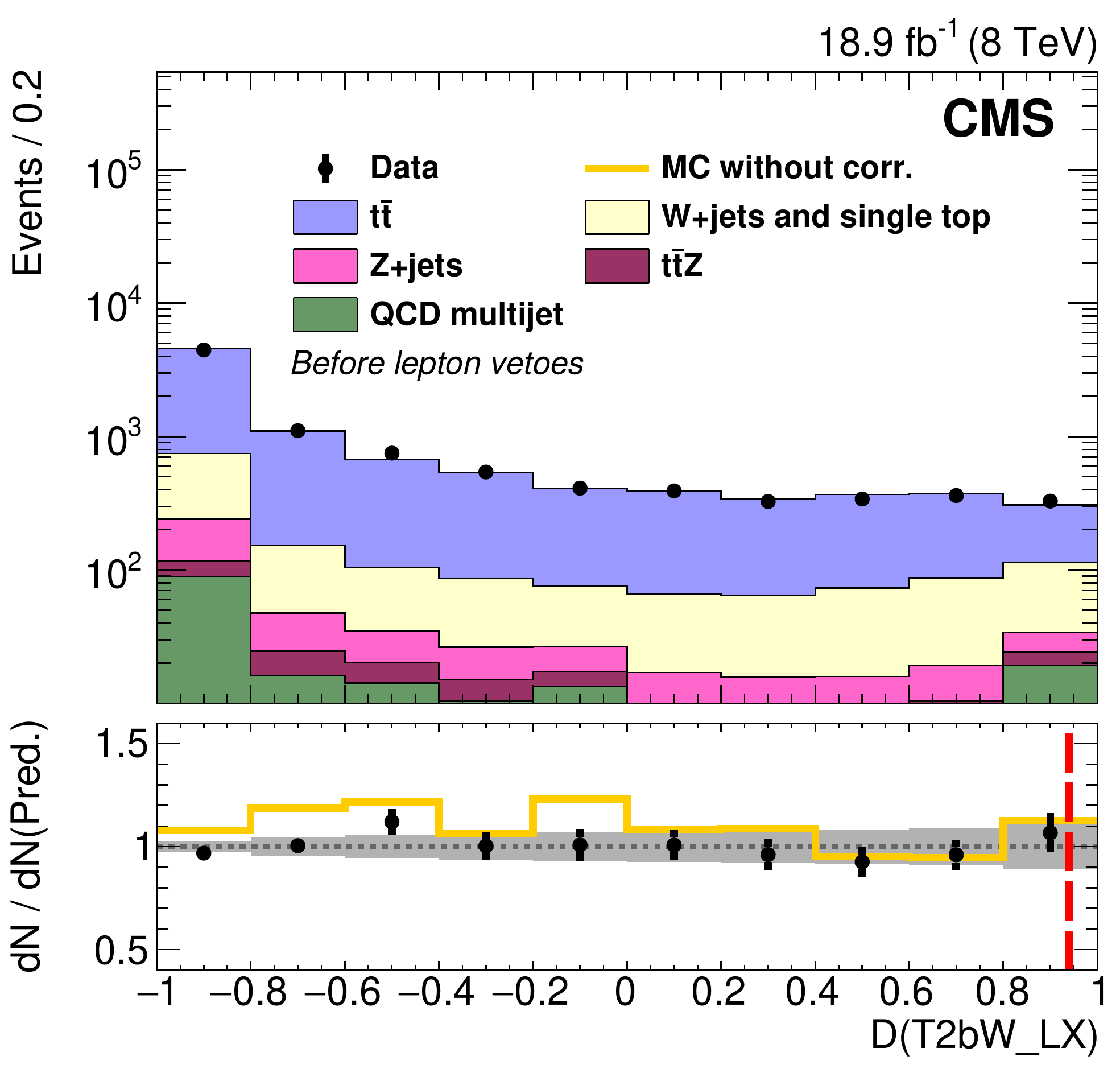}
\includegraphics[width=0.40\linewidth]{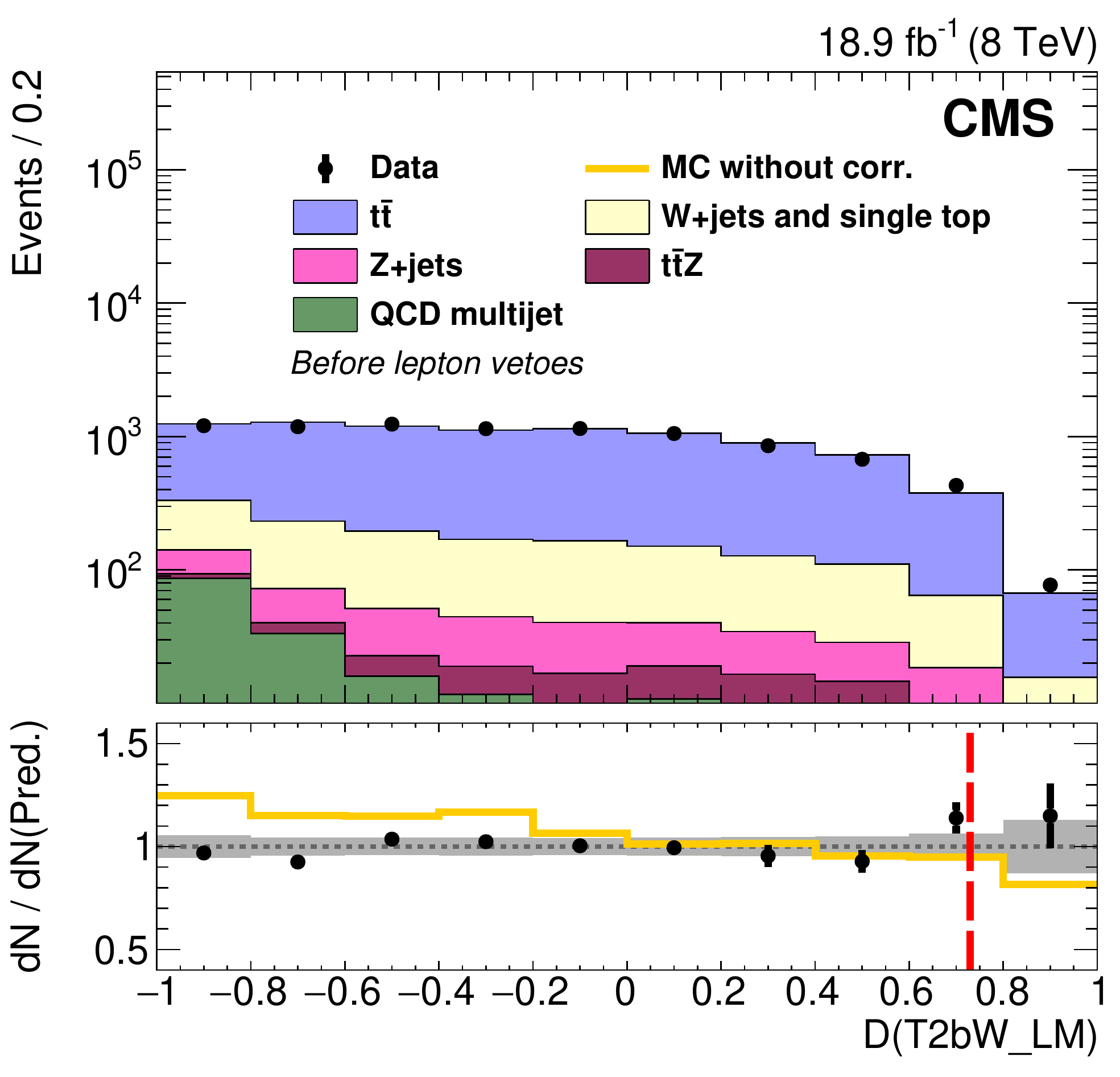}
\includegraphics[width=0.40\linewidth]{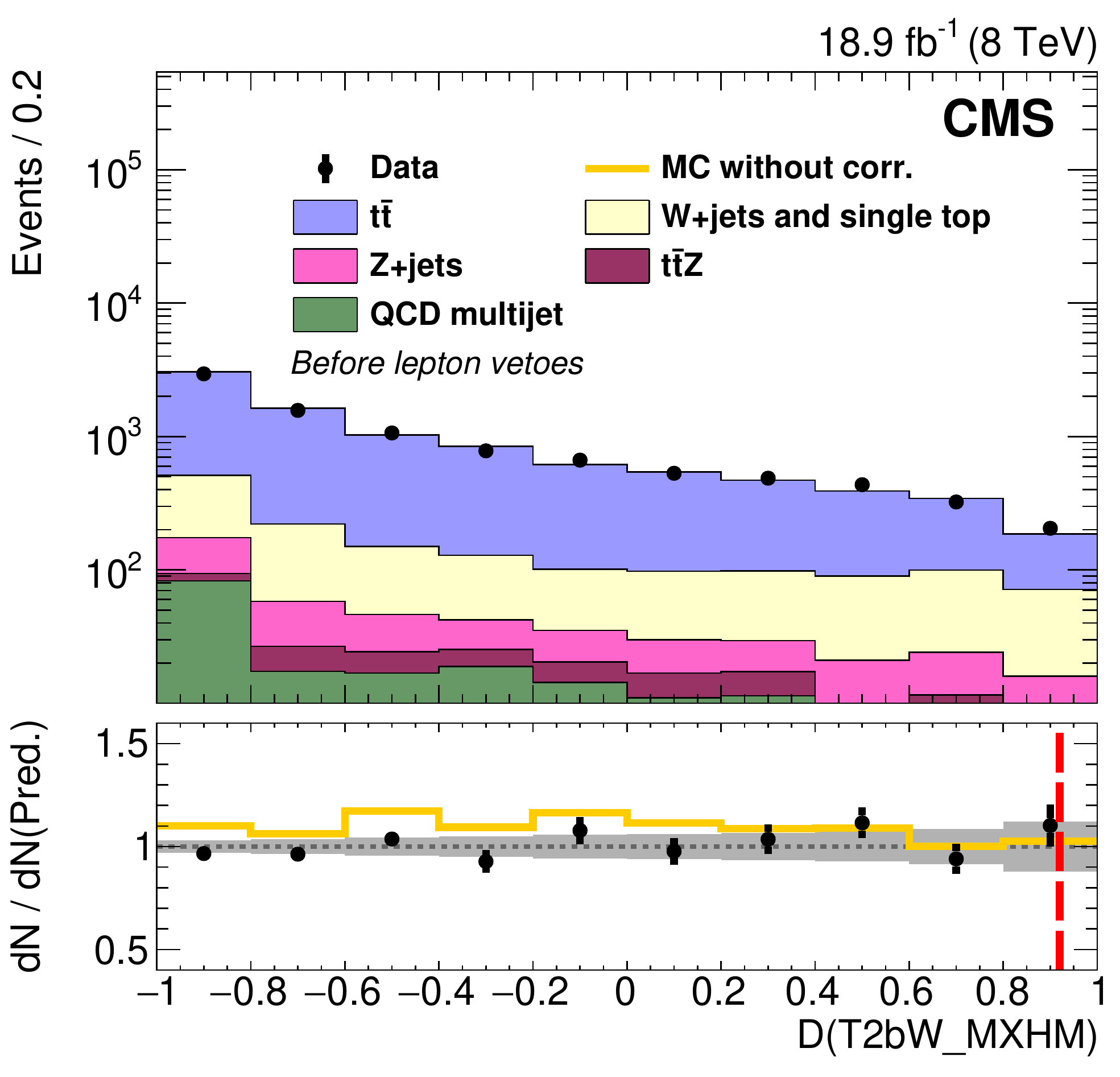}
\includegraphics[width=0.40\linewidth]{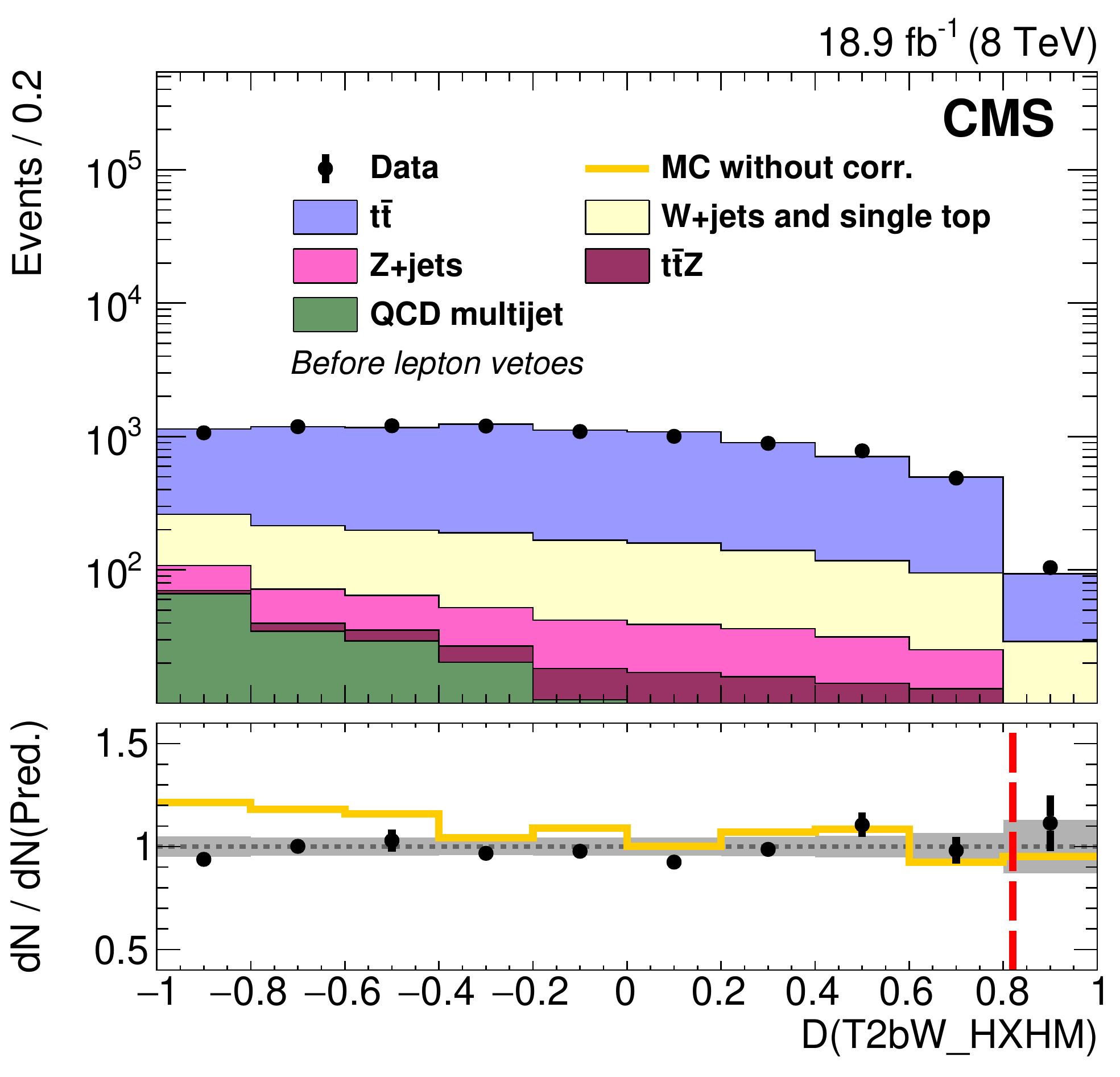}
\includegraphics[width=0.40\linewidth]{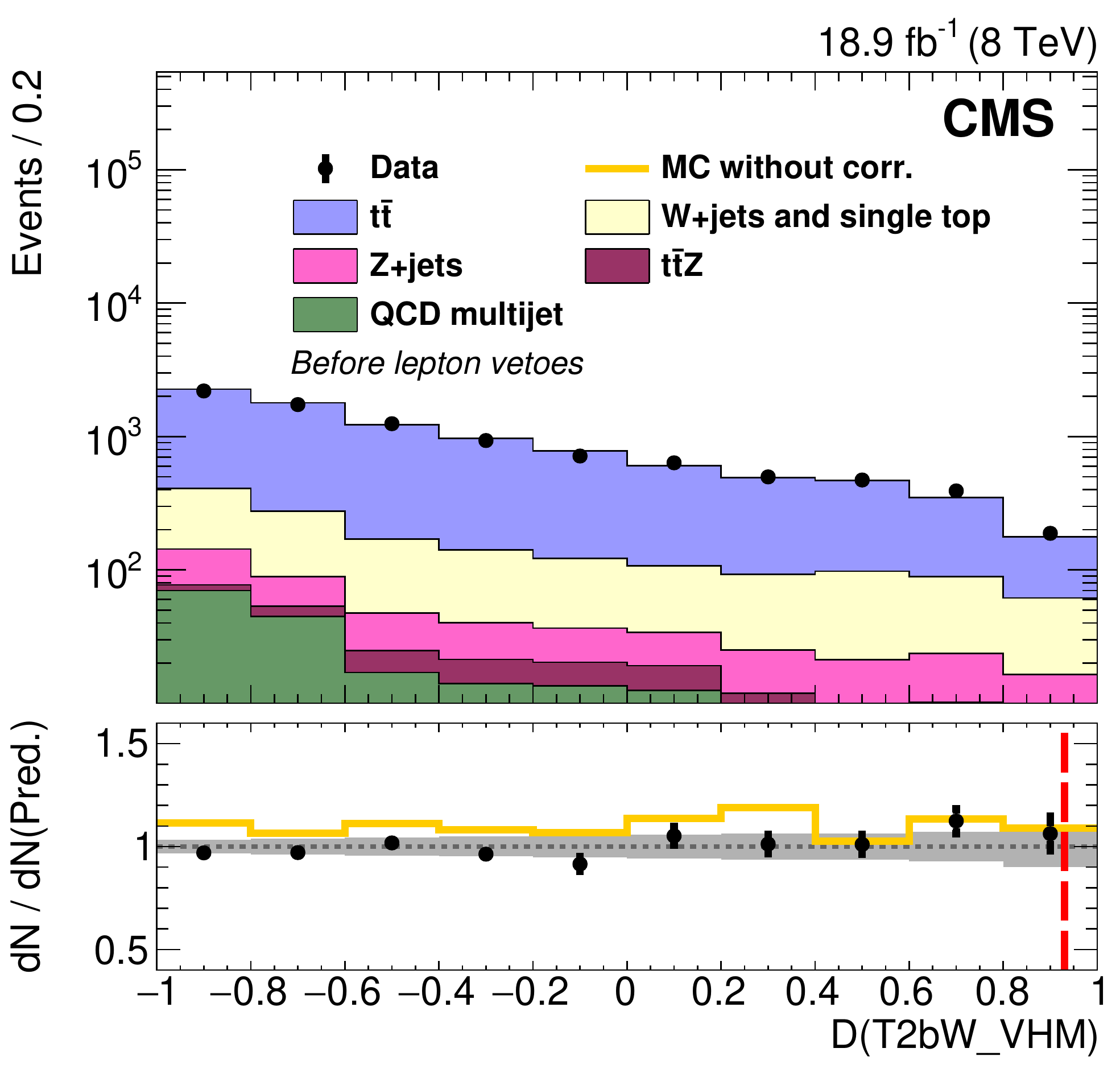}
\caption{Observed and predicted event yields for each T2bW search region discriminator (D) before lepton vetoes are applied, which are used for the cross-checks discussed in the text. The bottom pane of each plot shows the ratio of observed to predicted yields where the error bars on data points only include the statistical uncertainties in the data and MC event yields. The filled bands represent the relative systematic uncertainties in the predictions.}
\label{fig:results-T2bWDisc-incl}
\end{figure*}

The line in the lower pane of each plot in Figs.~\ref{fig:results-T2ttDisc-dist} and~\ref{fig:results-T2bWDisc-dist} labelled ``MC without corr.'' represents the sum of the MC contributions, relative to the prediction, prior to weighting by the  corrective scale factors discussed in the preceding sections. There are no statistically significant differences observed upon comparing the data with the uncorrected (or corrected) MC samples. Figures~\ref{fig:results-T2ttDisc-incl} and~\ref{fig:results-T2bWDisc-incl} provide a completely equivalent set of plots to those just described, but in this case, no lepton vetoes have been included in the selection of events. The event yields therefore are much higher in these cases. These data are used to provide a useful cross-check of the $\ttbar$, $\PW$+jets, and single top kinematic closure test. They also allow for a check of the agreement in event kinematics between MC simulation and data, without any potential biases that might arise in association with the application of the lepton vetoes to the simulation. Only those data with discriminator values less than 0.4 are used for these cross-checks because potential signal contamination could be non-negligible for larger discriminator values. Data and simulation agree within $\pm$20\% for all search regions.

The predicted and observed yields in the T2tt and T2bW search regions are summarized in Tables~\ref{tab:unblindresults_T2tt} and~\ref{tab:unblindresults_T2bW}. No statistically significant excess in data is observed. We therefore use these results to set upper bounds on the production cross sections for the T2tt and T2bW families of signal models.

\begin{table*}[h!tb]
\topcaption{\label{tab:unblindresults_T2tt}Predicted and observed data yields in the T2tt search regions. The uncertainties in the background predictions are the combined systematic and statistical uncertainties. The T2tt yields correspond to the simplified model points with  $(\mStop, \mLSP) = (500\GeV, 200 \GeV)$ and $(700\GeV, 0\GeV)$. The uncertainties in the signal yields are statistical only.}
\centering
\newcolumntype{x}{D{,}{\,\pm\,}{4.4}}
\begin{tabular}{lxxxx}
\hline
& \multicolumn{4}{c}{Search region yield}\\ \cline{2-5}\noalign{\smallskip}
& \multicolumn{1}{c}{T2tt\_LM} & \multicolumn{1}{c}{T2tt\_MM} & \multicolumn{1}{c}{T2tt\_HM} & \multicolumn{1}{c}{T2tt\_VHM}\\
\hline
$\ttbar$, $\PW$+jets, and single top  &19.8,3.3&8.53,1.74&3.22,0.95&1.11,0.50 \\
$\Z$+jets                          &0.69,0.23&2.30,0.90&1.92,0.84&0.59,0.28    \\
$\ttbar\Z$                         &1.34,0.49&2.66,1.27&1.62,0.75&0.99,0.49    \\
QCD multijet                       &0.91,0.58&0.17,0.07&0.04,0.02&0.01,0.01    \\[\cmsTabSkip]
{All SM backgrounds}               &22.7,3.4  &13.7,2.3  &6.8,1.5  &2.7,0.8      \\
{Observed data}             &  \multicolumn{1}{c}{16}            &  \multicolumn{1}{c}{18}            &  \multicolumn{1}{c}{7}            &            \multicolumn{1}{c}{2}      \\
\hline
T2tt (500, 200) &  10.9 , 0.4  &  27.2 , 0.6  &  12.0 , 0.4  &  5.53 , 0.27 \\
T2tt (700, 0) &  1.04 , 0.04  &  7.11 , 0.09  &  11.2 , 0.1  &  8.50 , 0.10 \\
\hline
\end{tabular}
\end{table*}

\begin{table*}[h!tb]
\topcaption{\label{tab:unblindresults_T2bW}Predicted and observed data yields in the T2bW search regions. The uncertainties in the background predictions are the combined systematic and statistical uncertainties. The T2bW yields correspond to the simplified model points with $(\mStop, \mLSP; x) = (500\GeV, 175\GeV; 0.25)$ and $(600\GeV, 0\GeV; 0.75)$. The uncertainties in the signal yields are statistical only.}
\centering
\cmsTableResize{
\newcolumntype{x}{D{,}{\,\pm\,}{4.4}}
\begin{tabular}{lxxxxx}
\hline
& \multicolumn{5}{c}{Search region yield}\\ \cline{2-6} \noalign{\smallskip}
& \multicolumn{1}{c}{T2bW\_LX} & \multicolumn{1}{c}{T2bW\_LM} & \multicolumn{1}{c}{T2bW\_MXHM} & \multicolumn{1}{c}{T2bW\_HXHM} & \multicolumn{1}{c}{T2bW\_VHM}\\
\hline
$\ttbar$, $\PW$+jets, and single top  &  6.88,2.18 & 31.3,4.1 & 3.89,1.19 & 12.7,3.0 & 2.31,0.85  \\
$\Z$+jets                          &  1.88,0.93 & 4.57,1.67 & 1.66,0.72 & 1.77,0.73 & 1.24,0.54  \\
$\ttbar\Z$                         &  0.59,0.30 & 2.46,1.11 & 0.83,0.39 & 1.72,0.79 & 0.62,0.26  \\
QCD multijet                       &  0.71,0.35 & 0.36,0.19 & 0.10,0.12 & 0.01,0.01 & 0.01,0.01  \\[\cmsTabSkip]
{All SM backgrounds}        &  10.1,2.4   & 38.7,4.6   & 6.5,1.4   & 16.2,3.2   & 4.2,1.0    \\
{Observed data}             & \multicolumn{1}{c}{12}             &  \multicolumn{1}{c}{47}            &  \multicolumn{1}{c}{6}            &  \multicolumn{1}{c}{14}            &  \multicolumn{1}{c}{4}             \\
\hline
T2bW (500, 175; 0.25) &  13.8 , 1.1  &  3.49 , 0.58  &  6.70 , 0.76  &  3.12 , 0.54  &  1.36 , 0.33 \\
T2bW (600, 0; 0.75)   &  4.66 , 0.13  &  7.21 , 0.16  &  8.79 , 0.18  &  8.77 , 0.18  &  8.99 , 0.18 \\
\hline
\end{tabular}}
\end{table*}

The signal yields and their corresponding efficiencies are estimated by applying the event selection criteria to simulated data samples. Systematic uncertainties in the signal selection efficiencies are assessed as a function of the $\PSQt$ and $\PSGczDo$ masses, and as a function of the mass splitting parameter $x$ in the case of the T2bW signal. The uncertainty in the jet energy scale (JES) has the largest impact on signal yield, followed by the b tagging efficiency uncertainty. The uncertainty associated with the parton distribution functions is evaluated by following the recommendation of the PDF4LHC group~\cite{Botje:2011sn,Alekhin:2011sk,Lai:2010vv,Martin:2009iq,Ball:2012cx}. Uncertainties in the jet energy resolution, initial-state radiation, and integrated luminosity~\cite{CMS:2013gfa} are also included. For the T2tt channel, we assign three additional uncertainties. The first accounts for the difference observed in the performance of the \CORRAL algorithm between the standard CMS full and fast detector simulations. This difference decreases with increasing top quark $\pt$ and so depends on the difference between $\mStop$ and $\mLSP$, reaching 20\% for cases where $\mLSP$ is close to $\mStop$. The other two uncertainties each have a magnitude of 5\% and cover the differences observed in parton shower (PS) algorithms (\PYTHIA versus \HERWIG) and top quark reconstruction efficiencies in data versus simulation. Table~\ref{tab:signaluncertainty} lists the magnitude of each systematic uncertainty in signal points for which this search has sensitivity. For T2tt, the total systematic uncertainty is less than 15\% for $\mStop-\mLSP>300\GeV$.

\begin{table}[h!tb]
\centering
\topcaption{\label{tab:signaluncertainty}Summary of the systematic uncertainties in the signal selection efficiencies. The uncertainties can depend on signal topology, mass values, and search region. The quoted value ranges capture the variations associated with these dependencies. In all cases, the upper bound corresponds to the region in which $\mLSP$ is close to $\mStop$. }
\begin{tabular}{lc}
\hline
Systematics source & Magnitude [\%] \\  \hline
      b tagging       & 5--10 \\
      Jet energy scale             & 5--20 \\
      Jet energy resolution             & $<$5 \\
      Initial-state radiation     	      & 1--20 \\
      Parton distribution functions     	      & 1--15 \\
      Integrated luminosity	  & 2.6 \\
      \CORRAL FastSim (T2tt) & 1--20 \\
      \CORRAL dependence on PS (T2tt) & 5 \\
      \CORRAL reconstruction (T2tt) & 5 \\
\hline
    \end{tabular}

\end{table}

In the absence of any significant observed excesses of events over predicted backgrounds in the various search regions, the modified frequentist CL$_\mathrm{S}$ method~\cite{Junk:1999kv,Read:2002hq,LHC-HCG} with a one-sided profile likelihood ratio test statistic is used to define 95\% confidence level (CL) upper limits on the production cross section for both the T2tt and T2bW simplified models as a function of the masses of the SUSY particles involved. Statistical uncertainties related to the observed numbers of events are modelled as Poisson distributions. Systematic uncertainties in the background predictions and signal selection efficiencies are assumed to be multiplicative and are modelled with log-normal distributions.

\begin{figure*}[htb]
\centering
      \includegraphics[width=0.49\textwidth]{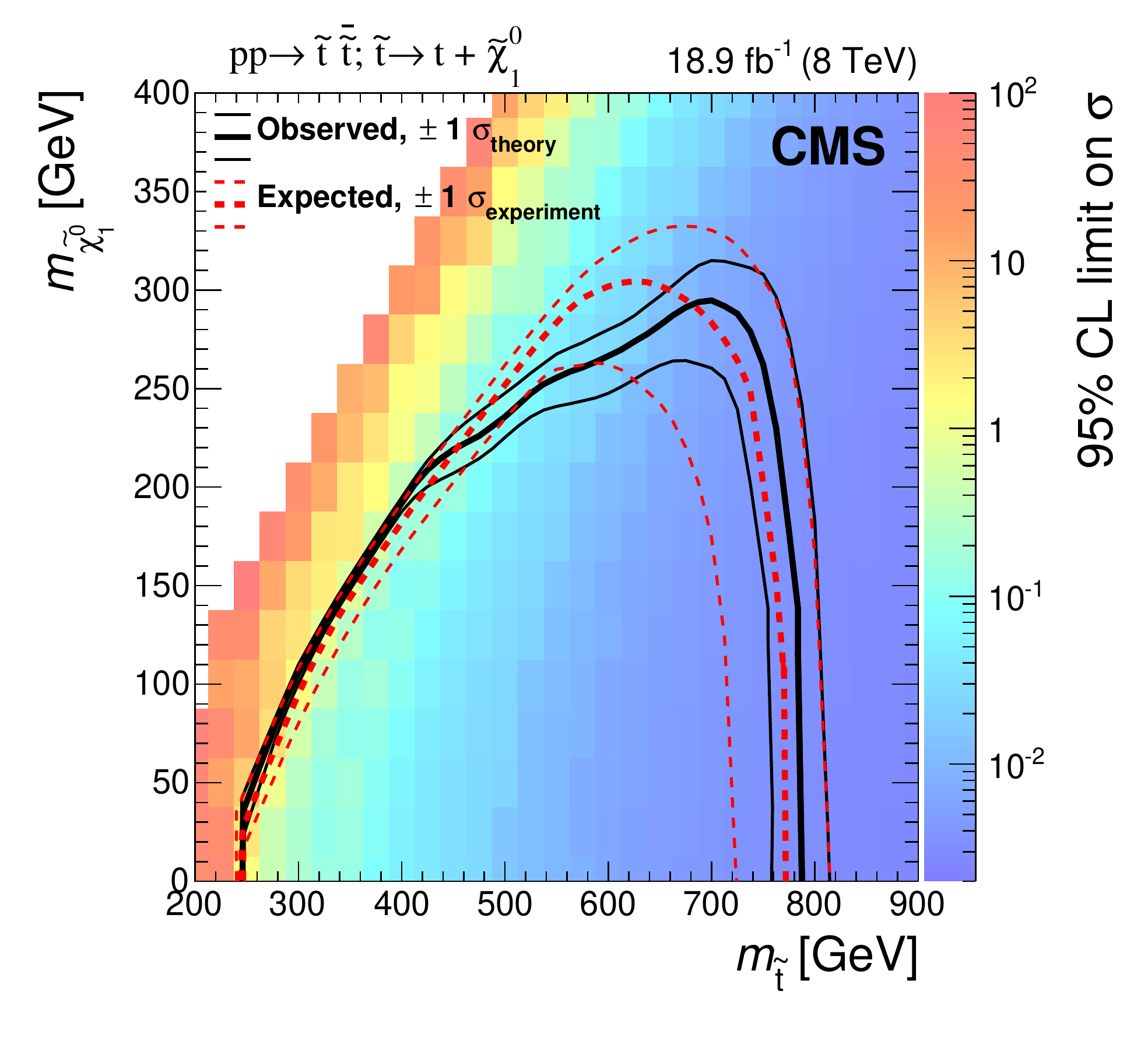}
      \includegraphics[width=0.49\textwidth]{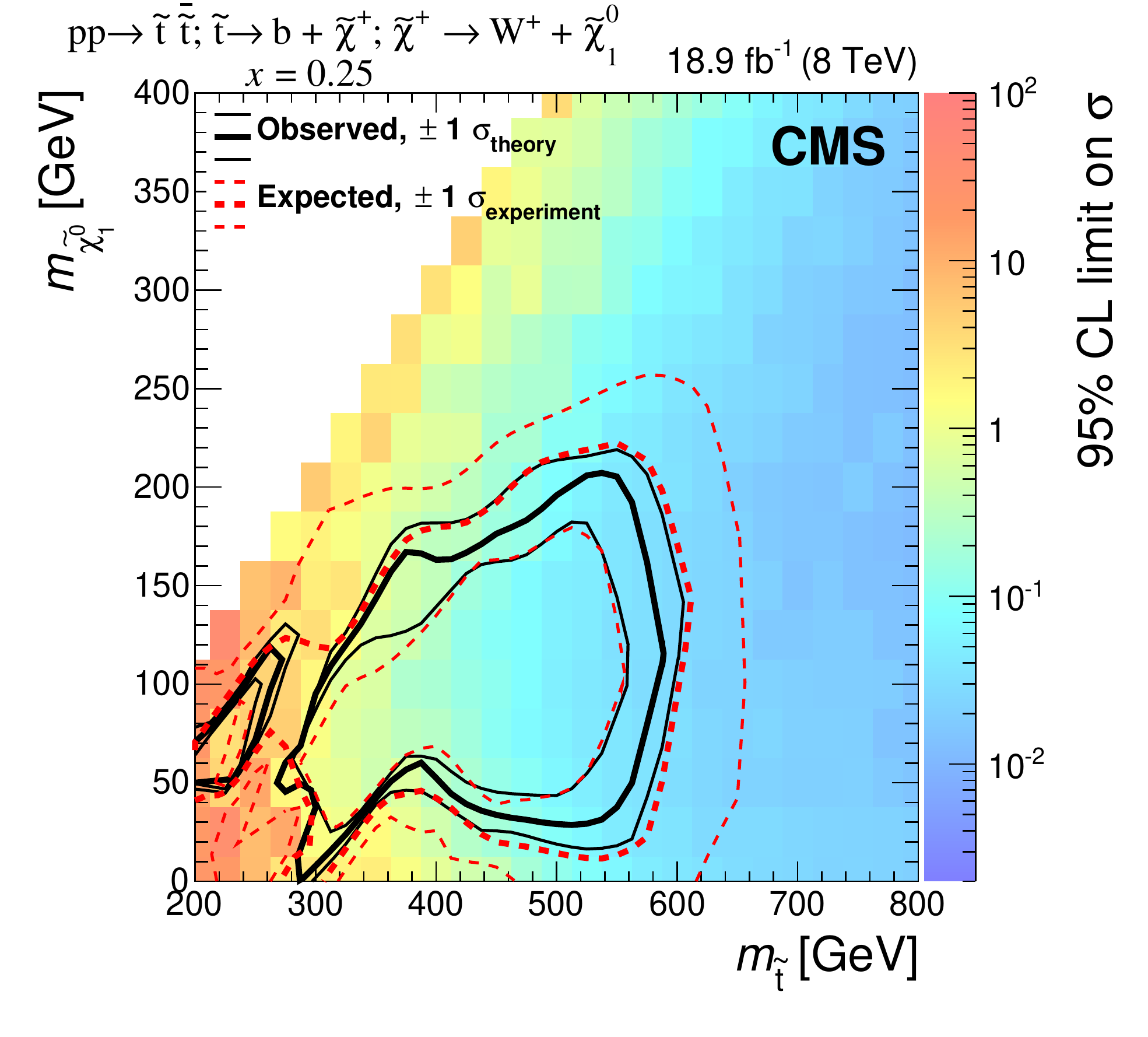}
      \includegraphics[width=0.49\textwidth]{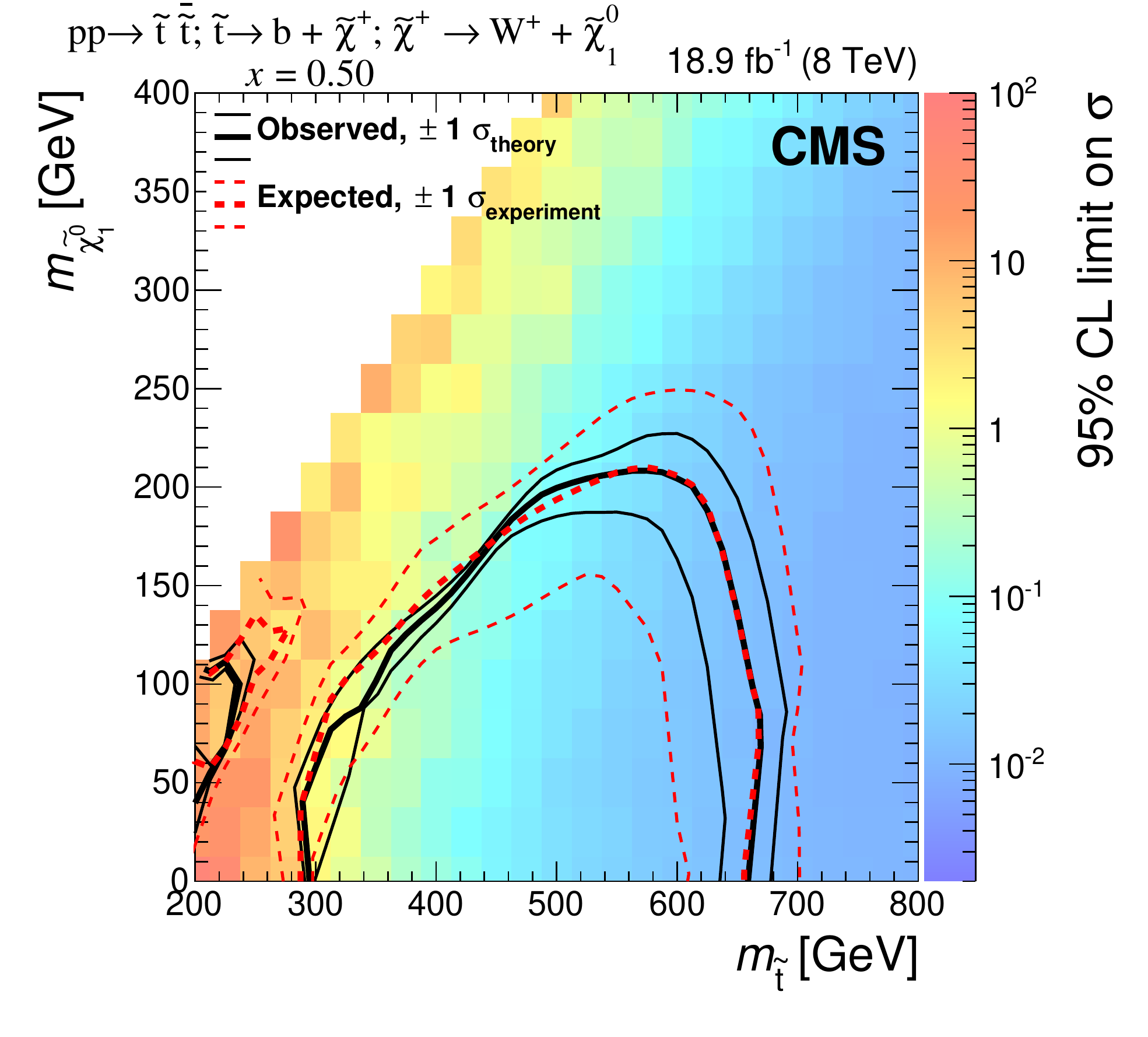}
      \includegraphics[width=0.49\textwidth]{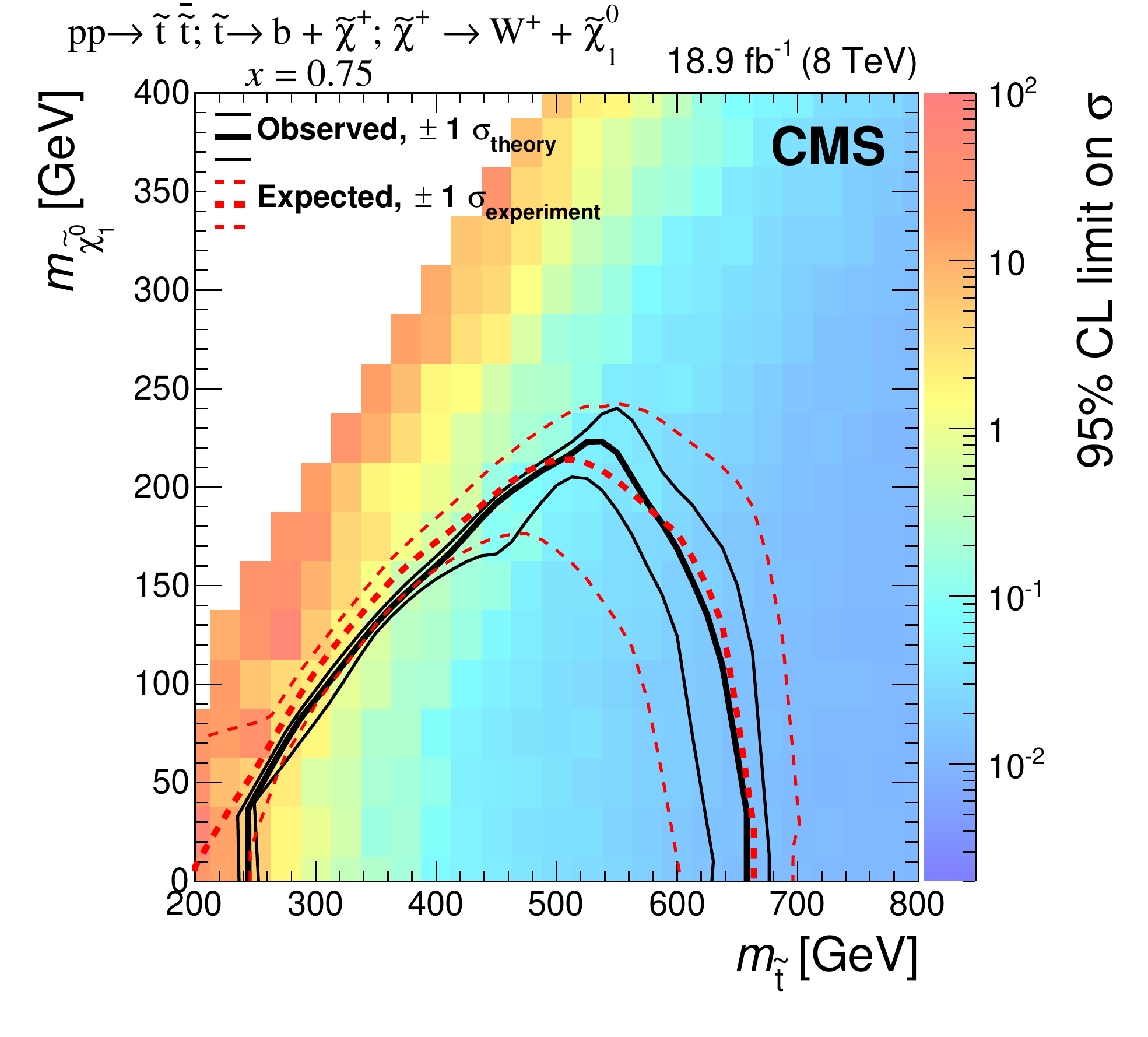}
  \caption{  \label{fig:expectedLimits}Observed and expected 95\% CL limits on the $\PSQt\PASQt$ production cross section and exclusion areas in the $\mStop-\mLSP$ plane for the T2tt (top left) and T2bW signal topologies (with  $x=0.25$, 0.50,  0.75). In the rare cases in which a statistical fluctuation leads to zero signal events for a particular set of masses, the limit is taken to be the average of the limits obtained for the neighboring bins. The $\pm1\sigma_{\text{theory}}$ lines indicate the variations in the excluded region due to the uncertainty in the theoretical prediction of the signal cross section.
}
\end{figure*}

For each choice of SUSY particle masses, the search region with the highest expected sensitivity (Fig.~\ref{fig:BestRegion}) is chosen to calculate an upper limit for the production cross section. The expected and observed upper limits in the production cross section for both the T2tt and T2bW topologies in the $\mStop-\mLSP$ plane are displayed in Fig.\ref{fig:expectedLimits}. For the T2tt topology this search is sensitive to models with $\mStop < 775\GeV$, or 755 $\GeV$ when conservatively subtracting one standard deviation of the theoretical uncertainty, and provides the most stringent limit to date for proton-proton collisions at $\sqrt{s}=8$\TeV on this simplified model for $\mStop > 600\GeV$. Sensitivity extends to models with $\mLSP < 290\GeV$ and this search is especially sensitive to the case of large $\mStop$ and low $\mLSP$ for which events typically have both large $\ETm$ and a high \CORRAL top pair reconstruction efficiency. In contrast, the analysis has no sensitivity to models with $\mStop - \mLSP < 200\GeV$ despite the large cross section of some signal scenarios.

This search is considerably less sensitive to the T2bW topology because that model does not feature on-shell top quark decays. The sensitivity in this case applies to scenarios with $\mStop < 650\GeV$, with the strongest results for large $x$ models for which $\mCHG$ is closer to $\mStop$ than $\mLSP$, resulting in a harder $\ETm$ spectrum. For scenarios with $x = 0.25$ the search has less sensitivity to models with $\mLSP \approx 0\GeV$ than to those with moderate $\mLSP$. In the former case the $\PSGcpm$ and W boson are close in mass and the signal has a low efficiency to pass the baseline selection's $\ETm$ criterion. The search also has less sensitivity to models with $\mLSP + m_{\PW} \approx \mCHG$ because in this scenario the signal has a low efficiency to pass the baseline selection's jet-multiplicity criterion.

\section{Summary}
\label{sec:conclusion}
We report a search for the direct pair production of top squarks in an all-hadronic final state containing jets and large missing transverse momentum. Two decay channels for the top squarks are considered. In the first channel, each top squark decays to a top quark and a neutralino, whereas in the second channel they each decay to a bottom quark and a chargino, with the chargino subsequently decaying to a $\PW$ boson and a neutralino. A dedicated top quark pair reconstruction algorithm provides efficient identification of hadronically decaying top quarks. The search is carried out in several search regions based on the output of multivariate discriminators, where the standard model background yield is estimated with corrected simulation samples and validated in data control regions. The observed yields are statistically compatible with the standard model  estimates and are used to restrict the allowed parameter space for these two signal topologies. The search is particularly sensitive to the production of top squarks that decay via an on-shell top quark. For models predicting such decays, a 95\% CL lower limit of 755\GeV is found for the top squark mass when the neutralino is lighter than 200\GeV, extending the current limits based on Run 1 searches at the LHC on these models by 50--100\GeV. In models with top squarks that decay via a chargino, scenarios with a top squark mass up to 620\GeV are excluded.

\begin{acknowledgments}
\hyphenation{Bundes-ministerium Forschungs-gemeinschaft Forschungs-zentren} We congratulate our colleagues in the CERN accelerator departments for the excellent performance of the LHC and thank the technical and administrative staffs at CERN and at other CMS institutes for their contributions to the success of the CMS effort. In addition, we gratefully acknowledge the computing centres and personnel of the Worldwide LHC Computing Grid for delivering so effectively the computing infrastructure essential to our analyses. Finally, we acknowledge the enduring support for the construction and operation of the LHC and the CMS detector provided by the following funding agencies: the Austrian Federal Ministry of Science, Research and Economy and the Austrian Science Fund; the Belgian Fonds de la Recherche Scientifique, and Fonds voor Wetenschappelijk Onderzoek; the Brazilian Funding Agencies (CNPq, CAPES, FAPERJ, and FAPESP); the Bulgarian Ministry of Education and Science; CERN; the Chinese Academy of Sciences, Ministry of Science and Technology, and National Natural Science Foundation of China; the Colombian Funding Agency (COLCIENCIAS); the Croatian Ministry of Science, Education and Sport, and the Croatian Science Foundation; the Research Promotion Foundation, Cyprus; the Ministry of Education and Research, Estonian Research Council via IUT23-4 and IUT23-6 and European Regional Development Fund, Estonia; the Academy of Finland, Finnish Ministry of Education and Culture, and Helsinki Institute of Physics; the Institut National de Physique Nucl\'eaire et de Physique des Particules~/~CNRS, and Commissariat \`a l'\'Energie Atomique et aux \'Energies Alternatives~/~CEA, France; the Bundesministerium f\"ur Bildung und Forschung, Deutsche Forschungsgemeinschaft, and Helmholtz-Gemeinschaft Deutscher Forschungszentren, Germany; the General Secretariat for Research and Technology, Greece; the National Scientific Research Foundation, and National Innovation Office, Hungary; the Department of Atomic Energy and the Department of Science and Technology, India; the Institute for Studies in Theoretical Physics and Mathematics, Iran; the Science Foundation, Ireland; the Istituto Nazionale di Fisica Nucleare, Italy; the Ministry of Science, ICT and Future Planning, and National Research Foundation (NRF), Republic of Korea; the Lithuanian Academy of Sciences; the Ministry of Education, and University of Malaya (Malaysia); the Mexican Funding Agencies (CINVESTAV, CONACYT, SEP, and UASLP-FAI); the Ministry of Business, Innovation and Employment, New Zealand; the Pakistan Atomic Energy Commission; the Ministry of Science and Higher Education and the National Science Centre, Poland; the Funda\c{c}\~ao para a Ci\^encia e a Tecnologia, Portugal; JINR, Dubna; the Ministry of Education and Science of the Russian Federation, the Federal Agency of Atomic Energy of the Russian Federation, Russian Academy of Sciences, and the Russian Foundation for Basic Research; the Ministry of Education, Science and Technological Development of Serbia; the Secretar\'{\i}a de Estado de Investigaci\'on, Desarrollo e Innovaci\'on and Programa Consolider-Ingenio 2010, Spain; the Swiss Funding Agencies (ETH Board, ETH Zurich, PSI, SNF, UniZH, Canton Zurich, and SER); the Ministry of Science and Technology, Taipei; the Thailand Center of Excellence in Physics, the Institute for the Promotion of Teaching Science and Technology of Thailand, Special Task Force for Activating Research and the National Science and Technology Development Agency of Thailand; the Scientific and Technical Research Council of Turkey, and Turkish Atomic Energy Authority; the National Academy of Sciences of Ukraine, and State Fund for Fundamental Researches, Ukraine; the Science and Technology Facilities Council, UK; the US Department of Energy, and the US National Science Foundation.

Individuals have received support from the Marie-Curie programme and the European Research Council and EPLANET (European Union); the Leventis Foundation; the A. P. Sloan Foundation; the Alexander von Humboldt Foundation; the Belgian Federal Science Policy Office; the Fonds pour la Formation \`a la Recherche dans l'Industrie et dans l'Agriculture (FRIA-Belgium); the Agentschap voor Innovatie door Wetenschap en Technologie (IWT-Belgium); the Ministry of Education, Youth and Sports (MEYS) of the Czech Republic; the Council of Science and Industrial Research, India; the HOMING PLUS programme of the Foundation for Polish Science, cofinanced from European Union, Regional Development Fund; the OPUS programme of the National Science Center (Poland); the Compagnia di San Paolo (Torino); MIUR project 20108T4XTM (Italy); the Thalis and Aristeia programmes cofinanced by EU-ESF and the Greek NSRF; the National Priorities Research Program by Qatar National Research Fund; the Rachadapisek Sompot Fund for Postdoctoral Fellowship, Chulalongkorn University (Thailand); the Chulalongkorn Academic into Its 2nd Century Project Advancement Project (Thailand); and the Welch Foundation, contract C-1845.
\end{acknowledgments}

\bibliography{auto_generated}   

\cleardoublepage \appendix\section{The CMS Collaboration \label{app:collab}}\begin{sloppypar}\hyphenpenalty=5000\widowpenalty=500\clubpenalty=5000\textbf{Yerevan Physics Institute,  Yerevan,  Armenia}\\*[0pt]
V.~Khachatryan, A.M.~Sirunyan, A.~Tumasyan
\vskip\cmsinstskip
\textbf{Institut f\"{u}r Hochenergiephysik der OeAW,  Wien,  Austria}\\*[0pt]
W.~Adam, E.~Asilar, T.~Bergauer, J.~Brandstetter, E.~Brondolin, M.~Dragicevic, J.~Er\"{o}, M.~Flechl, M.~Friedl, R.~Fr\"{u}hwirth\cmsAuthorMark{1}, V.M.~Ghete, C.~Hartl, N.~H\"{o}rmann, J.~Hrubec, M.~Jeitler\cmsAuthorMark{1}, V.~Kn\"{u}nz, A.~K\"{o}nig, M.~Krammer\cmsAuthorMark{1}, I.~Kr\"{a}tschmer, D.~Liko, T.~Matsushita, I.~Mikulec, D.~Rabady\cmsAuthorMark{2}, B.~Rahbaran, H.~Rohringer, J.~Schieck\cmsAuthorMark{1}, R.~Sch\"{o}fbeck, J.~Strauss, W.~Treberer-Treberspurg, W.~Waltenberger, C.-E.~Wulz\cmsAuthorMark{1}
% ...
\vskip\cmsinstskip
% ...
1:~~Also at Vienna University of Technology, Vienna, Austria\\
2:~~Also at CERN, European Organization for Nuclear Research, Geneva, Switzerland\\
% ...

\end{sloppypar}
\end{document}